\documentclass[prd, aps, nofootinbib, floatfix, superscriptaddress]{revtex4}
\usepackage{graphicx}
\usepackage{epsfig}
\usepackage{rotating}
\usepackage{amssymb}
\usepackage{subfigure}
\usepackage{dsfont}
\usepackage{psfrag}
\usepackage{amsmath,euscript,array,mathrsfs}
\usepackage{slashed}
\usepackage{array}
\usepackage{youngtab}
\usepackage{color}
\usepackage{bbold}
\usepackage{orcidlink}
\usepackage{hyperref}
\hypersetup{
colorlinks = true,
linkcolor = blue,
linktocpage = true,
citecolor = blue
}

\newcommand{\beq}{\begin{equation}}
\newcommand{\eeq}{\end{equation}}
\newcommand{\beqs}{\begin{eqnarray}}
\newcommand{\eeqs}{\end{eqnarray}}

\newcommand{\dd}{\mbox{d}}

\newcommand{\be}{\begin{equation}}
\newcommand{\ee}{\end{equation}}
\newcommand{\ba}{\begin{array}}
\newcommand{\ea}{\end{array}}

\newcommand{\orcidauthorPIAI}{0000-0002-2251-0111} 
\newcommand{\orcidauthorELANDER}{0000-0001-6348-8021}

\begin{document}

\title{Dilatonic states, phase transitions, and criticality in holography}

\author{Daniel Elander\,\orcidlink{\orcidauthorELANDER}}
\email{daniel.elander@gmail.com}
\affiliation{Porto, Portugal}

\author{Maurizio Piai\,\orcidlink{\orcidauthorPIAI}}
\email{m.piai@swansea.ac.uk}
\affiliation{Department of Physics, Faculty  of Science and Engineering, Swansea University, Singleton Park, SA2 8PP, Swansea, Wales, UK}
\affiliation{Centre for Quantum Fields and Gravity, Faculty of Science and Engineering, Swansea University, Singleton Park, Swansea, SA2 8PP, Wales, UK}

\date{\today}

\begin{abstract}
The dilaton is the hypothetical scalar particle associated with spontaneous breaking of approximate scale invariance. It has been predicted to arise dynamically, as a bound state, in special extensions of the standard model of particle physics based on composite dynamics. In confining gauge theories, generic arguments suggest that the scale of explicit symmetry breaking coincides with the confinement scale, that governs also its spontaneous breaking, by setting the size of all the condensates, as well as the masses of bound states and their excitations. Whether a light dilaton can nevertheless exist in such a context is an open question in quantum field theory, answering which has potentially transformative implications from a conceptual viewpoint, besides having striking phenomenological ramifications. It has been suggested that the mass of the dilaton might be suppressed, in respect to the confinement scale, in the special case in which the underlying strongly coupled dynamics undergoes a zero-temperature (weak) phase transition. Identifying the special conditions under which this happens is the subject of ongoing investigations. A dedicated programme of exploration, carried out within the context of gauge-gravity dualities (holography), has been set up to test the viability of this mechanism. We review the relevant technology within holography, and the status of such a programme, by comparing results obtained in a survey of explicit examples.  We do so both within the bottom-up, simplified version of gauge-gravity dualities, but also within the more refined, and complete, top-down approach to holography, in which the gravity theory is known to have a more fundamental origin. We highlight the first examples of confining theories in which a parametrically light holographic dilaton forms as a bound state in a region of parameter space in proximity to a critical point of the field theory.
\end{abstract}

\maketitle

\tableofcontents

\section{Introduction and motivation}

The dilaton is a spin-0, neutral, flavourless, parity and charge-conjugation even particle, that acts as the  pseudo-Nambu-Goldstone boson (PNGB) associated with the spontaneous breaking of (approximate) dilatation symmetry (or scale invariance). Its postulated existence as one of the long distance phenomena of special interacting quantum field theories (QFTs) in four space-time dimensions has a long history~\cite{Migdal:1982jp,Coleman:1985rnk}.  A number of striking implications of its emergence are independent of the microscopic dynamics, being governed by symmetry properties, which restrict its couplings. A good case in point is provided by the Standard Model (SM) of particle physics, in which one doublet of elementary Higgs fields is responsible for electroweak symmetry breaking (EWSB).  At the classical level, the only source of explicit symmetry breaking of scale invariance arises from the quadratic term in the Higgs potential itself, and quantum corrections are perturbatively small in the regime in which the Higgs particle is light.  Hence, the Higgs particle that originates from small fluctuations of the scalar fields around the minimum of the scalar potential can be identified as a dilaton, and its decay constant as the electroweak scale, $V_W\sim 246$ GeV. A consequence of the dilaton nature of the Higgs particle is that its couplings to SM fermions and gauge bosons, both those arising at tree and 1-loop level, are rigidly determined by scale invariance, and the size of its breaking effects, as explained for instance in Ref.~\cite{Goldberger:2007zk} and references therein.  This property underpins the huge predictive power of the Higgs sector of the standard model, that made it possible to tabulate the results of accurate calculations of production and decay rates (see, e.g., Ref.~\cite{Djouadi:2005gi} and references therein) before the first direct evidence of its existence came to be. This endeavour proved instrumental to the subsequent discovery of the Higgs boson at the Large Hadron Collider (LHC) experiments~\cite{ATLAS:2012yve,CMS:2012qbp}.

The relevance of the dilaton in models of new physics was discussed already in time-honoured proposals of dynamical electroweak symmetry breaking~\cite{Leung:1985sn,Bardeen:1985sm,Yamawaki:1985zg}. Interest on the dilaton as a feature of extensions of the standard model has undergone several periods of resurgence, in association with the periodical introduction of new ideas in the field, over the past five decades. For example, when the proposal of addressing the hierarchy problem with the Randall-Sundrum~\cite{Randall:1999ee} and Goldberger-Wise mechanisms~\cite{Goldberger:1999uk} was put forward, the realisation that such scenarios predict that the dilaton would emerge as a radion, the scalar related to the size of the warped extra dimension, spurred a number of interesting studies~\cite{DeWolfe:1999cp,Goldberger:1999un,Csaki:2000zn,Arkani-Hamed:2000ijo,Rattazzi:2000hs,Kofman:2004tk}. As a further example, Higgs searches at the LHC motivated the development of a whole body of work on phenomenological and field theoretical grounds~\cite{Hong:2004td,Dietrich:2005jn,Goldberger:2007zk,Vecchi:2010gj,Hashimoto:2010nw,Appelquist:2010gy,Chacko:2012sy,Bellazzini:2012vz,Abe:2012eu,Eichten:2012qb,Elander:2012fk,Hernandez-Leon:2017kea,Bellazzini:2013fga}.

The dilaton might arise in a number of possible scenarios of composite dynamics---see for example the reference lists in the review in Ref.~\cite{Cacciapaglia:2020kgq}. It might be one of the features of confining gauge theories with near-conformal dynamics, together with other properties important for  model-building,  such as the possible emergence of large anomalous dimensions for some of the effective operators controlling low-energy phenomena~\cite{Cohen:1988sq}. These theories have field content chosen to approach (from below) the lower edge of the conformal window. Their study requires to extend the analysis by Caswell, Banks and Zaks~\cite{Caswell:1974gg,Banks:1981nn} beyond perturbation theory.  The non-perturbative nature of the associated phenomena make them challenging to test with traditional field theory techniques~\cite{Holdom:1986ub,Holdom:1987yu,Appelquist:2010gy,Grinstein:2011dq}, though encouraging results suggest they might be realised in realistic settings---see for example  Refs.~\cite{Sannino:2004qp,Dietrich:2006cm,Ryttov:2007cx,Pica:2010mt,Pica:2010xq,Kim:2020yvr,Lee:2020ihn} as well as Refs.~\cite{Ryttov:2016ner,Ryttov:2016hdp,Ryttov:2016asb,Ryttov:2016hal,Ryttov:2017toz,Ryttov:2017kmx,Ryttov:2017dhd,Gracey:2018oym,Ryttov:2018uue,Ryttov:2020scx,Ryttov:2023uzc}, which build on the multi-loop calculations reported in Refs.~\cite{Chetyrkin:1997dh,Vermaseren:1997fq,Baikov:2016tgj,Herzog:2017ohr}.

In order to gain control of  non-perturbative effects, a natural choice  of investigative tools for  strongly coupled systems involves the use of the numerical techniques of  lattice field theory---see for example the review in Ref.~\cite{Rummukainen:2022ekh} and references therein. The possibility that a light dilaton be present in special cases has been actively investigated, in a number of candidate theories. These include the $SU(3)$ theory coupled to  $N_{\rm f}=8$ fermions transforming in the fundamental representation~\cite{LatKMI:2014xoh,Appelquist:2016viq,LatKMI:2016xxi,Gasbarro:2017fmi,LatticeStrongDynamics:2018hun,LatticeStrongDynamicsLSD:2021gmp,Hasenfratz:2022qan,LatticeStrongDynamics:2023bqp,LSD:2023uzj,LatKMI:2025kti}, or other choices of $N_{\rm f}$~\cite{Kotov:2021mgp,Chung:2023mgr,Nogradi:2023wnf}, the $SU(3)$ theory with  $N_{\rm s}=2$ fermions in the two-index symmetric representation~\cite{Fodor:2012ty,	Fodor:2015vwa,	Fodor:2016pls,Fodor:2017nlp,Fodor:2019vmw,Fodor:2020niv}, the $SU(2)$ theory coupled to $N_{\rm Adj}=1,\,2$ fermions in the adjoint representation~\cite{Athenodorou:2014eua,Athenodorou:2016ndx,Athenodorou:2021wom,Athenodorou:2024rba}, the 	$Sp(4)$ theory coupled to field content consisting of real or pseudoreal fermions~\cite{Bennett:2023gbe,Bennett:2024tex}. 

Some of these calculations provide indications of a light scalar singlet, which in turn stimulated renewed interest in dilaton effective field theory (dEFT), the long-distance description combining the dilaton with the PNGBs associated with internal global symmetries~\cite{Matsuzaki:2013eva,Li:2016uzn,Golterman:2016lsd,Kasai:2016ifi,Hansen:2016fri,Golterman:2016cdd,Appelquist:2017wcg,Appelquist:2017vyy,Cata:2018wzl,Golterman:2018mfm,Cata:2019edh,Cata:2019edh,Appelquist:2019lgk,Golterman:2020tdq,Golterman:2020utm,Appelquist:2020bqj,Appelquist:2022qgl,Zwicky:2023fay,Appelquist:2025tol}---see also Ref.~\cite{Appelquist:2022mjb} and references therein. This development further provided the background for new phenomenological  applications~\cite{Appelquist:2020bqj,Appelquist:2022qgl,Cacciapaglia:2023kat,Appelquist:2024koa,Bruggisser:2022ofg,Alonso:2025ksv,Wu:2025hfp,Cao:2026htw} and new field-theory motivated investigations~\cite{DelDebbio:2021xwu,Zwicky:2023bzk,Zwicky:2023krx,Cresswell-Hogg:2025kvr,Stegeman:2025sca,Stegeman:2025tdl}.
	
The programmatic goal of dEFT is to provide a calculable framework for the long distance properties of the dilaton in combination with other PNGBs, based only on symmetry principles, and without direct reference to its microscopic origin. It aim at doing so by emulating and generalising the  construction of the chiral Lagrangian~\cite{Weinberg:1966kf,Weinberg:1968de,Dashen:1969eg,Dashen:1969ez,Dashen:1970et,Pagels:1974se,Gasser:1982ap,Gasser:1983kx,Gasser:1983yg} (or the electroweak chiral Lagrangian~\cite{Appelquist:1980vg,Longhitano:1980iz,Appelquist:1993ka,Appelquist:1994qz}). The powerful tools of the chiral Lagrangian descend from the existence of a limit of  enhanced symmetry, in which one recovers a simple, yet interacting, weakly coupled field theory. If chiral symmetry is non-linearly realised, but exact, all the properties of the pions, the ${\cal O}(N_f^2)$ PNGBs described by the chiral Lagrangian, depend on the decay constant, $f_{\pi}$, and on an expansion in  derivatives, suppressed by appropriate powers of the cutoff, which is usually estimated to be $\Lambda^2\sim {\cal O}\left((4\pi f_{\pi})^2/N_f\right)$. As they are all proportional to powers of the momentum, such interactions are suppressed, in the long distance regime. Away from this symmetric point, additional interactions have to be added, that are controlled by small parameters,  reflecting the smallness of the symmetry breaking effects. 
	
If calculability in strongly coupled theories is a challenge, because of the non-perturbative dynamics, even the weakly coupled description of the dilaton that originates from it, which is provided by its long-distance description in dEFT, is more subtle than in other effective field theories (EFTs), and has its intrinsic limitations.  In contrast with the chiral Lagrangian, in dEFT even in the limit of exact dilatation invariance, besides the derivative couplings controlled by the decay constant of the PNGBs, $F_{\pi}$, and of the dilaton, $F_{\rm d}$, 
 there is also a potential term, $V(\chi)\propto \lambda \chi^4$, which does not violate any symmetry and involves no derivatives of the dilaton field, $\chi$. The natural size of this coupling can be estimated, using naive dimensional analysis, to be  $\lambda\sim {\cal O}\left(\Lambda^2/F_d^2\right) \sim  {\cal O}\left({(4\pi)^2}/{N_f^2}\right)$~\cite{Georgi:1992dw,Soldate:1989fh} (see also Ref.~\cite{Bijnens:2009qm}). Because this coupling is not protected (suppressed) by any considerations that are part of the dEFT construction itself, this is problematic, and requires tuning it to unnaturally small values. Depending on the context, the problems that arise from this observation can be more or less severe,  go under a plethora of different names, and can be articulated in a more precise way in several equivalent forms---see Ref.~\cite{Appelquist:2022mjb} and references therein for an example of such systematic analysis, based on introducing a spurion and using  power-counting rules along the lines of naive dimensional analysis (NDA)~\cite{Georgi:1992dw}. One extreme way of putting it (which we advise the reader to take with a grain of salt), is that the natural range of parameters suggested by NDA for the validity of dEFT as a weakly-coupled low-energy effective theory description of the behaviour of a light dilaton is, somewhat paradoxically, the one in which the dilaton is not light, and its self-couplings are not weak---not by chance, this observation is reminiscent of the hierarchy problem of the standard model.

The somewhat discouraging observation that closes the previous paragraph represents a generic, oversimplified expectation, rather than a rule. The underlying dynamics leading to a light dilaton is special, not generic. The major discovery within high energy physics and taking place in this millennium is the detection the Higgs boson, which can be interpreted as a dilaton, and is anticipated to be the main subject of precision experiments at colliders for the next two decades. It is hence of primary importance to understand whether it must have a weakly coupled origin, as in the minimal version of the standard model, or whether it can arise also as a composite state from new physics, and how to distinguish the two. These objectives have a special role in modern particle physics, and, as we explained, require overcoming two intertwined difficulties. One needs a computational tool appropriate to study non-perturbative effects in the strongly coupled microscopic theory. And one needs to identify a physical mechanism that stabilises (protects) the light mass of the dilaton in the regime of validity of the long-distance effective field theory.

This review is devoted to one possible way to address this twofold challenge. In order to provide calculability, we consider gauge-gravity dualities (holography)~\cite{Maldacena:1997re,Gubser:1998bc,Witten:1998qj,Aharony:1999ti}. We postulate the equivalence between observable quantities computed in a gauge theory of interest and in its corresponding gravity theory, that lives in a space-time with a higher number of dimensions. The effectiveness of this approach as an investigative tool descends from the fact that the dynamical regime in which the gauge theory is strongly coupled can be captured by the weakly coupled regime of the gravity theory.\footnote{Throughout this review, all the calculations we describe have been performed in the strict limit in which the classical gravity approximation holds, and tree-level effects provide a good description of the physics. We assume that the dynamical regime of interest is one in which corrections due to quantum gravity loops, or to the extended-object nature of the more fundamental theory of gravity, are suppressed enough that they can be ignored. In the original formulation of gauge-gravity dualities relating ten-dimensional type-IIB string theory living on a manifold with AdS$_5\times S^5$ geometry, and ${\cal N}=4$ Super-Yang-Mills theories in four dimensions,  with gauge group $SU(N_c)$, this dynamical regime of the gravity theory corresponds to the field-theory one in which both the 't Hooft coupling, $g_{\rm YM}^2 N_c$, and the number of colours, $N_c$, are large~\cite{Maldacena:1997re,Gubser:1998bc,Witten:1998qj,Aharony:1999ti}.  } Hence, perturbative techniques can be adapted to work with the gravity theory, in order to extract non-perturbative information about the field theory.

Holography finds a precise and prescriptive realisation, depending on the theory of interest, in the central concept of the dictionary of the correspondence. For example, holographic renormalisation~\cite{Bianchi:2001kw,Skenderis:2002wp,Papadimitriou:2004ap} is used to compute the free energy of a QFT in the presence of sources (control parameters), and the corresponding condensates (response functions) involving local field theory operators---though there exist limitations and subtleties, as exposed for instance in Ref.~\cite{Bobev:2013cja}. Non-local QFT operators can be studied by embedding extended objects into the gravity theory. For example, one can compute the expectation value of the Wilson loop~\cite{Maldacena:1998im,Rey:1998ik}, by introducing open strings and analysing the associated Nambu-Goto action. Where appropriate, the results yield linear confinement---see for example Refs.~\cite{Rey:1998bq,Brandhuber:1998bs,Brandhuber:1998er,Gross:1998gk,Brandhuber:1999jr,Nunez:2009da,Faedo:2013ota}. Spectra of bound states of the field theory (e.g., glueballs) can be computed systematically, by studying the behaviour of gauge-invariant combinations of the fluctuations of the classical fields of the gravity theory~\cite{Bianchi:2003ug,Berg:2005pd,Berg:2006xy,Elander:2009bm, Elander:2010wd,Elander:2010wn, Elander:2014ola,Elander:2018aub,Elander:2020csd}, that makes use of the Arnowitt-Deser-Misner (ADM) formalism ~\cite{Arnowitt:1959ah,Arnowitt:1962hi}. 

The tools provided by gauge-gravity dualities can be applied to a variety of physical problems, provided the gravity theory identified as the dual to the QFT of interest meets some special restrictions. Among other requirements, the gravity backgrounds must be regular, and have small curvature everywhere. For example, the physics of confinement originates in the gravity dual by the shrinking of a sub-manifold of the space-time, which can be a circle in the simplest models~\cite{Witten:1998zw,Wen:2004qh,Kuperstein:2004yf,Brower:2000rp,Elander:2013jqa}, or, in more sophisticated backgrounds, a 2-cycle ~\cite{Chamseddine:1997nm,Klebanov:2000hb,Maldacena:2000yy,Butti:2004pk,Dymarsky:2005xt,Andrews:2006aw,Hoyos-Badajoz:2008znk,Nunez:2008wi,Elander:2009pk,Cassani:2010na,Bena:2010pr,Bennett:2011va,Dymarsky:2011ve,Maldacena:2009mw,Gaillard:2010qg,Caceres:2011zn,Elander:2011mh,Elander:2012yh,Elander:2017hyr,Elander:2017cle},  contained within the base of the conifold~\cite{Candelas:1989js,Klebanov:1998hh,Klebanov:2000nc,Papadopoulos:2000gj}.  The breaking of chiral symmetry and other global symmetry breaking phenomena can be discussed by including D$p$-branes in the gravity theory~\cite{Karch:2002sh,Babington:2003vm,Kruczenski:2003be,Nunez:2003cf,Sakai:2004cn,Sakai:2005yt,Erdmenger:2007cm}, and studying their non-trivial embedding by minimising the DBI action~\cite{Leigh:1989jq}. By combining such techniques, the deconfinement phase transition and the out-of-equilibrium dynamics in its proximity can be characterised quantitatively~\cite{Bigazzi:2020phm,Ares:2020lbt,Bea:2021zsu, Bigazzi:2021ucw,Henriksson:2021zei,Ares:2021ntv,Ares:2021nap,Morgante:2022zvc,Bea:2021zol, Bea:2022mfb,Bea:2024xgv,Bea:2024bxu,Bea:2024bls,Casalderrey-Solana:2025cdy,Hoyos:2026eoq}. This is the case both at finite temperature, but also in the presence of a chemical potential, which requires sourcing a non-trivial profile for one of the gauge fields in the gravity theory---see for example Refs.~\cite{Chamblin:1999tk,Chamblin:1999hg,Gubser:1998jb,Cai:1998ji,Cvetic:1999ne,Cvetic:1999rb,Kim:2006gp,Horigome:2006xu,Kobayashi:2006sb,Mateos:2007vc,Nakamura:2006xk, Karch:2007pd}, as well as the more recent Ref.~\cite{Casalderrey-Solana:2011dxg}.

As for the second part of the problem, the stabilisation mechanism that renders the mass of the dilaton light, a number of ideas have been explored with gauge-gravity dualities, and it is customary to refer to the resulting light scalar particle as a holographic dilaton. We mention only briefly in this introduction the existence of a substantial body of work on the holographic dilaton, both  within the bottom-up~\cite{Elander:2011aa,Kutasov:2012uq,Lawrance:2012cg,Evans:2013vca,Hoyos:2013gma,Megias:2014iwa,Elander:2015asa,Megias:2015qqh,Athenodorou:2016ndx,Pomarol:2019aae,CruzRojas:2023jhw,Pomarol:2023xcc,CruzRojas:2025qcj} and the top-down~\cite{Elander:2009pk,Elander:2012yh,Elander:2013jqa,Elander:2017cle,Elander:2017hyr,Elander:2018aub,Elander:2018gte,Roughley:2021suu} approaches to holography---we return later to the characterisation of these two approaches.  The formation of large condensates for some of the operators of the gauge theory is one of the mechanisms  tested in this body of work.  We refer the reader to the literature also for  the idea that spontaneous breaking of approximate scale invariance might be related to the merging of fixed points of the renormalisation group (RG) flow~\cite{Kaplan:2009kr,Jensen:2010ga,Gorbenko:2018ncu,Gorbenko:2018dtm,Bea:2018whf,Faedo:2019nxw,Bea:2022mfb,Bea:2020ees,Bea:2021ieq}, for example in theories lying in the proximity of the Breitenlohner-Freedman (BF) bound~\cite{Breitenlohner:1982jf}.

This review does not cover all such promising ideas, even within holography. We focus our attention on one, narrowly defined, elegant dynamical mechanism to control the mass of the dilaton. We discuss only the case in which holography is used to describe field theories  that undergo a zero-temperature deconfinement phase transition for some value of  their control parameters (couplings, or sources). If the transition is of second order, or very weak, then in the region of parameter space in its proximity one expects the dilaton to appear and to be light, compared to the dynamically generated scale of confinement.
We summarise and discuss the results of testing these ideas, that have been published in Refs.~\cite{Elander:2020ial,Elander:2020fmv,Elander:2021wkc,Elander:2022ebt,Fatemiabhari:2024lct,Faedo:2024zib,Elander:2025fpk,Piai:2026rst,Fatemiabhari:2026rju}.

The ideas and techniques we present in this review can be applied both in the top-down as in the bottom-up approach to holography. The flexibility of the latter, in which simple sigma-models coupled to gravity are proposed, on the basis of phenomenological arguments, allows to collect useful information and test technical developments in a comparatively simple environment. Yet, in order to make direct reference to predictive field theories, and correlate observables measured at different scales, the former approach is our framework of choice. It requires embedding the gravity theories into known supergravities, that can be realised as appropriate limits of candidates for quantum gravity theories, and hence admit an equally complete interpretation in dual field theory terms.  In the body of this review, we exploit the vast body of knowledge cumulated in the literature about classical supergravity theories in various dimensions and their intercorrelation, referring the reader to more pedagogical presentations, for example those in Refs~\cite{Samtleben:2008pe,Freedman:2012zz}, for further details. Four broad classes of classical supergravity actions play a central, though non-exclusive, role in this research programme.

\begin{itemize}

\item The maximal gauged supergravity in $\hat{D}=7$ dimensions has $SO(5)$ gauge symmetry, and its lift to maximal supergravity in eleven dimensions involves a four-sphere, $S^4$~\cite{Pilch:1984xy,Pernici:1984xx,Pernici:1984zw,Nastase:1999cb,Cvetic:1999xp,Lu:1999bc,Cvetic:2000ah,Campos:2000yu,Samtleben:2005bp}. Notably, this theory provides the context in which the proposal by Witten for the holographic  dual description of confinement in gauge theories has its natural realisation~\cite{Witten:1998zw}. It is also the basis for the celebrated Sakai-Sugimoto modelling of chiral symmetry breaking~\cite{Sakai:2004cn,Sakai:2005yt}.

\item Romans half-maximal, ${\cal N}=(2,2)$,  supergravity theory in $\hat{D}=6$ dimensions has $F(4)$ superalgebra, and $SU(2)$ gauge symmetry. It lifts to massive type-IIA supergravity in ten dimensions~\cite{Romans:1985tw,DeWitt:1981wm,Giani:1984dw,Romans:1985tz}, and the internal space involves a  portion of $S^4$~\cite{Ferrara:1998gv,Brandhuber:1999np}. This is a remarkably simple gravity theory. In addition, its sigma-model coset and  uplifts to higher dimensional theories can be generalised in an equally remarkable plethora of different ways---see for instance Refs.~\cite{
Brandhuber:1999np,Cvetic:1999un,Legramandi:2021aqv,Hong:2018amk,Jeong:2013jfc,DAuria:2000afl,Andrianopoli:2001rs,
Nishimura:2000wj,Gursoy:2002tx,Nunez:2001pt,Karndumri:2012vh,Lozano:2012au,Karndumri:2014lba,
Chang:2017mxc,Gutperle:2018axv,Suh:2018tul,Suh:2018szn,Kim:2019fsg,Chen:2019qib} and references therein.

\item The maximal  supergravity theory in $\hat{D}=5$ dimensions~\cite{Pernici:1985ju,Gunaydin:1984qu,
 Gunaydin:1985cu} has $SO(6)\sim SU(4)$ gauge symmetry. It lifts to ten dimensions, the internal space being related to $S^5$~\cite{Gunaydin:1984fk,Kim:1985ez,Lee:2014mla,Baguet:2015sma,Hohm:2013vpa,Baguet:2015xha}. Its background solutions include the first example of gauge-gravity duality discovered by Maldacena~\cite{Maldacena:1997re,Aharony:1999ti}. Generalisations of such background provide a rich environment in which to test highly non-trivial ideas---see for instance Refs.~\cite{Distler:1998gb,Cvetic:2000nc,Bakas:1999ax}, 
 and, for the  Coulomb branch which will be mentioned later in the paper, Refs.~\cite{Freedman:1999gk,Kraus:1998hv,Brandhuber:1999jr,Cvetic:1999xx,Hernandez:2005xd}, as well as  Ref.~\cite{Pilch:2000ue}. Interesting work can also be found in Refs.~\cite{Pilch:2000fu,Khavaev:1998fb,Freedman:1999gp}, on the truncations of this maximal supergravity,  and in Ref.~\cite{
 Brandhuber:1999hb,Bianchi:2000sm,Brandhuber:2000fr,
 Bianchi:2001kw,Papadimitriou:2004rz}, in reference to the spectra of fluctuations of the theory.

\item The maximal supergravity in  $\hat{D}=4$ dimensions has $SO(8)$ gauge symmetry, and can be obtained from the  reduction on $S^7$ of maximal supergravity in eleven dimensions~\cite{deWit:1982bul,deWit:1986oxb,deWit:1981sst,deWit:1984nz,Nicolai:1985hs,Bobev:2009ms,Bobev:2010ib,Biran:1982eg,Nicolai:2011cy,deWit:2007kvg,Hull:1984wa,Pope:1984jj,Pope:1984bd,Englert:1983qe,Page:1984fu,Awada:1982pk,Englert:1982vs,Bremer:1998zp}---see also more recent applications to condensed matter physics~\cite{Gauntlett:2009bh,Gauntlett:2009zw}. One special property of this theory is its rich phase space, comprising of over one hundred distinct critical points, which can be explored analytically, but are also the subject of  machine-learning exercises~\cite{Comsa:2019rcz}.

\end{itemize}

In the context of classical supergravity, a rich literature exists on the development, study, and application of solution generating techniques, intended to overcome the technical challenges presented by the search for regular background solutions to the relevant systems of coupled, non-linear differential equations. This subject in itself would be broad and interesting enough to deserve its own dedicated review. We make extensive use of one such tool, by borrowing the ideas proposed in Ref.~\cite{Anabalon:2021tua} and exploited in Refs.~\cite{Nunez:2023xgl, Nunez:2023nnl,Fatemiabhari:2024aua,Chatzis:2024top,Giliberti:2024eii,Chatzis:2025wfv,Chatzis:2024kdu,Kumar:2024pcz,Macpherson:2024qfi,Chatzis:2025dnu,Chatzis:2025hek,Anabalon:2025sok,Macpherson:2025pqi,Fatemiabhari:2025usn,Fatemiabhari:2026goj,Anabalon:2026yxk}. The mechanism forming its basis is a generalisation of the Melvin flux-tube solutions~\cite{Melvin:1963qx} studied in Refs.~\cite{Astorino:2012zm,Lim:2018vbq,Kastor:2020wsm}---see also the discussions of the field theory interpretation in Refs.~\cite{Kumar:2024pcz,Castellani:2024ial}, and the precursor in Ref.~\cite{Cassani:2021fyv}. As we shall see, a number of non-trivial classes of background solutions exhibiting zero-temperature phase transitions can be generated this way, yielding a fertile terrain to test QFT ideas, including the mechanism for dilaton stabilisation of interest to this review. 

As anticipated, holography is most effective as a tool when the classical supergravity theory is weakly coupled. In this case, the background equations of all the aforementioned classes of supergravities, and their truncations, descend from a bosonic  sigma-model in which a number of scalars parametrise a (non-compact)  coset, $G/H$, coupled to gravity and to a gauge theory. In some cases, it is useful to dispense with the underpinning supergravity origin of such theory, and the technical complications of the sigma-model dynamics, and instead write a simple model, built on the basis of general phenomenological considerations. By doing so, one avoids the painstaking difficulty of finding regular background solutions of a complicated system of coupled non-linear differential equations, which are replaced by systems that are more readily solvable. The advantage of this bottom-up approach to holography is that it provides a simple and effective way of testing new ideas. A wealth of information  has been collected in the literature in this way. Beside the aforementioned Refs.~\cite{Randall:1999ee,Goldberger:1999uk,DeWolfe:1999cp,Goldberger:1999un,Csaki:2000zn,Arkani-Hamed:2000ijo,Rattazzi:2000hs,Kofman:2004tk}, we quote also the applications to composite Higgs models (CHMs), for example those based on the $SO(5)/SO(4)$ coset~\cite{Contino:2003ve, Agashe:2004rs,Agashe:2005dk,Agashe:2006at,Contino:2006qr,Falkowski:2008fz,Contino:2010rs,Contino:2011np}, and the more recent holographic studies of the dual of gauge theories as the completion of  composite Higgs/composite dilaton physics, in Refs.~\cite{Erdmenger:2020lvq,Erdmenger:2020flu,Elander:2020nyd,Elander:2021bmt,Erdmenger:2023hkl,Elander:2023aow,Elander:2024lir,Erdmenger:2024dxf}. Intrinsically, this framework is intended as a testing ground for new ideas and their phenomenological impact, deferring to the future the search for genuine microscopic realisations of such ideas, in the top-down context---for comparison, embedding the $SO(5)/SO(4)$ coset relevant to minimal CHMs and studying vacuum misalignment within the top-down context is highly challenging, even in the simplest examples, as exposed in Ref.~\cite{Elander:2021kxk}.

The core of the research we summarise in this review focuses on characterising the relation between the existence of zero-temperature phase transitions in strongly coupled, confining field theories, and the emergence of a light dilaton in their spectrum of bound states, over a relevant portion of parameter space.   Existing literature reports evidence of several distinct behaviours, that can be broadly classified according to three main groups, depending on the strength of the phase transition and the resulting mass of the dilaton.

\begin{itemize}

\item In a number of examples of first-order phase transitions, a parametrically light dilaton appears in the spectrum of fluctuations, but only along metastable branches of the gravity solution, that are not physically realised as stable equilibrium states in the dual field theories, which is demonstrated by the calculation of the free energy as a function of the control parameters. This happens in the top-down constructions exhibited in Refs.~\cite{Elander:2020ial,Elander:2020fmv,Elander:2021wkc}, and the bottom-up models of Refs.~\cite{Elander:2022ebt}. 

\item A second class of theories exhibit a more nuanced relation between the characteristics of the phase transition and the dilaton. Special cases~\cite{Fatemiabhari:2024lct,Piai:2026rst,Fatemiabhari:2026rju} show the emergence of a light dilaton also in some of the physically realised portion of parameter space in proximity of a first-order phase transition. 

\item In the few known examples in which a line of first-order transitions ends at a critical point, in proximity of which the transition is weak (close to second order), a physically realised, parametrically light dilaton is part of the QFT spectrum of bound states. This possibility has been demonstrated both in the bottom-up~\cite{Faedo:2024zib}, and, more convincingly, in the top-down holographic contexts~\cite{Elander:2025fpk}.\footnote{For complementary approaches to a similar problem, see also the field theory study of special super-renormalisable lower-dimensional field theories exposed in Ref.~\cite{Cresswell-Hogg:2025kvr}, and the lattice field theory studies reported in Refs.~\cite{Lucini:2013wsa,Bennett:2022yfa}.}

\end{itemize}

This paper is organised as follows.  In Sect.~\ref{Sec:holography}, we provide a pedagogical introduction to a selection of fundamental concepts necessary to understand the context of this review. We complement them with technical details, specifically needed for the calculations we discuss later in the paper, in Sect.~\ref{Sec:moreholography}. These two initial sections are intended for readers who may not have previous operational experience with holography, yet we do not intend them to provide comprehensive overviews of its foundations and developments, for which we refer the reader elsewhere in the literature (e.g., to Ref.~\cite{Aharony:1999ti}). We take a rather pragmatic, operational attitude in the exposition, focusing on the process used in calculations, more than its conceptual origin. For example, in deriving actions and backgrounds of relevance, we do not mention fundamental concepts such as conformal symmetry, supersymmetry, string theory, D$p$-branes, the large-$N$ expansion, the 't Hooft coupling, and the like. While all such concepts are essential in order to understand the deeper origin of gauge-gravity dualities, the language and prescriptions we use in actual calculations are of more general applicability, and ultimately rely only on the basic rules of classical field theory and general relativity, for reasons that will be apparent in the body of the paper. We then dive into the core of the review, and describe the pertinent top-down theories in Sect.~\ref{Sec:Top}, as well as the related bottom-up models in Sect.~\ref{Sec:Bottom}. In all cases, we provide only a schematic description of the classical action and classical backgrounds characterising the gravity theory. We then summarise the main results of global and local stability analyses carried out by computing free energy and spectrum of bound states of the dual field theory. In exposing such results, we dispense with any additional details and their field-theoretical microscopic interpretations, which can be found elsewhere and do not play a central role in the narrative we are developing in these few pages. Of course, this does not mean such aspects are less important; they are just not what this review is about. Sect.~\ref{Sec:Outlook} contains a brief summary  and a list of potential future avenues for further research.

\section{Generalities}
\label{Sec:holography}

The dictionary of gauge-gravity correspondences sits at the core of the material covered by this review, as it turns gauge-gravity dualities into a useful tool in the study of strongly coupled field theories. Illustrating how it descends from more fundamental principles would take us beyond our purposes, hence we dispense with it and refer the reader to the literature, for instance to Ref.~\cite{Aharony:1999ti}. In this part of the introduction we follow closely Ref.~\cite{Skenderis:2002wp}; in particular, we use Euclidean space-time metric with positive signature, while the rest of the review adopts Lorentzian, mostly $+$,  signature.

For a generic, asymptotically safe, field theory, admitting the existence of a set of gauge-invariant local operators, $\left\{{\cal O}_i\right\}$, the first assertion defining the correspondence can written as follows:
\beqs
\label{Eq:dictionary}
\left\langle e^{-\int_{\partial M} J_i\,{\cal O}_i}\right\rangle_{\rm QFT}
&=&
\int_{\Phi_i\rightarrow J_i} {\cal D} \Phi e^{-{\cal S}[\Phi]}
\,.
\eeqs
The left-hand side of Eq.~(\ref{Eq:dictionary}) is the familiar expression of the field-theory expectation value defined by the path integral, in which for the local operators of interest, ${\cal O}_i$, we introduce a corresponding source, $J_i$. The field theory (with its local operators and  sources) lives in a space-time with one fewer dimensions than that of the gravity theory. The latter, higher-dimensional space, $M$, should be a regular Riemannian manifold, the geometry of which approaches  asymptotically anti-de-Sitter (AdS). For each of the operators, ${\cal O}_i$, corresponding  fields, $\Phi_i$, live in the higher-dimensional space, and are governed by the gravity theory action, ${\cal S}[\Phi]$. The right-hand side of Eq.~(\ref{Eq:dictionary}) is the standard formulation of the path integral in the higher dimensional theory, except for the restriction denoted as $\Phi_i\rightarrow J_i$, which states that the boundary values of the fields, $\Phi_i$, are related to the sources, $J_i$, in the lower-dimensional boundary field theory.

The second assertion is that, while Eq.~(\ref{Eq:dictionary}) is valid in general,  its two sides are related by a weak-strong correspondence, in the following sense.  If one computes the leading saddle-point approximation of the gravity theory on the right-hand side, in the limit of weak coupling,  one obtains the full quantum generating functional of connected diagrams in the limit of strong coupling, for the field theory in the left-hand side. In the following, we will only restrict our attention to this limiting case. 
This statement can be written as a relation between the classical on-shell gravity action and the quantum functional,  $W[J]$, that generates the connected correlation functions:
\beqs
\label{Eq:W}
W[J]
&=&
-{\cal S}_{\rm on-shell}\left[J\right]\,,
\eeqs
so that any $n$-point, connected, correlation function takes the form~\cite{Skenderis:2002wp}:
\beqs
\left\langle O_i(x_1) \frac{}{} \cdots \frac{}{}O_j(x_n)\right\rangle
&=&(-)^{n+1}\left.
\frac{\delta^n {\cal S}_{\rm on-shell}\left[J\right]}{\delta J_i(x_1)\cdots \delta J_j(x_n)}\right|_{J\rightarrow 0}\,.
\eeqs

There are two subtleties implicit in these relations, associated on the one hand to the normalisation of the fields, sources, and operators (which requires a more precise definition of the relation between source and boundary value of the fields), and on the other hand to the process of regularising and renormalising the field theory (which requires including in the gravity action appropriate boundary-localised contributions that introduce the necessary counter-terms). Addressing algorithmically these subtleties is the purpose of holographic renormalisation~\cite{Bianchi:2001kw,Skenderis:2002wp,Papadimitriou:2004ap}. We will introduce these steps for the calculations relevant to this review, in Sect.~\ref{Sec:moreholography}, with specific reference to the calculation of the free energy density, ${\cal F}$,  and of the spectra of bound states of the field theory, while referring the reader to the literature for other observables.

Given any quantum field theory of interest, one can always define as one of its operators the (improved) energy-momentum tensor~\cite{Callan:1970ze}, $T_{\mu\nu}$, that acts as the boundary-localised source of the metric, $g_{\mu\nu}$, in the gravity theory. Its trace is related to the divergence of the dilaton current, $T_{\mu}^{\,\,\,\mu}=\partial_{\mu}D^{\mu}$, in field theory, and hence measures  the size of effects that break scale invariance. It couples to the trace of the metric in the higher-dimensional theory. The properties of the dilaton operator, field, and associated particle (in the weakly coupled, low-energy effective description of the field theory) are then encoded in the trace of the metric of the higher-dimensional theory and its fluctuations. Doing so requires correctly identifying the physical, diffeomorphism-invariant, properties,  by using an appropriate gauge-invariant formalism.

Gauge-invariant, field-theory operators with other quantum numbers correspond to fields with the same quantum numbers in the gravity theory: scalars, $\Phi^a$, fermions, $\psi^{\alpha}$, gauge bosons, $A_{\mu}{}^A$. The gauge bosons are particularly interesting. As any continuous global symmetry of the field theory is related, by Noether theorem, to a conserved current, $J_{\mu}{}^A$, the corresponding field is the gauge boson, $A_{\mu}{}^A$. The global symmetry of the field theory is then realised as a gauge theory of the higher dimensional dual one.\footnote{The global symmetry can also be gauged weakly, which requires generalising the process of holographic renormalisation, by adding boundary-localised derivative terms in the classical gravity action, a process that can be shown to introduce scheme dependence and running of the couplings. It must be implemented to discuss vacuum misalignment and the Higgs mechanism---see for example Refs.~\cite{Elander:2023aow,Elander:2024lir} and references therein. We return to this subtlety when discussing the boundary conditions for the fluctuations of the gauge fields, in Eq.~(\ref{eq:spin1BCs}).}

The classical action of the gravity theory then involves the fields, their interactions and self-couplings, and  the coupling of gravity. It usually contains terms with up to two derivates. By varying the action (inclusive of boundary terms) with respect to the fields, one finds the classical equations of motion, that include Einstein equations. In the cases of interest, it is possible to identify one of the coordinates of the space, $\rho$, orthogonal to the boundary, $\partial M$, with the renormalisation scale of the field theory. If one can assume that the background fields do not depend on the coordinates of the field-theory space, but only on the holographic coordinate, $\rho$, then one can interpret the set of second-order coupled differential equations obeyed by the background functions as a proxy for the renormalisation group equations (RGEs) obeyed by the couplings and vacuum expectation values of the operators of the field theory. Approaching the asymptotically AdS geometry of the gravity theory near its boundary is related to analysing the theory in its asymptotically safe, ultraviolet (UV) regime. The deep infrared (IR) of the QFT, with its non-perturbative dynamics, is encoded in the background fields far away from the boundary, deep into the space. Solving the ordinary differential equations involving $\rho$-dependent derivatives of the classical fields is then equivalent to integrating over the RGE flow. This process requires imposing two additional  sets of constraints.

First, one imposes that the asymptotic behaviour near the boundary matches the active sources of the field theory. Second,  one must require the absence of singularities in the interior of the geometry defined by the background solutions. The latter requirement often introduces tuned relations between integration constants in the background, that ultimately encode the relation between sources (couplings) and response functions (condensates) in the field theory. In particular, while the values of the sources are encoded in the leading boundary asymptotic values of the background fields, their sub-leading behaviour encodes the sizes of the corresponding condensates, that are then generically fixed in terms of the former.\footnote{The identification of source and response function is ambiguous in a special range of dimension of the field-theory operators, and requires additional care~\cite{Klebanov:1999tb}. We return to this caveat when discussing the UV expansions, after Eq.~(\ref{Eq:UV}).} Consequently, these constraints remove the spurious freedom in the classical equations, that are of second order, while the RGEs are a system of first-order equations. And furthermore it introduces precise relations between the couplings and condensates of the operators in the theory.

\subsection{Confinement in holography}

The notion of confinement in Yang-Mills  theories with gauge group $SU(N_c)$ refers to the behaviour of the rectangular  Wilson loop~\cite{Wilson:1974sk}, which is the non-local operator 
\beqs
{\cal W}_{\cal C}&\equiv&
\frac{1}{N_c}{\rm Tr}\left\{{\rm P}\left[e^{i \oint_{\cal C} A_{\mu}\dd x^{\mu}}\right]\right\}\,,
\eeqs
where ${\cal C}$ is a rectangle with finite time extent, $T_{QQ}$, and spatial extent, $L_{QQ}$. Gauge invariance is ensured by the path-ordering, ${\rm P}$, of gauge fields evaluated  along the closed circuit, and the trace over the gauge indexes, ${\rm Tr}$. By computing the QFT average of this expression, and Wick-rotating to Euclidean time, in the limit of large $T_{QQ}\rightarrow +\infty$, one recovers the potential energy, $E_{QQ}$, experienced by a static quark-antiquark pair held at finite distance, $L_{QQ}$:
\beqs
\left\langle {\cal W}_{\cal C} \right\rangle_{\rm QFT} & \rightarrow & e^{-E_{QQ} T_{QQ}}\,.
\eeqs
The theory is said to confine if, for large $L_{QQ}$, one recovers the linear potential,  $E_{QQ} \simeq \sigma L_{QQ} +\cdots$. The constant $\sigma$ is called the string tension, and the ellipsis stands for corrections to the string effective field theory description of the system~\cite{Luscher:2002qv,Aharony:2009gg}, that are suppressed for large $L_{QQ}$, and some of which are universal~\cite{Luscher:1980fr}.

In the context of gauge-gravity dualities, the Wilson loop being a non-local operator, its dual description requires the introduction of extended objects,  strings and branes, hence it can be computed only if the gravity dual can be lifted to one of the known string theories, within which one follows the construction proposed in Refs.~\cite{Maldacena:1998im,Rey:1998ik}. The two ends of an open string are attached to the Wilson loop of interest, ${\cal C}$, on a D$p$-brane localised at the asymptotic boundary of space. The QFT expectation value is computed by finding the area of the minimal surface in the higher-dimensional curved space that is bound to ${\cal C}$, by evaluating the Nambu-Goto action. This process is described in detail in Refs.~\cite{Rey:1998bq,Brandhuber:1998bs,Brandhuber:1998er,Gross:1998gk,Brandhuber:1999jr,Nunez:2009da,Faedo:2013ota}, and references therein.

In the limit in which $T_{QQ}\gg L_{QQ}$, effectively the problem reduces to finding a local and globally stable, one-dimensional, $U$-shaped string configuration that minimises the Nambu-Goto action, by assuming that the end points of the string are kept fixed on the boundary, at distance $L_{QQ}$ from one another. In general, the string is allowed to fall into the interior of the geometry, subject to gravity, as the curvature in the holographic direction, $\rho$, increases. It has a turning point, located at a minimum value, $\hat{\rho}$, of the holographic direction. If the following conditions are met, one finds a linear potential, the slope of which can be interpreted in terms of the string tension, $\sigma$, of the dual, confining gauge theory.

\begin{itemize}

\item[1.] The background is completely regular, but the holographic direction has a minimum value, $\rho_o$, at which the geometry closes smoothly. 

\item[2.] The Nambu-Goto action, evaluated with the induced metric on the two-dimensional world-volume of the string, results in an effective potential that is compatible with $U$-shaped string configurations. In particular, the end-points for the string approach the asymptotic boundary orthogonally to it, as prescribed by strings attached to boundary-localised $Dp$-branes. 

\item[3.] For large enough values of $L_{QQ}$, while the portion of the string near its extrema falls almost vertically into the holographic direction, once the string  reaches the proximity of the end of space,  $\hat{\rho}\simeq \rho_o$, the middle portion of the string lays parallel to the boundary directions, near the end of the geometry. These properties are ensured by checking that certain monotonicity conditions of the functions of the holographic direction, $\rho$, that enter the effective string description are satisfied~\cite{Brandhuber:1999jr,Nunez:2009da,Faedo:2013ota}.
Furthermore, one requires that even at the end of space the string is not tensionless, $\sigma\neq 0$. 

\end{itemize}

The end of space in condition 1. defines a non-trivial physical scale, the confinement scale, that governs the typical mass of bound states in the dual field theory. If condition 3. is satisfied, by comparing two similar configurations with length $L_{QQ}$ and $L_{QQ}+\Delta L_{QQ}$, respectively, one can conclude that the energy difference between them is $\Delta E_{QQ} = \sigma \Delta L_{QQ}$, which is the expectation from confinement. The requirement encoded in condition 2. has a somewhat technical origin, and it is needed to ensure that the end points of the string can be treated as static quarks in the dual field theory. It is generally satisfied by backgrounds that are asymptotically AdS, so that it does not play a big role in this review.

Condition 1. is more difficult to satisfy. It is not too hard to find solutions in which the space ends, but that are singular at the end point. Known regular examples assume that the geometry of the higher-dimensional space-time includes a compact sub-manifold the size of which shrinks for $\rho\rightarrow \rho_o$, and does so in such a way as not to introduce curvature nor conical singularities. Well known examples involve shrinking circles~\cite{Witten:1998zw,Wen:2004qh,Kuperstein:2004yf,Brower:2000rp,Elander:2013jqa}, so that the geometry is that of a cigar, with the tip at $\rho_o$. Similar results can be obtained also with higher-dimensional sub-manifolds, as in the aforementioned Refs.~\cite{Chamseddine:1997nm,Klebanov:2000hb,Maldacena:2000yy,Butti:2004pk,Dymarsky:2005xt,Andrews:2006aw,Hoyos-Badajoz:2008znk,Nunez:2008wi,Elander:2009pk,Cassani:2010na,Bena:2010pr,Bennett:2011va,Dymarsky:2011ve,Maldacena:2009mw,Gaillard:2010qg,Caceres:2011zn,Elander:2011mh,Elander:2012yh,Elander:2017hyr,Elander:2017cle}, but these are much more difficult to find, as the curvature of such manifolds requires introducing non-trivial fluxes that support it. The complicated geometrical problems these requirements define can be addressed, but often at the price of  involving twisting of the space-time symmetries of these sub-manifolds with the internal symmetries of supergravity, so that the associated fields develop the necessary fluxes. We will not discuss these cases further, but rather refer the reader to the rich and fascinating  literature on the subject.

The adoption of this prescription for the study of confinement, though well established, and applicable to the top-down models presented in the body of this review, leaves open the problem of how to generalise it to the bottom-up approach to holography, in which the gravity theory is written as an ad hoc toy model, with no known embedding into a more fundamental gravity theory that includes extended objects. We take a pragmatic approach in this review, by attributing the character of confinement to any bottom-up backgrounds living in higher dimensions that admit a regular, smooth, cigar-shaped geometry with a finite end of space, $\rho_o$, despite the fact that the action does not admit the existence of extended objects (strings) that can be used to probe the geometry. It is worth noting that other authors in the literature take a more constructive, less conservative approach, and add to the bottom-up gravity description such extended objects, by making additional ad hoc assumptions about how to  generalise  the Nambu-Goto and DBI actions---see for example the interesting studies in Refs.~\cite{Gursoy:2007cb,Gursoy:2007er,Jarvinen:2011qe,Alho:2013dka} and references therein.

\subsection{Phase transitions, from Van der Waals gas to holographic confinement}
\label{Sec:VdW}

The Van der Waals model of a real gas provides a useful illustration of much of the physics of first- and second-order phase transitions, that has applicability also to the context of gauge-gravity dualities. (See also Ref.~\cite{Aarts:2023vsf} for a review of phase transitions in particle physics.) It is a phenomenological model, that can be defined by assuming that the Helmholtz free energy, $F=F(V, T, N)$, expressed as a function of the thermodynamic quantities (temperature, $T$, number of particles, $N$, and volume, $V$),  reads as follows:
\beqs
F&=&-N k T \log\left[c \left(\frac{V}{N}-b\right) T^{3/2}\right] -\frac{a N^2}{V}\,,
\eeqs
where $a>0$, $b>0$, and $c>0$ are model-dependent 
constants that characterise the properties of the system (and descend from a truncated expansion of the microscopic dynamics in the presence of interactions),
while $k$ is the universal Boltzmann constant.
One must also impose the physical constraints that  $T>0$ and $V> N b$.  The  exponent of the temperature, $3/2$, appearing inside the logarithm, ensures that in the limit of extreme rarefaction, $V/N\rightarrow +\infty$, this expression for $F$ matches the one of the  monoatomic ideal gas.

By identifying the pressure, $P$, as the conjugate variable  to the volume, $V$,
\beqs
\label{Eq:Maxwell}
P&\equiv&-\left(\frac{\partial F}{\partial V}\right)_{N,T}\,,
\eeqs
 one can perform a Legendre transform and arrive at the Gibbs free energy, $G=G(P, T, N)$, defined as follows:
\beqs
G\,=\,G(P, T, N)&\equiv&F\,+\,P\,V\,.
\eeqs
By performing the derivative in Eq.~(\ref{Eq:Maxwell}), one finds the famous equation of state: 
\beqs
\label{Eq:VdW}
P&=&\frac{Nk T}{V-b N}\,-\,\frac{aN^2}{V^2}\,.
\eeqs
Holding fixed $N$ and $T>T_c\equiv \frac{8a}{27 k b}$, the isotherm function $P(V)$ is monotonically decreasing,  and similar  in shape to the ideal gas law, but for the presence of small corrections due to the effects encoded in $a$ and $b$.

\begin{figure}[t]
\centering
\includegraphics[width=0.4\textwidth]{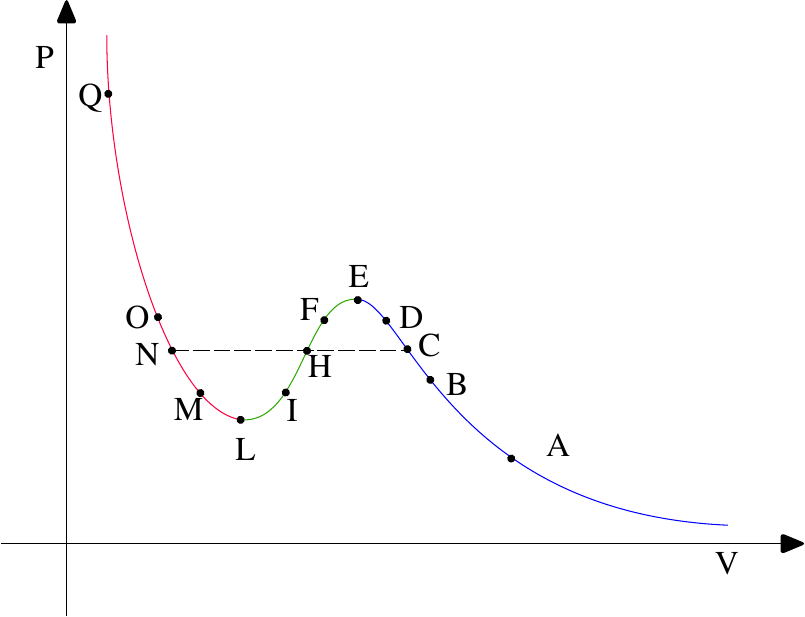}        
\includegraphics[width=0.4\textwidth]{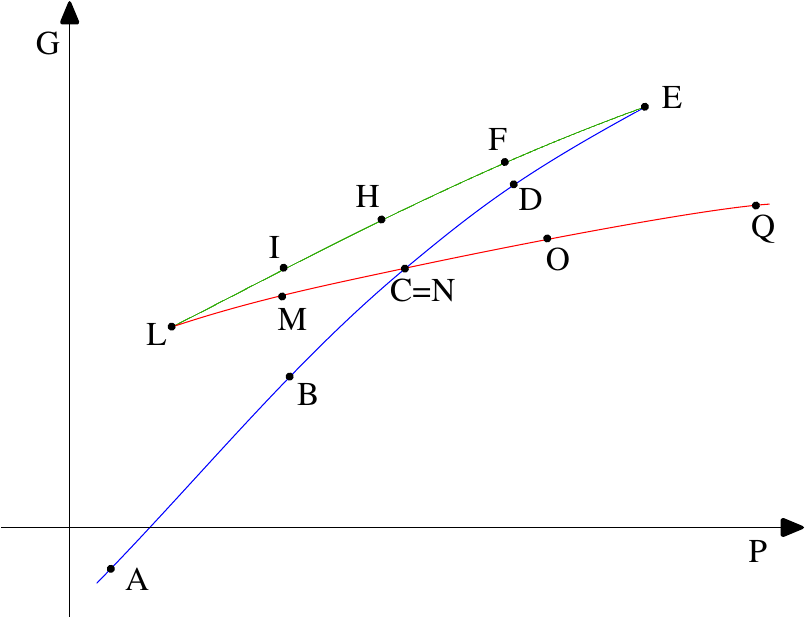}    
   \hfill
	\caption{Cartoons illustrating the behaviour of the Van der Waals gas. Left panel: the pressure, $P$, as a function of the volume, $V$, for fixed choices of number of particles, $N$, and temperature, $T$. The temperature chosen is below the critical value, $T<T_c$, and as a result the function $P(V)$ is not globally invertible, the inverse existing only locally. Right panel: the Gibbs free energy, $G$, as a function of the pressure, $P$, for the same values of $N$ and $T$ as in the left panel. Because of the non-invertibility of $P(V)$, the Gibbs free energy, which is computed as the Legendre transform of the Helmholtz free energy, $F$, is multivalued.  Figures taken from Ref.~\cite{Nunez:2009da}.
  \label{fig:VdW}}
\end{figure}

The schematic cartoons in Fig.~\ref{fig:VdW} are lifted from Ref.~\cite{Nunez:2009da}, though similar diagrams can be found in thermodynamics textbooks. They show $P(V)$ and $G(P)$, for fixed $N$ and a representative value of temperature, chosen below criticality, $T<T_c$. The function $P(V)$ in this case is not invertible and the Gibbs free energy is multivalued in $P$.  These findings demonstrate the existence of a line of first-order phase transitions in the $(P,T)$ plane. For fixed $T$, in a finite interval of values of pressure, $P$, there appear the phenomena of phase coexistence and metastability. Two (meta-)stable solutions ($A$ to $E$ and $L$ to $Q$, in the diagrams) compete to minimise the free energy, $G$, and define a non-analyticity in the minimum value of $G$, which is not differentiable at the intersection of the two branches (at the point denoted as $C=N$), as expected in a first-order phase transition. A third branch of solutions, connecting points $E$ and $L$, exists  in the same  range of pressure, $P$, but with larger values of free energy, $G$. The swallow-tail behaviour of the free energy is commonly discussed in textbooks on thermodynamics, and has recently been found to emerge even in the numerical lattice study of Yang-Mills theories,  using    Linear  Logarithmic Relaxation (LLR)~\cite{Lucini:2023irm,Bennett:2024bhy,Bennett:2025whm}.

At the transition value of the pressure, $P$ (between the points denoted  $C$ and $N$ in the figure),
the physically realised configuration does not follow the equilibrium isotherm given in Eq.~(\ref{Eq:VdW}). Instead, the system moves along the horizontal segment shown in the left panel of Fig.~\ref{fig:VdW}. This can be  obtained with Maxwell's construction. It depicts the fact that the system fragments into an admixture of patches in different regions of space (bubbles), containing either of the two phases, realised at the points denoted by $N$ (the liquid phase) and $C$ (the vapor phase), separated by walls.

The analytical properties of the free energy obey the concavity theorems. The derivative $\partial P/\partial V$ (taken holding fixed $N$ and $T$) is negative along the stable branches, and positive along the unstable branch. This is equivalent, via Maxwell's relation in Eq.~(\ref{Eq:Maxwell}), to the statement of positivity of the second derivative, $\partial^2F/\partial V^2 >0$, along the stable branches. The change of sign of the second derivative of the free energy along the unstable branch is the main signal of local, tachyonic instability in the system. Negative values of the second derivative appear only in (physically not realised) regions of parameter space, hidden beyond the phase transition, while the metastable branches are locally stable, even if globally unstable.

When raising the temperature, approaching $T_c$ from below, the first-order transition becomes weaker, the two stationary points visible in the left panel of Fig.~\ref{fig:VdW} coming closer to one another. Eventually, the transition disappears. To determine the critical value of the parameters at which this happens, one can impose the requirement that the two stationary points merge,  so that $P(V)$ develops an inflexion point. Setting the first and second derivative of $P$ with respect to $V$ to vanish, one can identify the critical values of the thermodynamic parameters at which the line of first-order phase transitions ends. One finds the critical values to be
\beqs
T_c\,=\,\frac{8a}{27 k b}\,,\quad P_c\,=\,\frac{a}{27 b^2}\,,\quad V_c=3N b\,.
\eeqs
At this end point, one finds the appearance of a second-order transition. The inflection in $P(V)$ implies that  $\partial^2F/\partial V^2 = 0$. The Gibbs free energy, $G(P)$, is continuous and differentiable, for $T=T_c$, but the second derivative of $G$ with respect to $P$ diverges at $P=P_c$. In this case, the unstable branch as well as the phenomena of phase coexistence disappear. The correlation length of the system diverges. 

Qualitatively, most of the properties of phase transition observed in the Van der Waals gas are true also for more general systems undergoing phase transitions, including the strongly coupled, confining field theories of interest in this review. Any such field theory admits a set of composite operators, and their sources, or couplings, act as control parameters,  in a similar way to the pressure, $P$, and temperature, $T$, in the Van der Waals case. The path integral evaluation of the QFT average in Eq.~(\ref{Eq:dictionary})---or  $W[J]$ in Eq.~(\ref{Eq:W})---provides the equivalent of the Gibbs free energy, and its functional derivatives yield QFT averages, $\langle {\cal O}_i \rangle_J$, that play the role of the response functions, as the volume $V$, and the entropy, $S$, in the real gas.

The physics of first-order phase transition in the holographic description of confining field theories  is extracted via  global stability analysis that is similar to what we illustrated for the Van der Waals gas: having computed the free energy as a function of the QFT sources, one looks for regions of parameter space displaying coexistence of phases, and by identifying the minimal free energy as a function of $J$, one finds that the first derivative of the absolute minimum is discontinuous at the transition. A local stability analysis shows that besides metastable solutions, often an unstable branch of solutions exists as well, along which the spectrum of local fluctuations contains a tachyon. Some additional refinement is needed in holography, because of the scheme-dependence of the free energy (and its derivatives) in interesting strongly coupled field theories admitting a gravity dual---see also the discussion in Ref.~\cite{Bobev:2013cja}. 

In some cases, one also finds that lines of first-order phase transition exist, in theories that admit multiple deformations. If one such line ends at a critical point, the transition becomes second order: phase coexistence disappears, the first derivatives of the free energy is continuous, but the second derivative is discontinuous. The divergence of the correlation length that characterises the modern definition of a second-order phase transition in thermodynamics, in the holographic description of strongly coupled field theories is associated with the appearance of a massless bound state in the spectrum of fluctuations at the transition.

\section{Technology}
\label{Sec:moreholography}

Here, we present the formalism that is used to study the holographic models summarised in Sects.~\ref{Sec:Top} and~\ref{Sec:Bottom}. We focus our attention on the gravity theories and on the prescriptions that allow to extract quantities  that are interpreted as field theory observables. The notation adopted throughout this section and the amount of detail we provide are chosen in order to make the presentation of this material self-contained. Yet, we alert the reader of the fact that in Sects.~\ref{Sec:Top} and~\ref{Sec:Bottom} we present results lifted from the original publications, which use different (and diverse) sets of conventions, hence the comparison requires to adopt some caution.

The gravity theories of interest are defined in $\hat D = D + D_T$ number of space-time dimensions, $D_T$ of which are compactified on a torus, $T^{D_T}$. They consist of gravity coupled to a matter field content that generically can include a sigma-model of real scalar fields, and a number of gauge fields. After presenting the action of the theory in $\hat D$ dimensions, we dimensionally reduce it on $T^{D_T}$, exemplifying  the explicit treatment for the cases $D_T = 1,\,2$. We discuss how to obtain classical, background solutions to the system of coupled ordinary differential equations, resulting from the ansatz that the backgrounds take either the domain-wall or soliton form for the metric, and in which the background functions depend explicitly only on the holographic direction in space-time.  The soliton ansatz encompasses the case in which  one of the compact dimensions shrinks smoothly in the IR of the geometry, thus modelling confinement in the dual field theory. The $D$-dimensional system is especially suited for the computation of mass spectra of bound states, obtained through the study of gauge-invariant fluctuations around the background solutions of interest. We describe the gauge-invariant formalism used to efficiently treat the mixing between scalar and gravity sectors, giving the spin-0 part of the spectrum, as well as the formalism used to compute spin-1 and spin-2 spectra. Finally, we present general formulas for the computation of the free energy, using the formalism of holographic renormalisation.

\subsection{Higher-dimensional gravity theory and dimensional reduction}

We start by defining the gravity theory in $\hat D$ dimensions. In general, we use hats to denote $\hat D$-dimensional quantities and indexes. The field content consists of gravity, coupled to $\hat n$ real scalar fields $\hat \Phi^{\hat a}$, with $\hat a = 1, \,\cdots,\, \hat n$. The space-time coordinates are denoted as  $x^{\hat M}$, where $\hat M$ includes non-compact directions parallel to the boundary ($\hat M = 0,1, \,\cdots\,, D-2$), the holographic radial direction ($\hat M = D$), and the coordinates on the torus ($\hat M = D+1, \,\cdots\,, \hat D$). Depending on the model, there may be also  a  gauge symmetry, with $\hat m$ gauge fields, $\hat{\mathcal A}_{\hat M}{}^{\hat A}$ with $\hat A = 1, \,\cdots,\, \hat m$. In this review all gauge symmetries will be Abelian: while the generalisation to the non-Abelian case follows the time-honoured rules, none of the solutions of interest for this review involve non-trivial flux for non-Abelian gauge fields, hence restricting attention to the Abelian case simplifies the notation without loss of generality.

The bulk action is given by\footnote{The complete action contains also boundary terms that we will discuss in due time.}
\beqs
\label{eq:SDhat}
	\mathcal S_{\hat D} &=& \frac{1}{4\pi G_{\hat D}} \int \dd^{\hat D} x \, \sqrt{-\hat g} \, \left\{ \frac{\hat R}{4}
- \frac{1}{2} \hat g^{\hat M \hat N} \hat G_{\hat a \hat b}(\hat \Phi) \partial_{\hat M} \hat \Phi^{\hat a} \partial_{\hat N} \hat \Phi^{\hat b} - \hat V(\hat \Phi) - \frac{1}{4} \hat g^{\hat M \hat O} \hat g^{\hat N \hat P} \hat H_{\hat A \hat B}(\hat \Phi) \hat F_{\hat M \hat N}{}^{\hat A} \hat F_{\hat O \hat P}{}^{\hat B} \right\} \,,
\eeqs
where $\hat g_{\hat M \hat N}$ is the metric in $\hat D$ space-time dimensions, $\hat g^{\hat M \hat N}$ its inverse, $\hat g$ its determinant, and $\hat R$ the Ricci scalar. We adopt mostly $+$ conventions for the signature of the metric. The field-strength tensors for the gauge fields are denoted as $\hat F_{\hat M \hat N}{}^{\hat A}$. The sigma-model metric, $\hat G_{\hat a \hat b}(\hat \Phi)$, the scalar potential, $\hat V(\hat \Phi)$, and the kinetic tensor for the gauge fields, $\hat H_{\hat A \hat B}(\hat \Phi)$, are in general functions of the sigma-model scalars, $\hat \Phi^{\hat a}$. We anticipate here that we conventionally choose the range of the toroidal coordinates to be $[0,\,2\pi[$, and set our units so that the Newtonian constant in  $\hat D$ dimensions is given by $G_{\hat D} = \frac{(2\pi)^{D_T}}{4\pi}$. These convenient choices will result in a simple form for the reduced action in $D$ dimensions.

\subsubsection{Reduction on one circle}

Let us consider the case $D_T = 1$, and reduce the theory on one circle,  parameterised by an angular variable, $\eta \equiv x^{\hat D}$, with $0 \leq \eta < 2 \pi$. In the reduction ansatz, all fields are independent of $\eta$. The metric in $\hat D$ dimensions is given by
\beq
\label{eq:metric-onecircle}
	\dd s_{\hat D}^2 = e^{-2\chi} \dd s_D^2 + e^{2(D - 2) \chi} \left(\dd \eta + \mathcal U_M \dd x^M \right)^2 \,,
\eeq
where $\chi$ and $\mathcal U_M$ are a scalar and a vector field, respectively, while $M$ is the $D$-dimensional space-time index, and $\dd s_D^2$ is the metric in $D$ space-time dimensions. Indexes and quantities without hats denote $D$-dimensional objects, to distinguish them from analogous quantities in $\hat{D}$ space-time dimensions..
The gauge fields, with $\hat{A}=1,\,\cdots,\,\hat{m}$, are parametrised as
\beq
	\hat{\mathcal A}^{\hat A} = \hat{\mathcal A}_M{}^{\hat A} \dd x^M + \hat{\mathcal A}_{\hat D}{}^{\hat A} \dd \eta \,,
\eeq
decomposing into one vector and one scalar field, $\hat{\mathcal A}_M{}^{\hat A}$ and $\hat{\mathcal A}_{\hat D}{}^{\hat A}$, respectively, for each $\hat{A}^{\hat A}$. Up to total derivatives and boundary terms, the reduced action in $D$ dimensions take a form analogous to the one in  $\hat D$ space-time dimensions:
\beq
\label{eq:Sreduced}
	\mathcal S_D = \int \dd^D x \, \sqrt{-g} \, \Bigg\{ \frac{R}{4} - \frac{1}{2} g^{MN} G_{ab}(\Phi) \partial_M \Phi^a \partial_N \Phi^b - V(\Phi) - \frac{1}{4} g^{MO} g^{NP} H_{AB}(\Phi) F_{MN}{}^A F_{OP}{}^B \Bigg\} \,.
\eeq

The scalars are given by $\Phi^a = (\hat \Phi^{\hat a}, \chi, \hat{\mathcal A}_{\hat D}{}^{\hat A})$. The scalar potential is
\beq
	V(\Phi) = e^{-2\chi} \hat V(\hat \Phi) \,,
\eeq
and the sigma-model metric, in the basis $(\hat \Phi^{\hat a}, \chi, \hat{\mathcal A}_{\hat D}{}^{\hat A})$, takes a  block-diagonal form:
\beqs
G_{ab} &=&
\left(
\begin{array}{c|c|c}
	\hat G_{\hat a \hat b}(\hat \Phi) && \cr
	\hline
	& \frac{1}{2} (D-1)(D-2) & \cr
	\hline
	&& e^{-2(D-2)\chi} \hat H_{\hat A \hat B}(\hat \Phi)
\end{array}\right)\,.
\eeqs

The vectors are given by $\mathcal A_M{}^A = (\hat{\mathcal A}_M{}^{\hat A}, \mathcal U_M)$. Their field strengths are in general defined by
\beq
	F_{MN}{}^A = \partial_M \mathcal A_N{}^A - \partial_N \mathcal A_M{}^A + Z^A{}_{Ba} \left( \mathcal A_M{}^B \partial_N \Phi^a - \mathcal A_N{}^B \partial_M \Phi^a \right) \,,
\eeq
where $Z^A{}_{Ba}$, in the present case, reduces to 
\beq
	Z^A{}_{Ba} \, \mathcal A_M{}^B \partial_N \Phi^a = \mathcal U_M \partial_N \hat{\mathcal A}_{\hat D}{}^{\hat B} \, \delta^A{}_{\hat B} \,,
\eeq
The  kinetic terms for the vectors are governed by the block-diagonal matrix  given by
\beqs
H_{AB} &=&
\left(
\begin{array}{c|c}
	e^{2\chi} \hat H_{\hat A \hat B}(\hat \Phi) & \cr
	\hline
	& \frac{1}{4} e^{2(D-1)\chi}
\end{array}\right)\,.
\eeqs

\subsubsection{Reduction on two circles}

For the case of a 2-torus,  $D_T = 2$, we make the simplifying assumption that no gauge field is present in the action written in $\hat D$ dimensions. This assumption holds in the applications discussed in Sects.~\ref{Sec:Top} and~\ref{Sec:Bottom}.  We reduce the theory on a circle twice, following the procedure of the previous subsection. We denote the coordinates parametrising the torus by $\eta_1 \equiv x^{D+1}$ and $\eta_2 \equiv x^{D+2}$, with $0 \leq \eta_{1,2} < 2 \pi$, and adopt an ansatz where all fields are independent of $\eta_{1,2}$. The metric is given by
\beq
\label{eq:metric-twocircles}
	\dd s_{\hat D}^2 = e^{-2\chi} \dd s_D^2 + e^{(D-2)\chi+2\omega} \left(\dd \eta_1 + \mathcal U_M \dd x^M + \upsilon \, \dd \eta_2 \right)^2 + e^{(D-2)\chi-2\omega} \left(\dd \eta_2 + \mathcal V_M \dd x^M \right)^2 \,,
\eeq
where $\chi$, $\omega$, and $\upsilon$ are scalar fields, while $\mathcal U_M$ and $\mathcal V_M$ are vector fields, with $M$ a $D$-dimensional spacetime index, and $\dd s_D^2$ is the metric in $D$ dimensions.

The resulting $D$-dimensional action takes the same form as in Eq.~\eqref{eq:Sreduced}. The scalars are given by $\Phi^a = (\hat \Phi^{\hat a}, \omega, \chi, \upsilon)$. The scalar potential is
\beq
	V(\Phi) = e^{-2\chi} \hat V(\hat \Phi) \,,
\eeq
and the sigma-model metric is
\beqs
G_{ab} &=&
\left(
\begin{array}{c|c|c|c}
	\hat G_{\hat a \hat b}(\hat \Phi) && & \cr
	\hline
	& 1 && \cr
	\hline
	&& \frac{1}{4} D(D-2) & \cr
	\hline
	&&& \frac{1}{4} e^{4\omega}
\end{array}\right)\,.
\eeqs

The vectors are given by $\mathcal A_M{}^A = (\mathcal U_M, \mathcal V_M)$, with
\beq
	Z^A{}_{Ba} \, \mathcal A_M{}^B \partial_N \Phi^a = \mathcal V_M \partial_N \upsilon \, \delta^A{}_{\mathcal U} \,,
\eeq
and
\beqs
H_{AB} &=&
\left(
\begin{array}{c|c}
	\frac{1}{4} e^{D \chi + 2\omega}& \cr
	\hline
	& \frac{1}{4} e^{D \chi - 2\omega}
\end{array}\right)\,.
\eeqs

\subsection{Background equations of motion and solutions}

In order to find background solutions to the gravity system, we write the equations of motion for the fields, in the language of the reduced $D$-dimensional theory. We use an ansatz in which the scalar fields, $\Phi^a(r)$, have profile that only depend on the holographic radial coordinate, $r$, while the vectors $\mathcal A_M{}^A$ vanish on the background. We find it convenient to introduce also an alternative choice of radial coordinate, $\rho$, related to $r$ as $\dd r = e^\chi \dd \rho$. The metric takes the domain-wall (DW) form in $D$ space-time dimensions, given by
\beq
\label{eq:metric-reduced}
	\dd s_D^2 = \dd r^2 + e^{2A(r)} \dd s_{1,D-2}^2 = e^{2\chi(\rho)} \dd \rho^2 + e^{2A(\rho)}\dd s_{1,D-2}^2 \,,
\eeq
where $A=A(r)=A(\rho)$ is a warp factor, and $\dd s_{1,D-2}^2$ is the $(D-1)$-dimensional Minkowski metric. 

The background equations of motion are given as
\begin{align}
\label{eq:BackgroundEOMs1}
	\partial_r^2\Phi^a + (D-1)\partial_r A \partial_r \Phi^a + \mathcal G^a{}_{bc} \partial_r \Phi^b \partial_r \Phi^c - V^a &= 0 \,, \\
\label{eq:BackgroundEOMs2}
	(D-1)(D-2) (\partial_r A)^2 - 2 G_{ab} \partial_r \Phi^a \partial_r \Phi^b + 4V &= 0 \,.
\end{align}
In these expressions, derivatives with respect to the sigma-model scalars, $\Phi^a$, are denoted as
\beq
	V_a \equiv \partial_a V = \frac{\partial V}{\partial \Phi^a} \,,
\eeq
and the sigma-model metric, $G_{ab}$, (and its inverse, $G^{ab}$) are used to lower (raise) sigma-model indices. The sigma-model connection is given by an expression that is analogous to the Christoffel symbols in general relativity:
\beq
	\mathcal G^a{}_{bc} = \frac{1}{2} G^{ad} \left( \partial_b G_{cd} + \partial_c G_{db} - \partial_d G_{bc} \right) \,.
\eeq

\subsubsection{Domain-wall solutions}

We refer to the first class of solutions of interest as domain-wall (DW) solutions in $\hat D$ space-time dimensions. The metric is of the form
\beq
	\dd s_{\hat D}^2 = \dd \rho^2 + e^{2 \hat A(\rho)} \dd s_{1,\hat D-2}^2 \,,
\eeq
where $\hat A(\rho)=A(\rho)-\chi(\rho)$ is the warp factor of the higher-dimensional theory, and $\dd s_{1,\hat D-2}^2$ is the $(\hat D - 1)$-dimensional Minkowski metric. While the vectors, $\hat{\mathcal A}_{\hat M}{}^{\hat A}$, vanish on the background, the scalars, $\hat \Phi^{\hat a}(\rho)$, in general have non-trivial radial profiles. 

We generically assume that the scalar potential of the theory in $\hat D$ dimensions, $\hat V(\hat \Phi)$, has a local maximum at $\hat \Phi^{\hat a} = 0$. This is true for the majority of the models considered in Sects.~\ref{Sec:Top}~and~\ref{Sec:Bottom}, and ensures the existence of an AdS background solution. Such solution describes the dual of a CFT that admits a deformation. Without loss of generality, we require that  $\hat V(0) = - \frac{(\hat D - 2)(\hat D -1)}{4}$. As this is negative, there exist DW solutions in which the geometry is locally AdS$_{\hat{D}}$, and the choice of normalisation is such that $A$ and $\chi$ are linear in the variable $\rho$, and $\hat{A}^{\prime}(\rho)=1$, while all other background functions vanish identically.

 By comparing to the parametrisations introduced earlier on, in Eqs.~\eqref{eq:metric-onecircle},~\eqref{eq:metric-twocircles},~and~\eqref{eq:metric-reduced}, for general DW solutions one obtains relations between $\hat{A}$, $A$, and $\chi$; for the one-circle reduction one has that
\beq
\label{eq:DWconstraint1}
	A = (D-1) \chi = \frac{D-1}{D-2} \hat A \,,
\eeq
while for the two-circle reduction one has that
\beq
\label{eq:DWconstraint2}
	A = \frac{D}{2}\chi = \frac{D}{D-2} \hat A \,,
\eeq
together with $\omega = 0 = \upsilon$.

\subsubsection{Of first-order equations}

In many applications, the scalar potential, $\hat V(\hat \Phi)$, that governs the ordinary differential equations in $\hat D$ dimensions can be obtained from a superpotential, $\hat W(\hat \Phi)$, as
\beq
\label{eq:VfromW}
	\hat V = \frac{1}{2} \hat G^{\hat a \hat b} \frac{\partial \hat W}{\partial \hat \Phi^{\hat a}} \frac{\partial \hat W}{\partial \hat \Phi^{\hat b}} - \frac{\hat D-1}{\hat D-2} \hat W^2 \,.
\eeq
When this is the case, a subclass of DW solutions can be found by solving the first-order equations
\beq
	\partial_\rho \hat \Phi^{\hat a} = \hat G^{\hat a \hat b} \frac{\partial \hat W}{\partial \hat \Phi^{\hat b}}  \,, \qquad \partial_\rho \hat A = - \frac{2}{\hat D-2} \hat W \,.
\eeq
It can be verified that any solution to these equations also solves the second-order system of background equations with DW ansatz, which hence makes it easier to find solutions. With abuse of language, these solutions are referred to as supersymmetric, due to the first-order nature of the ordinary differential equations controlling the backgrounds, though this is not necessarily connected to supersymmetry of the gravity theory (see for instance the discussion in Ref.~\cite{Freedman:2003ax}). For example, establishing whether the backgrounds preserve supersymmetry, in the supergravity theories that we consider, would involve reinstating the fermion field content (which we truncate). Conversely, the superpotential formalism is applicable also in bottom-up models with a purely bosonic field content.

\subsubsection{Regular soliton solutions and confinement}

The regular soliton solutions, taking the form of Eq.~(\ref{eq:metric-onecircle}) or~(\ref{eq:metric-twocircles}), but for which one of the circles shrinks smoothly in the inside of the geometry, provide a dual description of confining field theories.\footnote{The regularity of the solutions in $\hat{D}$ dimensions does not immediately imply that they admit, in a more fundamental theory in higher dimensions, a lift to backgrounds that are themselves free of singularities. A case in point is represented by the lifts to ten dimensions of the regular background solutions of Romans supergravity in $\hat{D}=6$ dimensions, described in Sect.~\ref{Sec:D6}, in which case the internal, compact,  four-dimensional space is not a smooth manifold, but rather it has a boundary.}  The end of space in the geometry, appearing at $\rho\rightarrow\rho_o$, along the holographic direction, corresponds to the dynamically generated scale that yields a mass gap, in the IR regime of the QFT. The solutions of this type do not satisfy the DW constraints, Eq.~\eqref{eq:DWconstraint1} or Eq.~\eqref{eq:DWconstraint2}, relating $A$ and $\chi$.  All curvature invariants remain finite in the complete range of the radial coordinate, and furthermore one requires that there is no conical singularity at the end of the space. 

Let us assume that all the scalars, $\hat \Phi^{\hat a}$, of the original $\hat D$-dimensional theory, and also all the scalars, $\hat A_{\hat D}{}^{\hat A}$, coming from higher-dimensional vectors, are allowed to have non-trivial  profiles in the  holographic direction, yet all reach finite values at the end of space. Together with this restriction, in the case of the reduction on a 2-torus, $D_T = 2$, the regular solutions that we consider also obey the additional constraint
\beq
	\omega = A - \frac{D}{2} \chi \,,
\eeq
which simplifies the system of background equations. With this in place, we may expand the solutions close to the end of space, for small $(\rho - \rho_o)$, leading to the following generic IR expansions:
\begin{align}
	\Phi^a &= \Phi_I^a + \,\cdots\, \,, \\
	\chi &= \chi_I + \frac{1}{D-2} \log(\rho - \rho_o) + \,\cdots\, \,, \\
	A &= A_I + \frac{1}{D-2} \log(\rho - \rho_o) + \,\cdots\, \,,
\end{align}
where $\Phi_I^a$, $\chi_I$, and $A_I$ are integration constants, and the sigma-model index, $a$, is restricted so as to not include $\chi$ or $\omega$. In order to avoid a conical singularity at $\rho = \rho_o$, for $D_T = 1$ one fixes $\chi_I = 0$, while for $D_T = 2$ one fixes $A_I = (D-1) \chi_I$.

In the much simpler case in which all the scalars (excepting  $\chi$ and $\omega$) vanish, $\hat \Phi^{\hat a} = 0 = \hat A_{\hat D}{}^{\hat A}$, it is possible to find analytic solutions. For $D_T = 1$, these are given by
\begin{align}
	\chi(\rho) &= \chi_I - \frac{\log \left( D/2 \right)}{D-2} - \frac{1}{D} \log \left( \cosh \left( \frac{D (\rho - \rho_o)}{2} \right) \right) +\frac{1}{D-2} \log \left( \sinh \left( \frac{D (\rho - \rho_o)}{2} \right) \right) \,, \\
	A(\rho) &= A_I - \frac{\log \left( D/2 \right)}{D-2} + \frac{1}{D} \log \left( \cosh \left( \frac{D (\rho - \rho_o)}{2} \right) \right) +\frac{1}{D-2} \log \left( \sinh \left( \frac{D (\rho - \rho_o)}{2} \right) \right) \,,
\end{align}
while for $D_T = 2$, one obtains
\begin{align}
	\chi(\rho) &= \chi_I - \frac{\log \left( \frac{D+1}{2} \right)}{D-2} - \frac{D-3}{(D-2)(D+1)} \log \left( \cosh \left( \frac{(D+1) (\rho - \rho_o)}{2} \right) \right) + \frac{1}{D-2} \log \left( \sinh \left( \frac{D (\rho - \rho_o)}{2} \right) \right) \,, \\
	A(\rho) &= A_I - \frac{\log \left( \frac{D+1}{2} \right)}{D-2} + \frac{D-1}{(D-2)(D+1)} \log \left( \cosh \left( \frac{(D+1) (\rho - \rho_o)}{2} \right) \right) + \frac{1}{D-2} \log \left( \sinh \left( \frac{D (\rho - \rho_o)}{2} \right) \right) \,.
\end{align}

In the presence of magnetic fluxes, because the functions $\hat{H}_{AB}$ controlling the kinetic terms of the gauge bosons may depend on the sigma-model scalars, one may find that the deep IR behaviour of some of the sigma-model scalars and some of the fluxes are closely related. Nevertheless, one must require that the solutions be regular, that there are no conical singularities, and that the flux vanishes exactly at the end of space, as the circle supporting it disappears. Examples of such solutions can be found in the literature, and we will recall them explicitly when needed, in Sects.~\ref{Sec:Top} and~\ref{Sec:Bottom}. While we are interested in classes of solutions that lead to regular backgrounds, it may still be the case that in certain regions of parameter space the curvature invariants reach large values, at least in proximity of the end of space. In limiting cases, one may even approach singular solutions. One therefore needs to be careful to assess when to trust the gravity theory as a dual description of a QFT.

\subsubsection{UV expansions}

As anticipated, for the majority of the cases that we consider, the solutions of interest asymptotically approach the AdS$_{\hat{D}}$ geometry, as all the scalars except $\chi$ (and, when present, $\omega$) vanish asymptotically.  We may then expand general solutions for the scalars close to the boundary, by writing a power expansion at small values of a new parameterisation of the  holographic direction, $z \equiv e^{-\rho}$. This expansion is equivalent to a  UV expansion in the dual field theory, and captures its renormalisation group (RG) flow in close proximity of the UV fixed point.

Let us consider the case of the one-circle reduction, $D_T = 1$ (it is straightforward to generalise to the case $D_T = 2$). The UV expansions for $\chi$ and $A$ take the following form:
\begin{align}
	\chi &= \chi_U - \frac{1}{D-2} \log(z) + \,\cdots\, \chi_D z^D + \,\cdots\, \,, \\
	A &= A_U - \frac{D-1}{D-2} \log(z) + \,\cdots\, \,,
\end{align}
with $\chi_U$, $\chi_D$, and $A_U$  integration constants.  When convenient, we set  the two constants  $\chi_U=0=A_U$, as they play no important role in the following, being related to an overall rescaling of the units. For a generic scalar of the sigma-model in ${D}$ dimensions (aside from $\chi$), one has the generic expansion
\beqs
\label{Eq:UV}
	\phi &=& \phi_J z^{\Delta_J} + \,\cdots\, + \phi_V z^{\Delta_V} + \,\cdots\, \,, 
\eeqs
where $\phi_J$ and $\phi_V$ integration constants.  The ellipses in Eq.~(\ref{Eq:UV}) refer to other powers, and possibly logarithms, that are not independent, their coefficients being related to the integration constants we already introduced.

Let us restrict ourselves, for the moment, to the scalars, $\hat \Phi^{\hat a}$, that have a scalar origin in the $\hat D$-dimensional model. In this case, the exponents, $\Delta_J$ and $\Delta_V$, satisfy the constraint $\Delta_J + \Delta_V = \hat{D}-1$,  the dimension of the QFT space-time. The exponents $\Delta_{J,V}$ are determined by the bulk mass of the scalar field, $m_{\phi}^2 = \partial_\phi^2 \hat V |_{\phi = 0}$ (for a canonically normalised $\phi$), computed for small perturbation around $\phi=0$.  As long as the mass squared of a bulk scalar field is above the BF stability bound~\cite{Breitenlohner:1982jf}, $m^2_{\phi}>-D^2/4$, the quadratic equation $\Delta(\Delta-D)=m^2_{\phi}$ admits two distinct real solutions. 

For $m^2_{\phi}>-D^2/4+1$, the interpretation of Eq.~(\ref{Eq:UV}) relates $\phi_J$ to the source of a local operator in the QFT, that has dimension $\Delta_J = \frac{D}{2} - \sqrt{\frac{D^2}{4} + m_{\phi}^2}$.  The operator itself has dimension $\Delta_V = \frac{D}{2} + \sqrt{\frac{D^2}{4} + m_{\phi}^2}$, and $\phi_V$ is related to its condensate, via holographic renormalisation~\cite{Bianchi:2001kw,Skenderis:2002wp,Papadimitriou:2004ap}. In the range $-D^2/4<m^2_{\phi}\leq-D^2/4+1$, the interpretation is similar, but there is a potential ambiguity in the identification of source and operator, leading to two different possible quantisations of the dual QFT~\cite{Klebanov:1999tb}. In the limiting case of the saturation of the BF bound, $m^2_{\phi}=-D^2/4$, the asymptotic solution for $\phi$ involves two terms with the same power of $z$, but one of them has also a logarithm, taking the form $z^{D/2}\log(z)$.  
When several scalars are present, they may mix non-trivially, which introduces additional subtleties, but without altering the main arguments summarised so far.

An important special case for the purposes of this review appears when a scalar, $\hat{\mathcal A}_{\hat D}{}^{\hat A}$, of the sigma-model in $D$ space-time dimensions has its origin in $\hat D$ space-time dimensions as a gauge field, with flux turned on along the compactified part of the geometry. In the double-Wick-rotated case in which the circle is the temporal one, related to the temperature of the dual theory, the leading and sub-leading modes  correspond to turning on a chemical potential and a resulting charge density, respectively. While this breaks the $(\hat D-1)$-dimensional Poincaré symmetry, it preserves that of $\hat D-2$ dimensions, and allows for an interpretation of the gravity theory as the holographic dual of a field theory in $\hat D-2$ dimensions.
In the case of interest, in which the circles are space-like, in the QFT there is a magnetic field. Due to the origin of the scalars, $\hat{\mathcal A}_{\hat D}{}^{\hat A}$, as components of vector fields in the $\hat D$-dimensional model, one finds that $\Delta_J=0$ and $\Delta_V=\hat D - 3$. We anticipate that the solutions of interest in Sects.~\ref{Sec:Flux-D7}, \ref{Sec:Flux-D6}, and~\ref{Sec:BU-flux}, are those in which both the corresponding coefficients in the UV expansions are non-zero.

\subsection{Mass spectra}
\label{Sec:mass}

We now turn to the holographic computation of mass spectra of bound states in the dual field theory. As explained in Sect.~\ref{Sec:holography}, general $n$-point correlation functions between local operators of the field theory are computed holographically by differentiating the on-shell action of the gravity theory with respect to the boundary values of the bulk fields, which act as sources for the operators of interest. The masses of the bound states, in particular, can be extracted from the poles of the two-point functions of the dual field theory. Since this involves the differentiation of the on-shell action twice,  for the purpose of computing mass spectra it is sufficient to expand the classical gravity action by including fluctuations upon the background classical fields, retaining terms up to quadratic order in the small fluctuations. As long as one is  interested only in the location of the poles, and hence the mass spectrum, one can follow a simple procedure, that takes advantage of gauge invariance, and that we describe in some detail in the following. Conversely, the full computation of the two-point functions requires to implement the procedure of holographic renormalisation, with careful treatment of local counter-terms, in order to cancel divergences and retain the finite contributions. We can dispense with this additional complication, for the purposes of extracting spectra. Yet, the calculation of the free energy, that we discuss in Sect.~\ref{Sec:free-energy}, follows the full holographic renormalisation prescription.

The algorithmic process we adopt to compute mass spectra can be described, in broad strokes, as follows.

\begin{itemize}
	\item
		We treat the fluctuations in the dimensionally reduced, $D$-dimensional gravity theory, in which all the backgrounds of interest have the domain-wall form. The angular variables describing the internal torus do not affect the fluctuations, that only depend on the coordinates, $x^M$, as we ignore Kaluza-Klein towers of states with masses controlled by the size of the compact dimensions. This allows us to use a standardised formalism, that does not require knowing the intricate details of the higher-dimensional origin of the fields in the lower-dimensional theory, and that is best suited to study the lightest bound states of the theory.

	\item
		We introduce cutoffs at $r = r_1>r_o$ and $ r= r_2\gg r_1$, along the holographic coordinate, corresponding in the dual QFT to IR and UV cutoffs, respectively. At these new boundaries, we add localised actions that include the Gibbons-Hawking-York (GHY) term, mass terms for the scalar fluctuations, and kinetic terms for the vectors.

	\item 
	We study the variational problem of the theory with IR and UV cutoffs, freely varying the fluctuations. From this, we obtain (linearised) equations of motion as well as boundary conditions for the fluctuations. 		We perform a Fourier transformation in the QFT space-time, along the directions parallel to the boundary, and write the fluctuations so that they are functions of the holographic radial direction, $r$, and the $(D-1)$-momenta, $q^\mu$, with $\mu = 0,\, 1,\,\cdots\,,\, D-2$. What results is an over-constrained system, so that the linearised equations of motion for the fluctuations usually admit solutions that satisfy both the IR and UV boundary conditions only for special values of the mass squared, $M^2 \equiv -q^2 = - \eta_{\mu\nu}q^\mu q^\nu$, where $\eta_{\mu\nu}$ is the Minkowski metric.

	\item
		 At this point, because the boundary conditions depend on various parameters present in the boundary actions in the IR and UV, so do the mass spectra. In the following, we comment only briefly on the suitable choices of these parameters in the context of the models we consider, referring the reader to the original papers for details.
    \item
        Finally, the physical mass spectrum is obtained by identifying the aforementioned $M^2 = -q^2$ for which solutions to the fluctuation equations exist, and taking the limits of the IR cutoff towards the end of space, $r_1\rightarrow r_o$, and the UV cutoff towards the boundary of the geometry, $r_2\rightarrow +\infty$.\footnote{In the unfortunate cases in which such limits do not converge, this is usually an  indication that there is a pathology in the gravity background solution, that is only an incomplete approximation of the dual to the field theory of interest. This pathology does not emerge in any of the models described in this review.}
\end{itemize}

The limiting procedure described above generically reproduces the standard prescription in gauge-gravity dualities of retaining the regular modes in the IR and the sub-leading modes in the UV. This process, in which boundary conditions are imposed at different points along the holographic direction,  can be implemented numerically either with the midpoint determinant method, as in Ref.~\cite{Berg:2006xy}, or, equivalently, by using a pseudospectral method based on approximating the solutions as a series of the first $K$ Chebyshev polynomials of the first kind~\cite{Boyd_book}. In both cases, it is useful to improve the convergence of the limits by computing the asymptotic expansions, in the IR and UV, of the fluctuations, then imposing the boundary conditions on the expansions, and finally using the results to set up boundary conditions in the numerical process. As discussed for example in Ref.~\cite{Athenodorou:2016ndx}, the presence of cutoffs plays a role similar to that of finite volume and finite discretisation effects in lattice field theories, and as suggested in Ref.~\cite{Elander:2017hyr}, the process of softening the cutoff effects by using short-distance expansions is closely reminiscent of what in lattice field theory goes under the name of improvement~\cite{Symanzik:1983dc,Symanzik:1983gh,Iwasaki:1985we,Luscher:1996sc,Luscher:1996ug}.

\subsubsection{Fluctuation equations}

We present here the linearised equations of motion and boundary conditions satisfied by the fluctuations. In $D$ space-time dimensions, the background solutions are characterised by the radial profiles of the scalars, $\Phi^a(r)$, and the warp factor, $A(r)$, appearing in the metric. We denote the fluctuations of the scalar fields as $\varphi^a(x^\mu,r)$, so that
\beq
	\Phi^a(x^\mu,r) =  \Phi^a(r) + \varphi^a(x^\mu,r) \,.
\eeq
Since the vectors vanish on the background, their fluctuations are simply given by $\mathcal A_M{}^A(x^\mu,r)$. Finally, along the lines of the ADM formalism~\cite{Arnowitt:1962hi}, the metric is written by slicing it along the holographic direction, and decomposed as follows:
\beqs
\label{eq:metricADM}
	\dd s_D^2 &=& \left( (1 + \nu)^2 + \nu_\sigma \nu^\sigma \right) \dd r^2 + 2 \nu_\mu \dd x^\mu \dd r 
	+ e^{2 {A}(r)} \left( \eta_{\mu\nu} + h_{\mu\nu} \right) \dd x^\mu \dd x^\nu \,, \\
    	\label{eq:metricADM2}
	h^\mu{}_\nu &=& \mathfrak e^\mu{}_\nu + \partial^\mu \epsilon_\nu + \partial_\nu \epsilon^\mu 
	+ \frac{\partial^\mu \partial_\nu}{\Box} H + \frac{1}{D-2} \delta^\mu{}_\nu h \,,
\eeqs
with the small fluctuation fields denoted as $\nu(x^\mu,r)$, $\nu^\mu(x^\mu,r)$, $\mathfrak e^\mu{}_\nu(x^\mu,r)$, $\epsilon^\mu(x^\mu,r)$, $H(x^\mu,r)$, and $h(x^\mu,r)$, where $\epsilon^\mu$ is transverse and $\mathfrak e^\mu{}_\nu$ is transverse and traceless. Boundary (field theory) indexes are lowered or raised by the Minkowski metric, $\eta_{\mu\nu}$, and its inverse, $\eta^{\mu\nu}$, respectively, and $\Box\equiv \partial_{\mu}\partial^{\mu}$.

The fluctuation equations for the spin-0 sector involves both the fluctuation of the scalar fields, $\varphi^a(x^\mu,r)$, as well as scalar parts of the metric. It is convenient to work in (diffeomorphism) gauge-invariant variables, which allows one to decouple the equations. The gauge-invariant variables are defined as~\cite{Bianchi:2003ug,Berg:2005pd,Berg:2006xy,Elander:2009bm, Elander:2010wd,Elander:2010wn, Elander:2014ola,Elander:2018aub,Elander:2020csd}
\beqs
\label{Eq:a}
	\mathfrak a^a &=& \varphi^a - \frac{\partial_r \Phi^a}{2(D-2)\partial_r A} h \,, \\
	\mathfrak b &=& \nu - \partial_r \left( \frac{h}{2(D-2)\partial_r A} \right) \,, \\
	\mathfrak c &=& e^{-2A} \partial_\mu \nu^\mu - \frac{e^{-2A} \Box h}{2(D-2) \partial_r A} 
	- \frac{1}{2} \partial_r H \,, \\
	\mathfrak d^\mu &=& e^{-2A} P^\mu{}_\nu \nu^\nu - \partial_r \epsilon^\mu \,,
\eeqs
together with the spin-2 fluctuation, $\mathfrak e^\mu{}_\nu$, defined in Eq.~(\ref{eq:metricADM2}). In these expressions, $\Phi^a$ and $A$ are understood to denote the background solutions, while $P^\mu{}_\nu \equiv \delta^\mu{}_\nu - \frac{\partial^\mu \partial_\nu}{\Box}$ projects to transverse components. The equations of motion for $\mathfrak b $, $\mathfrak c$, and $\mathfrak d^\mu$ are algebraic, leaving us with the decoupled fluctuation equations for $\mathfrak a^a$ and $\mathfrak e^\mu{}_\nu$, from which one can derive the spin-0 and spin-2 spectra, respectively.

The linearised equations of motion for the spin-2 fluctuations, $\mathfrak e^\mu{}_\nu(q^\mu,r)$, are given by
\beq
\label{eq:spin2EOMs}
	\left[ \partial_r^2 + (D-1) \partial_r {A} \partial_r - e^{-2A} q^2 \right] \mathfrak e^\mu_{\,\,\,\nu} = 0 \,,
\eeq
and are complemented by Neumann boundary conditions
\beq
\label{eq:spin2BCs}
	\partial_r \mathfrak e ^\mu{}_\nu \Big|_{r=r_i}= 0 \, .
\eeq

The linearised equations of motion for the spin-0 fluctuations, $\mathfrak a^a(q^\mu,r)$, are given by
\beqs
\label{eq:spin0EOMs}
	&& \Big[ \mathcal D_r^2 + (D-1) \partial_r A \mathcal D_r - e^{-2A} q^2 \Big] \mathfrak a ^a \nonumber \\
	&& - \Big[ V^a{}_{|c} - \mathcal R^a{}_{bcd} \partial_r \Phi^b \partial_r \Phi^d + 
	\frac{4 (\partial_r \Phi^a V^b + V^a 
	\partial_r \Phi^b) G_{bc}}{(D-2) \partial_r A} + 
	\frac{16 V \partial_r \Phi^a \partial_r \Phi^b G_{bc}}{(D-2)^2 (\partial_r A)^2} \Big] \mathfrak a^c = 0 \,,
\eeqs
with the associated boundary conditions reading as follows:
\beqs
\label{eq:spin0BCs}
 \frac{2 e^{2A}\partial_r \Phi^a}{(D-2)q^2 \partial_r A}
	\left[ \partial_r \Phi^b \mathcal D_r - \frac{4 V \partial_r \Phi^b}{(D-2) 
	\partial_r A} - V^b \right] \mathfrak a_b - \mathfrak a^a\Big|_{r_i} = 0 \,,
\eeqs
where the background covariant derivative is  $\mathcal D_r \mathfrak a^a \equiv \partial_r \mathfrak a^a +
 \mathcal G^a_{\ bc} \partial_r  \Phi^b \mathfrak a^c$, while $V^a{}_{|b} \equiv \frac{\partial V^a}{\partial \Phi^b} + \mathcal G^a_{\ bc} V^c$, and the Riemann tensor of the sigma-model metric is given by $\mathcal R^a{}_{bcd} \equiv \partial_c \mathcal G^a{}_{bd} -\partial_d \mathcal G^a{}_{bc}
+ \mathcal G^e{}_{bd} \mathcal G^a{}_{ce} - \mathcal G^e{}_{bc} \mathcal G^a{}_{de}$. These boundary conditions were obtained in Ref.~\cite{Elander:2010wd} from more general expressions, by taking the limit of sending boundary-localised mass terms  towards infinity, which is equivalent to imposing Dirichlet boundary conditions for $\varphi^a\big|_{r_i} = 0$.

Finally, the equations of motion for the transverse, physical parts of the vectors, $\mathcal A^{\mu A}(q^\mu,r)$, are given by
\beqs
\label{eq:spin1EOMs}
	\Big[ e^{-(D-3)A} \partial_r \left[ e^{(D-3)A} H_{CB} \left( \partial_r \mathcal A^{\nu B} - Z^B{}_{Da} \partial_r \Phi^a \mathcal A^{\nu D} \right) \right] - H_{CB} e^{-2A} q^2 \mathcal A^{\nu B} && \nonumber \\
	+ Z^{E}{}_{Cb} \partial_r \Phi^b H_{EB} \left( \partial_r \mathcal A^{\nu B} - Z^B{}_{Da} \partial_r \Phi^a \mathcal A^{\nu D} \right) \Big] P^\mu{}_\nu &=& 0 \,,
\eeqs
with Neumann boundary conditions\footnote{In the limit of removing the UV cutoff, these boundary conditions allows to identify the spin-1 spectrum of the dual field theory, for $M^2 > 0$, when such theory has a continuous, global symmetry. Alternatively, it is possible to (weakly) gauge the global symmetry, by appropriate generalisations of the boundary terms in the UV.}
\beq
\label{eq:spin1BCs}
	\partial_r \mathcal A^\nu{}^A P^\mu{}_\nu \Big|_{r_i} = 0 \,.
\eeq
We note in particular that the terms containing $Z^A{}_{Ba}$ in the equations of motion become important when flux is turned on along the internal part of the $\hat D$-dimensional geometry.

In summary, by solving the equations of motion \eqref{eq:spin2EOMs},~\eqref{eq:spin0EOMs},~and~\eqref{eq:spin1EOMs}, subject to boundary conditions \eqref{eq:spin2BCs},~\eqref{eq:spin0BCs},~and~\eqref{eq:spin1BCs}, identifying those $M^2 = -q^2$ for which solutions exist, and taking the limit of removing the IR and UV cutoffs, one is able to obtain, respectively, the mass spectrum of spin-2, spin-0, and spin-1 states of the dual field theory.

\subsection{Probe approximation: a diagnostic test for the holographic dilaton}

Suppose one has computed the mass spectrum, following the procedure outlined in the previous subsection, and one finds that a light spin-0 state is present in some region of parameter space. One would like to have a way to establish whether or not such state is a dilaton, at least approximately. A diagnostic tool to this purpose was devised and exposed in Ref.~\cite{Elander:2020csd}, to which we refer for details on the derivation of the process and an extensive set of tests. The key  observation this is based upon is that the QFT dilatation operator, $T_\mu{}^\mu$, couples to the scalar component of the metric, $h$, in Eqs.~\eqref{eq:metricADM} and~\eqref{eq:metricADM2}. One then proceeds to calculate the mass spectrum, implementing the so-called probe approximation, which consists of neglecting $h$ in the equations of motion and boundary conditions---see Eq.~(\ref{Eq:a}). Such approximation is only valid for states that do not significantly overlap with the dilatation operator, and hence if one finds that the light scalar state mentioned above no longer is present after the approximation has been made, or that its mass is greatly different, one can conclude that the state, computed correctly with the gauge-invariant formalism,  is a dilaton, as it is sourced by the dilatation operator. Conversely, states that are accurately reproduced in the probe approximation have a negligible dilaton component, and if they happen to be light, it is for some other reason, not related to scale invariance.

The probe approximation consists of replacing Eqs.~\eqref{eq:spin0EOMs}~and~\eqref{eq:spin0BCs} with the following equations:
\beq
	\Big[ \mathcal D_r^2 + (D-1) \partial_r A \mathcal D_r - e^{-2{A}} q^2 \Big] \mathfrak p^a - V^a{}_{|c} \, \mathfrak p^c = 0 \,,
\eeq
with boundary conditions
\beq
	 \mathfrak p^a\Big|_{r_i} = 0 \,,
\eeq
where we have denoted the scalar fluctuations in the probe approximation by $\mathfrak p^a$, to avoid confusion. We emphasise that the probe scalar spectrum, computed in this way, is not part of the actual, physical mass spectrum of the theory, but only serves as a diagnostic, that can be used to identify those states that have a significant dilaton component. In particular, in some of the plots in Sects.~\ref{Sec:Top} and~\ref{Sec:Bottom}, the spectrum computed in the probe approximation is displayed together with the results of the (correct) gauge-invariant treatment, and it should be stressed that these are not two separate sets of states.

\subsection{Free energy}
\label{Sec:free-energy}

Following Sect.~\ref{Sec:holography}, the free energy density is given by minus the on-shell action of the gravity theory. In this brief subsection, we include those parts of the boundary conditions that contribute and which we neglected to write earlier in Eq.~\eqref{eq:SDhat}. They are given by
\beq
	\mathcal S^{(boundary)}_{\hat D} = \frac{1}{4\pi G_{\hat D}} \sum_{i=1,2} (-)^i \int \dd^{\hat D - 1} x \, \sqrt{-\tilde{\hat g}} \left( \frac{\hat K}{2} + \hat \lambda_i(\hat \Phi) \right) \Bigg|_{\rho = \rho_i} \,,
\eeq
where $\hat K$ is the extrinsic curvature appearing in the GHY boundary action, while $\hat \lambda_i(\hat \Phi)$ are boundary potentials.

As for the calculation of the mass spectrum, the physical result is obtained in the limit of removing the IR and UV cutoffs at $\rho = \rho_i$. In the IR, we set
\beq
	\hat \lambda_1 = -\frac{D-2}{2} \partial_\rho A(\rho) \Big|_{\rho = \rho_1} \,.
\eeq
For the DW solutions, this provides a regularisation that renders the free energy finite. At the same time, it reproduces the expected result for the regular soliton solutions, in the limit of taking $\rho_1\rightarrow \rho_o$, towards the end of space.

 In the UV, the scalar potential, $\hat \lambda_2(\hat \Phi)$, contains the counter-terms necessary to cancel QFT divergences, following the procedure of holographic renormalisation~\cite{Bianchi:2001kw,Skenderis:2002wp,Papadimitriou:2004ap}. These may be obtained by solving Eq.~\eqref{eq:VfromW} in terms of $\hat W$, taking care to pick the solution of the equation to the partial derivatives  that has the smallest coefficient for the quadratic term (typically, the dimension of the deformation).\footnote{When one does not know the superpotential $\hat W$ in closed form, it is still possible to find such solution perturbatively in $\hat \Phi$, up to the necessary order that cancels the divergences of the theory and leaves a finite result. This is possible because of the finite number of divergences one expects to arise in the dual field theory.} We denote such solution by $\hat W_2$, and thus define
\beq
	\hat \lambda_2(\hat \Phi) = \hat W_2(\hat \Phi) \,.
\eeq

By making use of the equations of motion,  one can show that the on-shell action evaluates to
\beq
	\mathcal S^{(total)}_{\hat D} \Big|_{on-shell}= - \frac{1}{4\pi G_{\hat D}} \int \dd^{\hat D-1} x \, \mathcal F = - \int \dd^{D-1} x \, \mathcal F \,,
\eeq
where the free energy density, $\mathcal F$, is given by
\beq
	\mathcal F = - \lim_{\rho_2 \rightarrow +\infty} \left. e^{(D-1)A(\rho) - \chi(\rho)} \left( \frac{D-2}{2} \partial_\rho A(\rho) + \hat{W}_2(\hat\Phi(\rho))\right)\right|_{\rho=\rho_2}\,.
\eeq
Since this expression only involves quantities evaluated in the UV, in practice it is often convenient to use the previously discussed UV expansions to obtain explicit expressions in terms of the UV coefficients, including the sources and VEVs of the dual theory.

Finally, in order to be able to extract dimensionless quantities from our analysis, we find it useful to introduce  a scale setting procedure that defines a characteristic energy scale, $\Lambda$, for a given background solution, following Ref.~\cite{Csaki:2000cx}:
\beq
\label{Eq:Lambda}
	\Lambda^{-1} \equiv \int_{r_o}^{\infty} \dd \tilde r \, e^{- A(\tilde r)} = \int_{\rho_o}^{\infty} \dd \tilde \rho \, e^{\chi(\tilde \rho) - A(\tilde \rho)} \,.
\eeq
For instance, the combination $\mathcal F / \Lambda^{\hat D - 1}$ is dimensionless, and will be used in the global stability analysis in Sects.~\ref{Sec:Top} and~\ref{Sec:Bottom}.

\section{Top-down examples}
\label{Sec:Top}

We start the survey of relevant calculations by summarising examples originating in the context of top-down holographic models. We report here only the characterisation of the relevant backgrounds, and their embedding in supergravity, referring the reader to the original publications for details. We discuss the global and local stability analysis based on the holographic calculation of free energy and spectrum of bound states. We comment on the interpretation of one of the lightest bound states as a dilaton, when appropriate. Where useful, we provide also comments about the difference in notation between the original publications and the general prescriptions in Sect.~\ref{Sec:moreholography}.

\subsection{Holographic confinement from maximal supergravity in $\hat{D}=7$ dimensions}
\label{Sec:D7}

\begin{figure}[t]
\centering     
\includegraphics[width=0.4\textwidth]{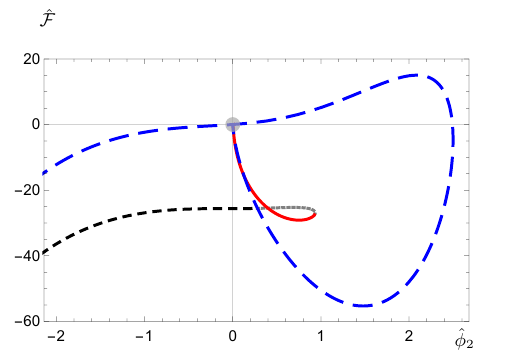}    
\includegraphics[width=0.4\textwidth]{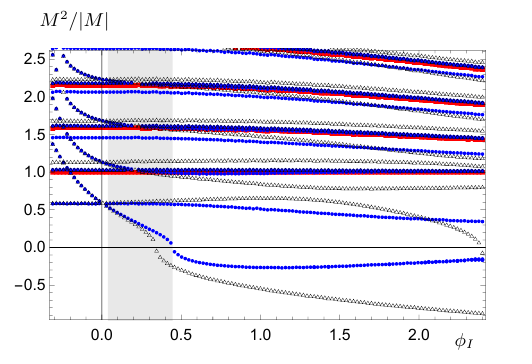}    
   \hfill
	\caption{In the theory discussed in Sec.~\ref{Sec:D7}, the classical gravity background is obtained within the maximal supergravity theory in $\hat{D}=7$ dimensions, truncating it to only one scalar field, $\phi$. 
	The left panel shows the free energy density, $\hat{\cal F}$, normalised to the holographic scale, $\Lambda$, as a function of the deformation parameter, $\hat{\phi}_2$ (itself expressed in units of $\Lambda$), for two distinct one-parameter families of solutions: the regular soliton solutions corresponding to confining field theories (dashed black, short-dashed grey, and solid red lines) and a class of singular domain wall solutions (long-dashed, blue line). The right panel shows the spectrum of masses, $M^2/|M|$, of gauge-invariant fluctuations computed along the one-parameter family of soliton solutions, identified by the asymptotic value of the scalar at the end of space, $\phi_I$. The blue disks denote the spin-0 states, corresponding to the fluctuations of the three scalars retained in the reduction to $D=5$ dimensions.  The red squares are the spin-2 fluctuations. The grey triangles represent the spectrum of scalars computed in the probe approximation, defined by neglecting mixing of the scalars with the trace of the metric.  The masses are normalised so that the lightest spin-2 state has mass $M_2=1$. The vertical grey band coincides with the grey portion of the curve in the left-hand panel, and corresponds to the metastable branch of the regular soliton/confining solutions. Figures taken from Ref.~\cite{Elander:2020fmv}.
  \label{fig:D7}}
\end{figure}

The theory studied in Ref.~\cite{Elander:2020fmv} is a truncation of maximal supergravity in $\hat{D}=7$ dimensions, retaining only one scalar, $\phi$, coupled to gravity.  The potential admits two critical points, one of which is stable, and the corresponding AdS$_7$ solutions are dual to the strongly coupled, ${\cal N}=(2,0)$, theory living in the $\hat{D}-1=6$  dimensions defined by a stack of M5-branes. We focus attention on gravity backgrounds that asymptotically approach this AdS$_7$ solution in the UV. The bulk scalar field corresponds, on the field theory side, to a composite operator of dimension $\Delta=4$. 

Among many possible classical solutions, we focus attention on a one-parameter class of domain-wall solutions that are constructed by varying the value of the scalar in such a way that the solutions approach the AdS$_7$ stable critical point in the UV, with non-vanishing coefficients $\phi_2$, corresponding to a non-trivial source, and $\phi_4$, signalling a condensate. The scalar diverges towards $\phi\rightarrow -\infty$ in the deep IR of the geometry. These solutions are badly singular, as they do not satisfy even the mild requirements of Gubser's criteria~\cite{Gubser:2000nd}, according to which the potential, evaluated along the solution, should be bounded from above. Yet, as we shall see, they are the one with minimal free energy in a portion of parameter space, which makes them particularly important.

Soliton solutions in the same theory are constructed by reducing $D_T=2$ of the dimensions on circles. The resulting theory can be described as a sigma-model in $D=5$ dimensions, consisting of three scalars coupled to gravity. Regular and smooth solutions can be found that generalise Witten's original proposal for the holographic description of confinement~\cite{Witten:1998zw}.  In the resulting one-parameter class of solutions, one of the circles shrinks smoothly to zero size at a finite value of the holographic direction. The non-trivial value of the sigma-model scalar, $\phi$, at this end of space, which is denoted as $\phi_I$, can be used to parametrise the family. The corresponding field theory deformation, $\phi_2$, and condensate, $\phi_4$, are non-trivial functions of $\phi_I$, that can be evaluated numerically.

The global stability analysis of the regular soliton solutions reveals close analogies with the case of the Van der Waals gas discussed in Sect.~\ref{Sec:VdW}. As can be seen in Fig.~\ref{fig:D7}, when one plots the free energy density, $\hat{\cal F}$ (normalised with $\Lambda$), as a function of the deformation parameter, $\hat{\phi}_2$ (also normalised with $\Lambda$), rather than $\phi_I$, one finds a range of $\hat{\phi}_2$ in which the result is multi-valued. This one-parameter class of solutions splits into three separate branches. Two of them cross each other, and compete with one another to provide the minimum of $\hat{\cal F}$ for the interesting range of $\hat{\phi}_2$. A third branch consists of solutions with larger free energy. Interestingly, the line of singular solutions of the domain-wall type intersects the line of soliton solutions as well. There is a first order phase transition between the soliton and domain wall solutions, taking place at a critical value  of the source, $\hat{\phi}_2=\hat{\phi}_2^c\simeq 0.281$, or, equivalently, $\phi_I< \phi_I^c\simeq 0.039$. Below it, the regular soliton solutions are realised, and the dual field theory interpretation in terms of confining theories is viable.  Above the critical value, the domain-wall solutions are energetically favoured. However, due to their singular nature, one has to exercise some caution regarding this interpretation. A complete analysis may require going beyond the regime in which the gravity theory is valid, and it might happen that, should such an analysis be possible, it would reveal additional branches not captured by our treatment.

The local stability analysis is conducted by computing the spectrum of gauge invariant fluctuations, as the concavity theorems cannot be applied, because the regularisation of the free energy makes its first and second derivatives with respect to the source be scheme dependent. The spectrum is displayed in Fig.~\ref{fig:D7} (see also Refs.~\cite{Brower:2000rp,Elander:2013jqa}), for the regular soliton solutions. For $\hat{\phi}_2<\hat{\phi}_2^c$, the mass spectrum shows no striking features. The lightest state is a scalar, having mass approximately half of that of the lightest tensor state. Past the critical value of the deformation, there is a region $\hat{\phi}_2^c<\hat{\phi}_2$ for which the soliton solutions are metastable: they have energy larger than the domain-wall singular ones, yet they do not support tachyons. In this region the mass of the lightest state is parametrically suppressed, and decreases (in respect to the other bound states) until it vanishes, at a special point $\phi_I^0\simeq 0.447$. Also, in this metastable region the mass of the lightest scalar is not captured well by the probe approximation, demonstrating that mixing of the scalar state with the trace of the energy-momentum tensor is large. This state is hence a dilaton, the PNGB associated with scale invariance. For even larger values of the deformation, corresponding to $\phi_I>\phi_I^0$, the lightest state becomes tachyonic. This third branch of soliton solutions is both locally and globally unstable, and cannot be realised physically.

In summary, this theory predicts the existence of a phase transition at zero temperature, triggered by dialling the strength of the source associated to an operator of dimension $\Delta=4$. The lightest state in the theory is a dilaton along the metastable branch of regular soliton solutions past the transition. There is a special value of the deformation for which the mass of the dilaton vanishes exactly, though this value marks the appearance of a tachyonic  instability, far past the transition itself. As the stable solutions can be interpreted as the dual of a confining QFT, while the branch of solutions that is energetically favoured past the transition are of the domain wall type, it is tempting to think of these phenomena in terms of a confinement/deconfinement transition, with the caveat that the singular behaviour of said solutions points to the incompleteness of the gravity description of the physics for positive and large values of the deformation, $\hat{\phi}_2$.

\subsection{Holographic confinement from half-maximal supergravity in $\hat{D}=6$ dimensions} 
\label{Sec:D6}

\begin{figure}[t]
\centering  
\includegraphics[width=0.48\textwidth]{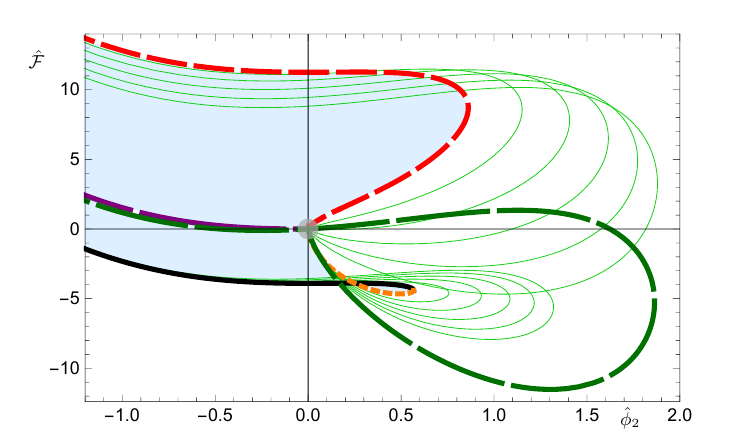}    
\includegraphics[width=0.48\textwidth]{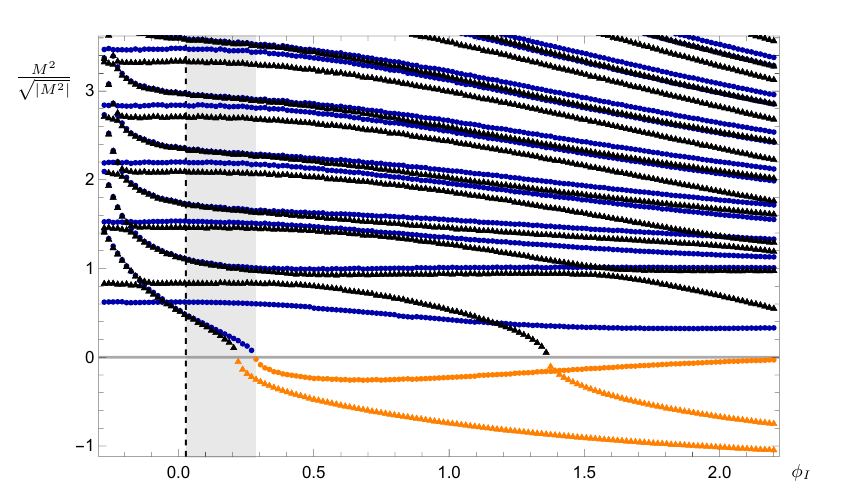}    
   \hfill
	\caption{ In the theory discussed in Sec.~\ref{Sec:D6}, the classical gravity backgrounds are obtained within the half-maximal supergravity theory in $\hat{D}=6$ dimensions, the coset space of the sigma model  of which contains only one scalar field, $\phi$. 
	The left panel shows the free energy, $\hat{\cal F}$, normalised to the holographic scale, $\Lambda$, as a function of the deformation parameter, $\hat{\phi}_2$ (itself expressed in units of $\Lambda$), for several distinct one-parameter and two-parameter families of solutions: the regular, soliton solutions, corresponding to confining field theories (orange and  black lines), the conformal solutions (purple), the singular skewed solutions (red), the singular domain wall solutions (green and cyan). The right panel shows the spectrum of masses, $M^2/|M|$, of gauge-invariant fluctuations computed along the one-parameter family  of regular soliton solutions, labelled by the asymptotic value of the scalar, $\phi$, at the end of space, $\phi_I$. The blue disks denote the spin-0 states, corresponding to the fluctuations of the two scalars retained in the reduction to $D=5$ dimensions, and the black triangles represent the spectrum of scalars computed in the probe approximation, defined by neglecting mixing with the trace of the metric.  The masses are normalised so that the lightest spin-2 state has mass $M_2=1$. Tachyonic states have been highlighted in orange, but retaining the shape of the symbol. The vertical grey band corresponds to the metastable branch of regular soliton solutions, the beginning of which is identified in the left panel by the intersection between the line corresponding to the regular soliton solutions, in black, and the singular domain-wall solutions, in green, and the end of which corresponds to the appearance of a tachyon (dashed orange line). Figures taken from Ref.~\cite{Elander:2020ial}.
  \label{fig:D6}}
\end{figure}

The top-down holographic theory discussed in Ref.~\cite{Elander:2020ial} is based on the half-maximal supergravity in $\hat{D}=6$ dimensions, the scalar coset of which is described by just one scalar, $\phi$, coupled to gravity, while we do not discuss here the other fields of the supergravity theory---see the calculation of the spectrum of bosons in Refs.~\cite{Elander:2018aub,Elander:2020ial}.  The potential admits two critical points, one of which is stable, and the corresponding AdS$_6$ solutions can be lifted to massive type-IIA supergravity in ten dimensions.  The lift involves a non-trivial internal space with a boundary, corresponding to an $O8$ plane~\cite{Ferrara:1998gv,Brandhuber:1999np}, that has AdS$_6\times S^3$ geometry. While locally the internal space is related to a portion of $S^4$,  the isometries of the internal space are  reduced. They realise the $SU(2)$ gauge symmetry of the gauged supergravity in $\hat{D}=6$ dimensions.  A $U(1)$ gauge symmetry that is Higgsed into a massive 2-form is also present. The non-trivial scalar, $\phi$, corresponds to a composite operator of dimension $\Delta=3$ in the dual QFT in five dimensions. 

Domain-wall solutions exist that can be grouped in several families, classified in detail in Ref.~\cite{Elander:2020ial}. A one parameter family of IR-conformal solutions interpolate between the two critical points~\cite{Gursoy:2002tx}. Two separate two-parameter families of singular domain wall solutions exist, one of which ends in good singularities and the other in bad singularities (according to Gubser's criteria~\cite{Gubser:2000nd}). These two classes are separated by singular supersymmetric solutions with vanishing free energy. A special one-parameter family represents the limiting case of the badly singular solutions,  has an especially simple IR expansion, and its complete functional form is known numerically. This limiting case is important in the global stability analysis, to which we return in a couple of paragraphs.

Soliton solutions are constructed by reducing $D_T=1$ of the dimensions on a circle. The resulting theory can be described as a sigma-model in $D=5$ dimensions, consisting of two scalars coupled to gravity (having truncated additional fields coming from tensors and forms in higher dimension). Similar to the model of Sect.~\ref{Sec:D7}, regular and smooth solutions exist that generalise Witten's original proposal for the holographic description of confinement~\cite{Witten:1998zw}, presently in the context of half-maximal supergravity in $\hat{D}=6$ dimensions---see also Refs.~\cite{Wen:2004qh,Kuperstein:2004yf,Elander:2013jqa}. The resulting one-parameter class of regular solutions 
can be parametrised by the value of the scalar, $\phi$, at this end of space, denoted $\phi_I$. The source of the corresponding operator, $\phi_2$, and its condensate, $\phi_3$, are non-trivial functions of $\phi_I$, that are extracted numerically. The analysis in Ref.~\cite{Elander:2020ial} details also the existence of an additional one-parameter class of soliton solutions, referred to as skewed, that are of some interest to the global stability analysis, but are singular and we do not further discuss here.

The global stability analysis is summarised by the left panel of Fig.~\ref{fig:D6}, and once more reveals close analogies with the case of the Van der Waals gas discussed in Sect.~\ref{Sec:VdW}. The free energy density, $\hat{\cal F}$ (normalised with the holographic scale $\Lambda$), is studied as a function of the deformation parameter, $\hat{\phi}_2$. Despite the rich classification of solutions presented above, only two classes are important. First, the line of regular soliton solutions, describing confinement in the dual field theory, splits into three connected branches: a stable one, a metastable one, and an unstable one. The stable and metastable portions are separated by the intersection with the special line of singular solutions of the domain-wall type mentioned at the end of the paragraph above, listing domain-wall solutions. It is displayed as a long-dashed (green) line in the figure. The phase transition takes place when $\hat{\phi}_2=\hat{\phi}_2^c\simeq 0.169$, which corresponds to $\phi_I= \phi_I^c\simeq 0.027$, below which the regular soliton solutions are realised, and the dual field theory interpretation in terms of confining theories is viable. Above this critical value, the true ground state of the theory is not known, and the free energy prefers singular (domain wall) solutions over the regular (soliton) ones, indicating that a more complete analysis may be required. All other solutions are disfavoured by the global stability analysis.

The regularisation of the free energy results in a scheme dependence of the first and second derivatives of ${\cal F}$ with respect to $\phi_2$, hence the local stability analysis is conducted by computing the spectrum of gauge invariant fluctuations. The spectrum is displayed in the right panel of Fig.~\ref{fig:D6} (see also Refs.~\cite{Wen:2004qh,Kuperstein:2004yf,Elander:2013jqa} for partial results). Along the stable branch of solutions, for $\hat{\phi}_2<\hat{\phi}_2^c$, the only non-trivial feature shown by the spectrum is that the mass of one of the two lightest states is not accurately captured by the probe approximation. This observation suggests that it could be interpreted as a dilaton, given that this is evidence of coupling to the trace of the energy-momentum tensor. This feature is somewhat unexpected as this is not the lightest scalar particle, and  its mass is not suppressed---this observation was discussed in Ref.~\cite{Elander:2020csd}. Interestingly, this becomes more clear along the metastable branch,  for $\hat{\phi}_2>\hat{\phi}_2^c$, in which case both of the two lightest states show discernible discrepancies in their masses, in respect to the results of the probe approximation.  One of these two states becomes parametrically light when dialling up $\hat{\phi}_2$.  If one analyses backgrounds with even larger values of $\hat{\phi}_2$, first this light scalar becomes tachyonic, both in the correct as in the probe approximation calculations. Further along the unstable branch of regular soliton solutions, the probe approximation, which never fully reproduces the second to lightest mass, anywhere in parameter space, yields another, unphysical tachyon. This behaviour of the probe calculation highlights the fact that also this state is sourced by the dilatation operator, and non-trivial mixing is present between the two lightest scalar states.

Summarising the salient results in this theory, we find results that are similar to those abridged in Sect.~\ref{Sec:D7}. A phase transition appears at zero temperature, triggered by dialling the strength of the source associated to a QFT operator of dimension $\Delta=3$. The lightest state in the theory behaves like a dilaton along the metastable branch of solutions past the transition, where it can be dialled to arbitrarily light mass, and eventually there is a special value of the deformation for which the mass of the dilaton vanishes exactly,  marking the appearance of an instability, but this happens far past the transition itself.  A distinctive feature, in respect to the theory in Sect.~\ref{Sec:D7}, is that  also a second light state shows evidence of coupling to the trace of the energy-momentum tensor, although it does not become parametrically light.

\subsection{The Coulomb branch: lower-dimensional confinement from maximal supergravity in $\hat{D}=5$ dimensions} 
\label{Sec:D5}

\begin{figure}[t]
\centering  
\includegraphics[width=0.43\textwidth]{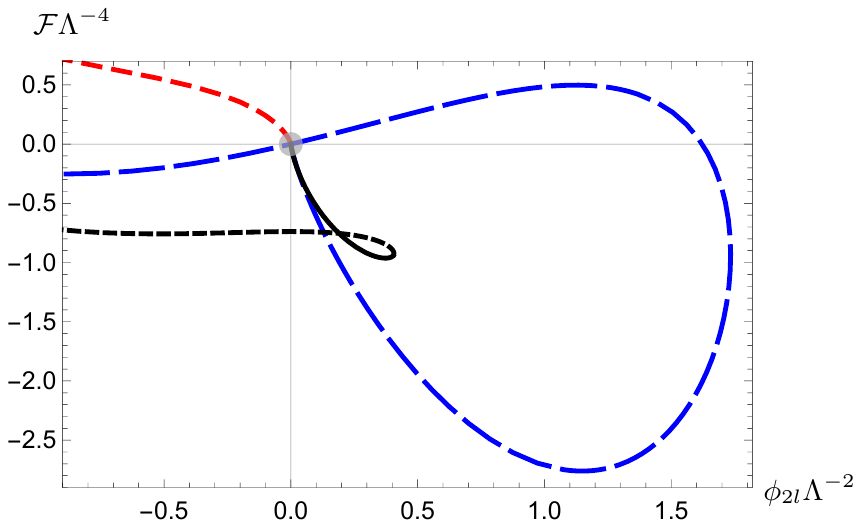}    
\includegraphics[width=0.43\textwidth]{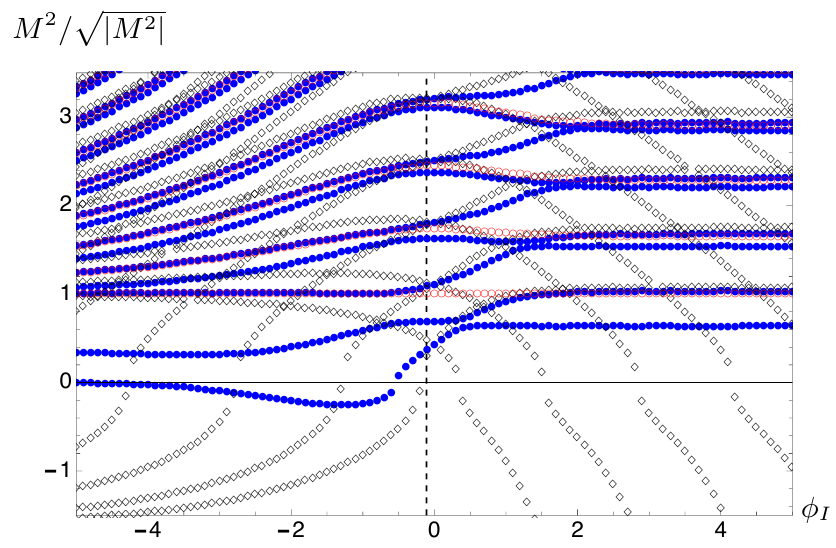}    
   \hfill
	\caption{ In the theory discussed in Sec.~\ref{Sec:D5}, the classical gravity background is obtained within the maximal supergravity theory in $\hat{D}=5$ dimensions, truncating it to only one scalar field, $\phi$. 	The left panel shows the free energy, $\hat{\cal F} = \mathcal F \Lambda^{-4}$ (normalised to the holographic scale $\Lambda$), as a function of the deformation parameter, $\hat{\phi}_{2l} =\phi_{2l} \Lambda^{-2}$ (itself expressed in units of $\Lambda$), for three distinct one-parameter families of solutions: the regular, soliton solutions, corresponding to confining field theories (black, continuous and dashed line), and two classes of singular DW solutions (negative DW in red and positive DW in blue). The right panel shows the spectrum of masses, $M^2/|M|$, of gauge-invariant fluctuations, computed along the one-parameter family of regular soliton solutions, labelled by the asymptotic value of the scalar field, $\phi$,  at the end of space, denoted as $\phi_I$. The blue disks represent the spin-0 states, corresponding to the fluctuations of the two scalars retained in the reduction to $D=4$ dimensions, the red circles correspond to the tensor, spin-2 states, and the diamonds represent the spectrum of scalars computed in the probe approximation, defined by neglecting mixing with the trace of the metric.  The masses are normalised so that the lightest spin-2 state has mass $M_2=1$. The vertical dashed line denotes the value of the parameter, $\phi_I=\phi_I^c\simeq -0.067$, at which a first-order phase transition takes place, and coincides with the intersection of the short-dashed black and long-dashed blue lines in the left plot. A tachyon appears for $\phi_I<\phi_I^{\ast}\simeq -0.52$. Figures taken from Ref.~\cite{Elander:2021wkc}.
  \label{fig:D5}}
\end{figure}

The top-down holographic theory discussed in Ref.~\cite{Elander:2021wkc} is rooted in the ${\cal N}=8$ maximal supergravity in $\hat{D}=5$ dimensions. The scalar coset consists of 42 scalars, that decompose as  $\mathbb{1}_{\mathbb{C}}\oplus 10_{\mathbb{C}}\oplus 20^{\prime}$ of the $SU(4)\sim SO(6)$ gauge symmetry that  corresponds to the isometries of the $S^5$ sphere appearing in the lift to type-IIB supergravity in ten dimensions. We retain only one scalar, $\phi$. Its  non-trivial profile yields to two inequivalent branches  of background solutions (depending on the sign of $\phi$), both of which break $SO(6)\rightarrow SO(2)\times SO(4)$. The scalar potential admits only one critical point, which is stable, and corresponds to the celebrated AdS$_5$ solution that  lifts to AdS$_5\times S^5$ in ten dimensions~\cite{Maldacena:1997re}.  The non-trivial scalar, $\phi$, corresponds to a composite operator of dimension $\Delta=2$ in the dual QFT. Because this is half of the dimensionality of the space-time in the dual QFT, the deformation parameter has itself dimension $4-\Delta=2$, and its presence is signalled by a logarithmic dependence of the dimension-2 coefficient in the UV asymptotic expansion of the background solutions.

Several classes of domain-wall solutions are known to exist, even restricting attention to the ones that leave the  $SO(2)\times SO(4)$ subgroup invariant---see for example the discussion in Refs.~\cite{Pilch:2000ue,Distler:1998gb,Cvetic:2000nc,Bakas:1999ax}. The Coulomb branch of the dual, ${\cal N}=4$ superconformal field theories with $SU(N)$ gauge theory, is identified with those solutions that preserve $16$ supercharges~\cite{Freedman:1999gk}---see also Refs.~\cite{Brandhuber:1999jr,Kraus:1998hv,Cvetic:1999xx}.
In Ref.~\cite{Elander:2021wkc}, special attention is devoted to two special branches of such singular solutions, dubbed positive and negative DW solutions, respectively, in which the divergences are soft enough to allow to perform both a global and local stability analysis, in combination with regular soliton solutions.

Soliton solutions  are built by reducing $D_T=1$ of the dimensions on a circle, and allowing for $\phi$ to be non-trivial, hence generalising the construction referred to as QCD$_3$ in Ref.~\cite{Brower:2000rp}, which studies the  spectrum of fluctuations  for $\phi=0$. After dimensional reduction, the sigma-model in $D=4$ dimensions consists of two scalars coupled to gravity. A  one-parameter class of regular solutions 
can be parametrised by the value of the scalar, $\phi$, at this end of space, denoted $\phi_I$. The corresponding coupling, $\phi_{2l}$, and condensate, $\phi_2$, are related to $\phi_I$, in a way that can be determined numerically.  

The global stability analysis is summarised by the left panel Fig.~\ref{fig:D5}, and again reveals analogies with the Van der Waals gas, discussed in Sect.~\ref{Sec:VdW}. The dimensionless combination, ${\cal F}\Lambda^{-4}$, of the free energy density and the scale $\Lambda$, can studied as a function of the coupling, ${\phi}_{2l}\Lambda^{-2}$. The one-parameter family of regular soliton solutions splits into three connected branches: a stable, a metastable, and an unstable one, respectively.
The stable and metastable portions are separated by the intersection with the line representing singular, positive DW solutions, the former providing the minimum free energy for positive small (or negative) values of $\phi_{2l}\Lambda^{-2}$, the latter for the larger values of $\phi_{2l}\Lambda^{-2}$, as  found in Ref.~\cite{Elander:2021wkc}. The negative DW solutions appear to exist only for negative values of $\phi_{2l}<0$, and their free energy is always higher than the soliton ones. The phase transition between regular soliton solutions and positive DW solutions appears for $\phi_I=\phi_I^c\simeq -0.067$.

The regularisation of the free energy results in the scheme dependence of the first and second derivatives of ${\cal F}$ in respect to $\phi_{2l}$, which is even more severe than for the theories in Sects.~\ref{Sec:D7} and ~\ref{Sec:D6}, as discussed in Ref.~\cite{Elander:2021wkc}, because of the presence of the aforementioned logarithmic dependencies. The mild singularity of the two classes of DW solutions allows us to compute the spectrum of fluctuations with the same gauge-invariant formalism and regularisation process as for the soliton solutions, though we reproduce here only the calculation for the regular solution, referring the reader to the original publication for more details. The negative DW solutions are always unstable, as for any allowed value of $\phi_{2l}$ the spectrum contains a tachyon. Tachyons appear also along the positive DW and regular soliton solutions, but in both cases only along portions of the branches that are disfavoured by the global stability analysis. Interestingly, the probe approximation fails quite spectacularly to capture correctly the spectrum of the scalar excitations along the soliton solutions, showing some level of agreement only for the heavy states, but limited to a region of small $\phi_I$ close to the transition. Even in this restricted region of parameter space, the lightest scalar, the mass of which is parametrically suppressed along the metastable branch of soliton solutions, is not captured by the probe approximation.

The results of this analysis are hence slightly more subtle than those reported in Sects.~\ref{Sec:D7} and~\ref{Sec:D6}, and generalise them. A phase transition is present, at zero temperature, triggered by the source associated with a QFT operator of dimension $\Delta=2$. The lightest state in the theory is a dilaton along the metastable branch of solutions, appearing  past the transition. The state becomes a tachyon at some point beyond the transition. The  nature of the lightest state as a dilaton persists also along the stable branch of soliton solutions, despite the fact that its mass in at the most suppressed by a factor of ${\cal O}(1/2)$, in respect to the lightest tensor state in the theory. A common limitation applies to the three theories analysed in Sects.~\ref{Sec:D7}, \ref{Sec:D6}, and~\ref{Sec:D5}: while the soliton solutions are regular, the domain-wall ones involved in the phase transition are singular, and hence provide only an incomplete description of the physics. Nevertheless, the main results of these studies, namely the fact that a dilaton appears, and its mass is parametrically suppressed along a metastable brunch of regular solution, are robust, as they are not affected by the aforementioned singularity. Even if  additional branches of solutions may exist that overtake the singular ones, the metastability of the branch along which the dilaton becomes massless would be unaffected.

\subsection{Confinement with magnetic fluxes in maximal supergravity in $\hat{D}=7$ dimensions} 
\label{Sec:Flux-D7}

\begin{figure}[t]
\centering  
\includegraphics[width=0.48\textwidth]{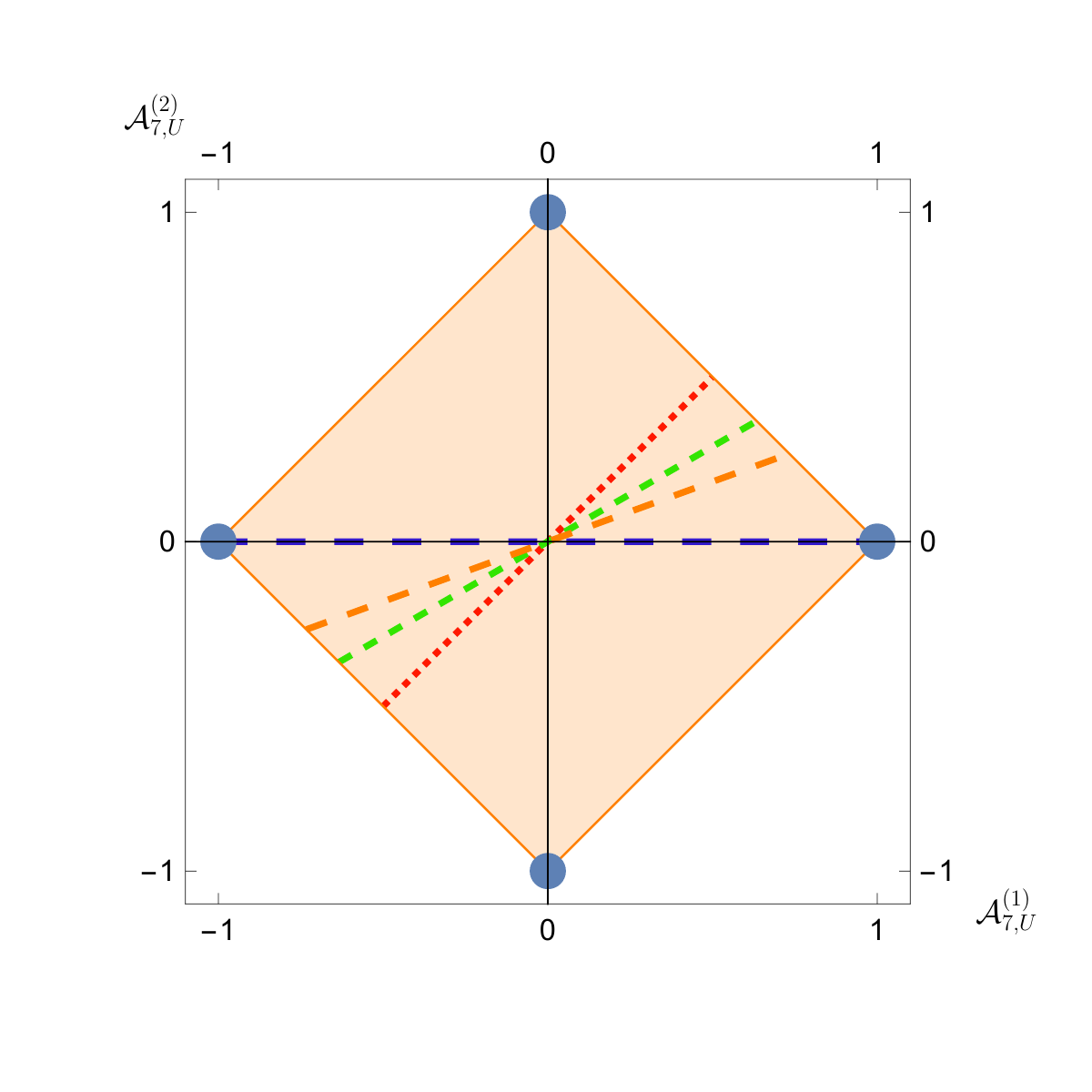}    
\includegraphics[width=0.48\textwidth]{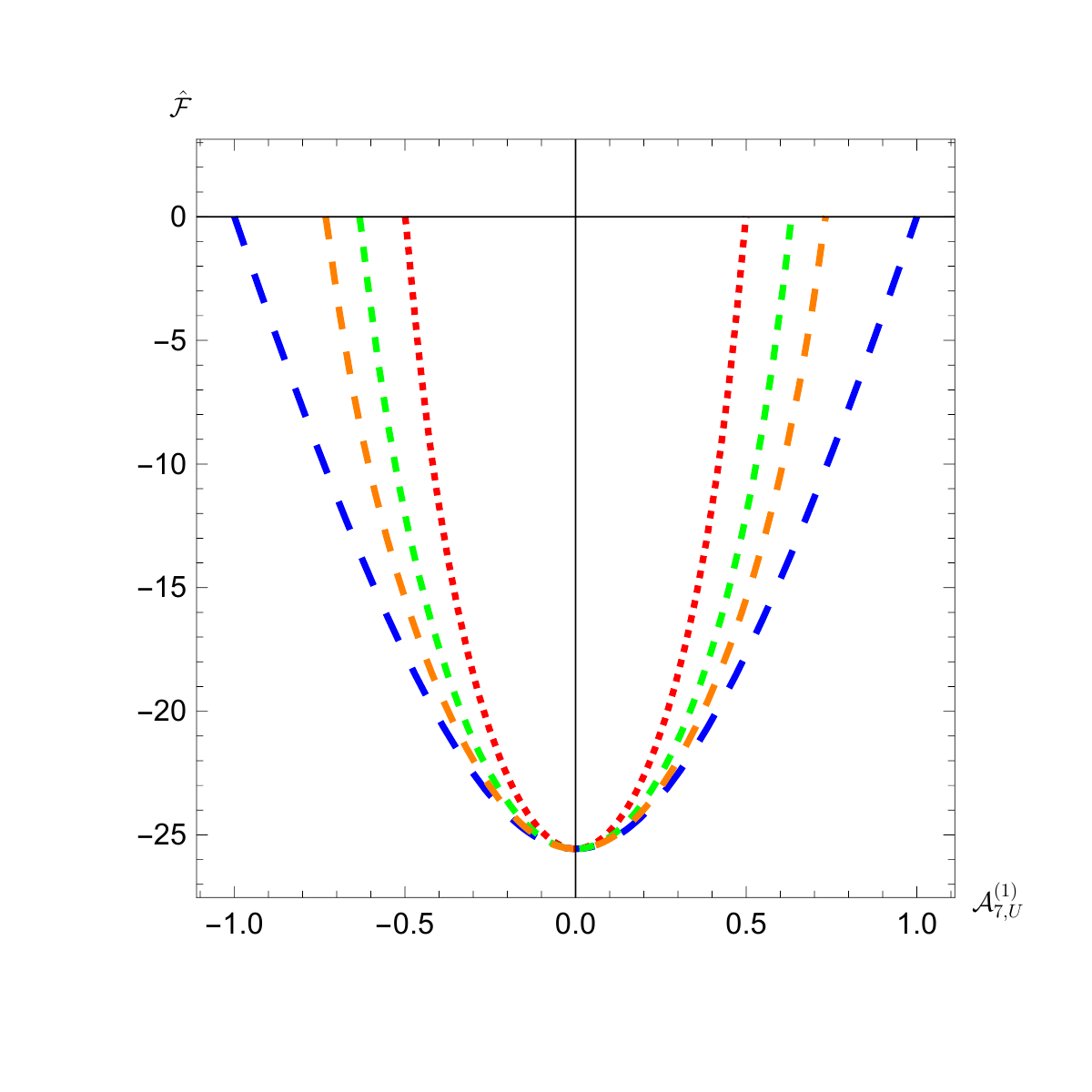}    
\includegraphics[width=0.48\textwidth]{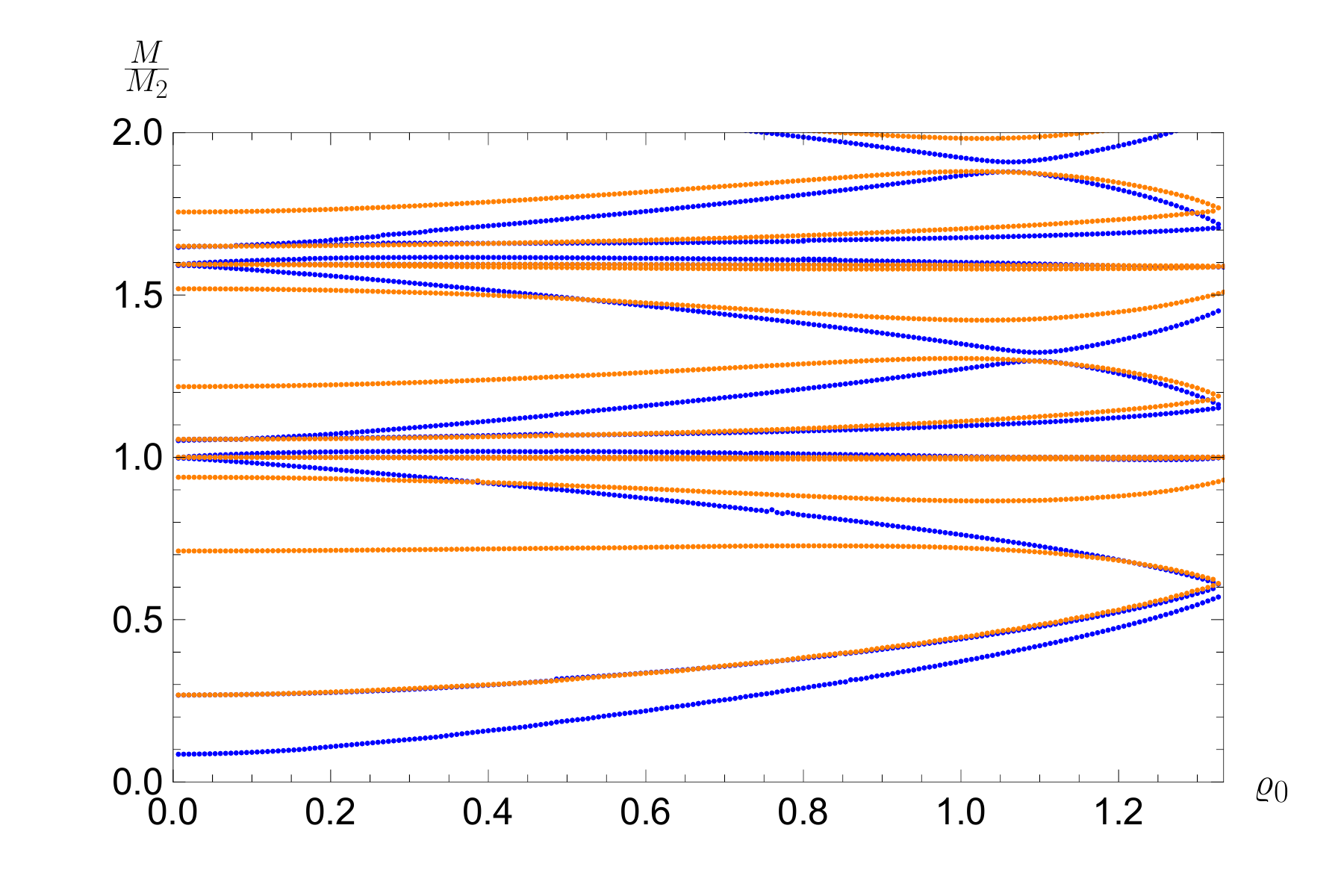}    
   \hfill
	\caption{ In the theory discussed in Sec.~\ref{Sec:Flux-D7}, the classical gravity background is obtained within the maximal supergravity theory in $\hat{D}=7$ dimensions, truncating it to retain only two scalar fields, $\phi_1$ and $\phi_2$, but including two $U(1)$ gauge fields, spanning a subgroup of the gauged $SO(5)$. The theory is then dimensionally reduced, under that assumption that  one dimension is a circle.
	The top-left panel shows the phase diagram of the theory, in the plane defined by the two fluxes, $({\cal A}_{7,U}^1, \,{\cal A}_{7,U}^2)$, of the two $U(1)$s, along the compact direction. Inside the square the soliton solutions have minimal energy. Outside the square the regular domain wall solutions are energetically favoured, and the edges and vertexes denote first-order, zero-temperature, deconfinement phase transitions.
	The top-right panel shows the free energy density, $\hat{\cal F}$ (normalised to the holographic scale $\Lambda$), as a function of the deformation parameters, for four distinct one-parameter families of solutions obtained by fixing the ratio of the two fluxes. The colour coding of the four curves is chosen to match the lines in the top-left panel, for illustrative purposes.
	The bottom panel shows the spectrum of masses, $M^2/|M|$, of gauge-invariant fluctuations computed along the one-parameter line of soliton solutions defined by setting one of the fluxes to zero, corresponding to the blue, horizontal line in the top-left panel. We choose to represent the family of solutions by the position, $\varrho_0$ of the end of space. The blue disks denote the spin-0 states, corresponding to the fluctuations of the five scalars retained in the reduction to $D=5$ dimensions. The orange disks represent the spectrum of scalars computed in the probe approximation, defined by neglecting mixing of the scalars with the trace of the metric.  The masses are normalised so that the lightest spin-2 state has mass $M_2=1$. Figures taken from Ref.~\cite{Piai:2026rst}.
  \label{fig:Flux-D7}}
\end{figure}

For the study reported in Ref.~\cite{Piai:2026rst}, we return to the maximal supergravity in $\hat{D}=7$ dimensions, and apply to it a generalisation of the ideas exposed in Ref.~\cite{Anabalon:2021tua}---see also Refs.~\cite{Nunez:2023xgl, Nunez:2023nnl,Fatemiabhari:2024aua,Chatzis:2024top,Giliberti:2024eii,Chatzis:2025wfv,Chatzis:2024kdu,Kumar:2024pcz,Macpherson:2024qfi,Chatzis:2025dnu,Chatzis:2025hek,Anabalon:2025sok,Macpherson:2025pqi,Fatemiabhari:2025usn,Fatemiabhari:2026goj,Anabalon:2026yxk}. Given an Abelian subgroup of  the gauged symmetries, and assuming one dimension is a circle, a magnetic flux can be turned on, in the form of a non-trivial value of the Abelian gauge field along the compact direction. The resulting solutions are the gravity duals of confining theories with magnetic flux. The existence of such regular solutions is related to the gravity dual of strongly coupled field theories in the presence of finite temperature and chemical potential---see for instance Refs.~\cite{Chamblin:1999tk,Chamblin:1999hg,Gubser:1998jb,Cai:1998ji,Cvetic:1999ne,Cvetic:1999rb,Kim:2006gp,Horigome:2006xu,Kobayashi:2006sb,Mateos:2007vc,Nakamura:2006xk, Karch:2007pd} and the pedagogical introduction in Ref.~\cite{Casalderrey-Solana:2011dxg}.

The construction in Ref.~\cite{Piai:2026rst} retains two scalars, $\phi_{1,2}$, coupled to gravity in $\hat{D}=7$ dimensions, chosen so that when both  develop a  non-trivial profile, the gauge symmetry is broken as $SO(5)\rightarrow SO(2)\times SO(2)$. The  truncation contains also the two Abelian gauge fields, ${\cal A}_{\hat{M}}^{(1),(2)}$, associated with the unbroken $SO(2)\times SO(2)$~\cite{Cvetic:1999ne,Liu:1999ai,Wu:2011gp}---see also the related studies  in Refs.~\cite{Cvetic:1999xp,Bobev:2023bxl,Chong:2004dy,Chow:2011fh}, related to the study of the QFT in the presence of finite chemical potential.  The scalar potential is a generalisation of the one exploited in Ref.~\cite{Elander:2020fmv}, and discussed in Sect.~\ref{Sec:D7}, as a function of $\phi$. One finds the identification $\phi_1=\phi$, but in the present case the second scalar, $\phi_2$, also appears non-trivially in the potential. The dual interpretation, in the context of the strongly coupled, ${\cal N}=(2,0)$, theory living on a stack of M5-branes,  involves two inequivalent composite operators of dimension $\Delta=4$, combining to break the global symmetry to $SO(2)\times SO(2)$. All solutions of interest define gravity backgrounds in which $\phi_1$ and $\phi_2$ vanish asymptotically, and the geometry approaches AdS$_7$ in the UV region of the holographic direction.  

Both domain-wall and soliton solution can be built by performing the reduction of the theory on a circle ($D_T=1$). The resulting theory is a sigma-model in $D=6$ dimensions, consisting of five scalars coupled to gravity: besides the two scalars already present in  $\hat{D}=7$ dimensions, $\phi_{1,2}$, and the scalar that appears in the soliton form of the metric, $\chi$, which controls the volume for the circle, the seventh components of the two gauge fields are also retained, ${\cal A}_{7}^{(1),(2)}$. Domain-wall solutions of interest have AdS geometry and constant  ${\cal A}_{7}^{(1),(2)}\equiv {\cal A}_{7,U}^{(1),(2)}$.
A two-parameter family of soliton solutions exists, in which ${\cal A}_{7}^{(1),(2)}$ asymptotically reach constants in the UV, but vanish exactly at the end of space in the IR, at a value, $\varrho_0$, of the holographic direction. The holographic direction denoted as $\varrho$ is related non-trivially to $\rho$, via a change of variable controlled by a monotonic functions, that is described in detail in Ref.~\cite{Piai:2026rst}. The two scalars, $\phi_{1,2}$,  are non-trivial, but vanish asymptotically. The value of the dimensionless constants ${\cal A}_{7,U}^{(1),(2)}$ serve unambiguously as a parametrisation of the whole two-parameter family. The non-linear relations between integration constants that ensure regularity of the geometry when approaching $\varrho_0$ are discussed in detail in Ref.~\cite{Piai:2026rst}, to which we refer the reader for details. We notice here that the space extends for $\varrho_0 \leq \varrho\leq +\infty$, but the value of $\varrho_0$ is different for each solution. It can be used as part of an alternative  parametrisation of the individual soliton solutions, in which case it is convenient to use the ratio of the asymptotic values ${\cal A}_{7,U}^{(2)}/{\cal A}_{7,U}^{(1)}$ as the second parameter.

The first panel of Fig.~\ref{fig:Flux-D7} shows the phase diagram. Domain-wall solutions exist for any values of ${\cal A}_{7,U}^{(1),(2)}$. Regular soliton solutions exist only inside a square, obtained by joining the vertexes, $\left({\cal A}_{7,U}^{(1)},\,{\cal A}_{7,U}^{(2)}\right)=(1,\,0),\,(0,\,1),\,(-1,\,0),\,(0,\,-1)$. The free energy density of the domain-wall solutions vanishes identically. The regular soliton solutions inside the square have negative-definite free energy, hence are always favoured over the domain-wall ones, for values of ${\cal A}_{7,U}^{(1),(2)}$ inside the square. Along the edges of the square, the soliton solutions have vanishing free energy, and are distinct from the domain-wall ones, a hint of  phase coexistence, characteristic of first-order phase transitions. Unfortunately, in this limit the soliton solutions are singular. The minimum of the free energy, measured in units of the holographic scale, $\Lambda$, is continuous everywhere, while its first derivative with respect to the two deformation parameters is discontinuous along the edges of the square, as shown in the second panel of Fig.~\ref{fig:Flux-D7}. This observation is compatible with the existence of a first-order phase transition taking place along the edges. We can refer to this phenomenon as a confinement/deconfinement transition, given that the two competing classes of solutions are both regular and both admit a clear QFT interpretation. Yet, the aforementioned appearance of a singularity at the boundary of the family of soliton solutions, and the absence of metastable branches appearing past it, suggest to adopt some caution in this interpretation.

The local stability analysis is conducted by computing the spectrum of gauge invariant fluctuations, for a few representative examples of one-parameter subfamilies of the soliton solutions, chosen by holding fixed the ratio of the deformation parameters, ${\cal A}_{7,U}^{(2)}/{\cal A}_{7,U}^{(1)}$. One example of such an analysis, in which ${\cal A}_{7,U}^{(2)}=0$,  is displayed in the third panel of Fig.~\ref{fig:Flux-D7}. The position along the branch is parametrized by the value of $\varrho_0$, which vanishes at the vertex of the square, and reaches its maximum, $\varrho_0=\frac{4}{3}$, at the geometric centre of the square.
The two lightest scalar states are never massless, yet their masses are numerically suppressed, when approaching the transition. the heaviest  of them is well captured by the probe approximation, while the lightest one is not. This suggests that the lightest state is a dilaton, and that in a significant portion of parameter space the suppression of its mass, in respect to the typical mass of other bound states in the dual QFT, is numerically significant, reaching over an order of magnitude in comparison to the lightest spin-2 state.

Recapitulating the results highlights several distinctive features of this theory. The two-parameter phase diagram shows a closed line of first-order  transitions between confined and deconfined phases, captured by regular backgrounds in the dual gravity description. The spectrum of bound states shows that there are sizeable regions of parameter space in which a dilaton is present, along physically realised branches of solutions. Its mass is numerically suppressed,  though not  parametrically. Furthermore, the presence of a singularity at the edges of the square, in correspondence to the line of phase transitions, indicates the presence of regions of space-time with large curvature, near the end of space of the soliton solutions. This might be an indication that the dual gravity description is incomplete, and this observation should be the subject of further studies in the future.

\subsection{Confinement with magnetic fluxes in half-maximal supergravity in $\hat{D}=6$ dimensions} 
\label{Sec:Flux-D6}

\begin{figure}[t]
\centering  
\includegraphics[width=0.4\textwidth]{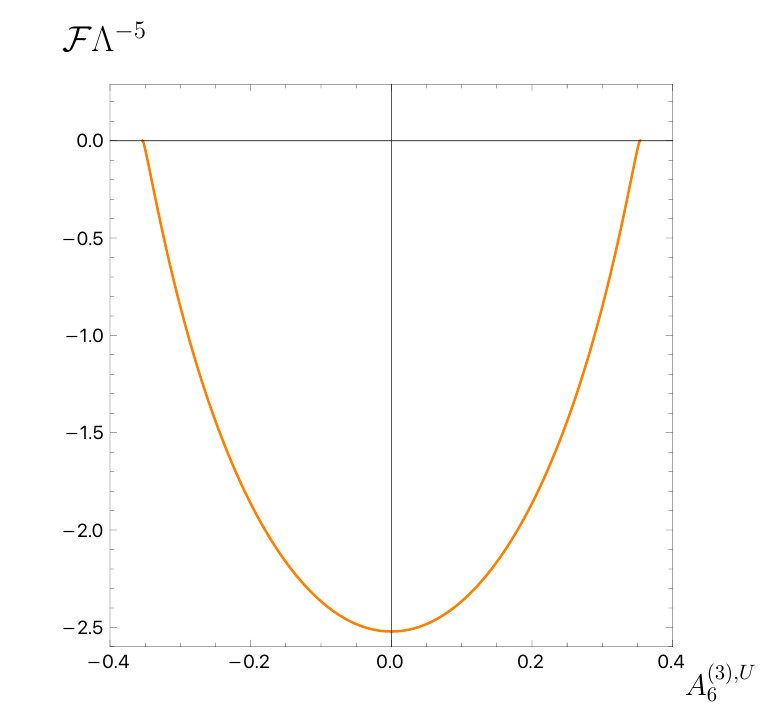}    
\includegraphics[width=0.4\textwidth]{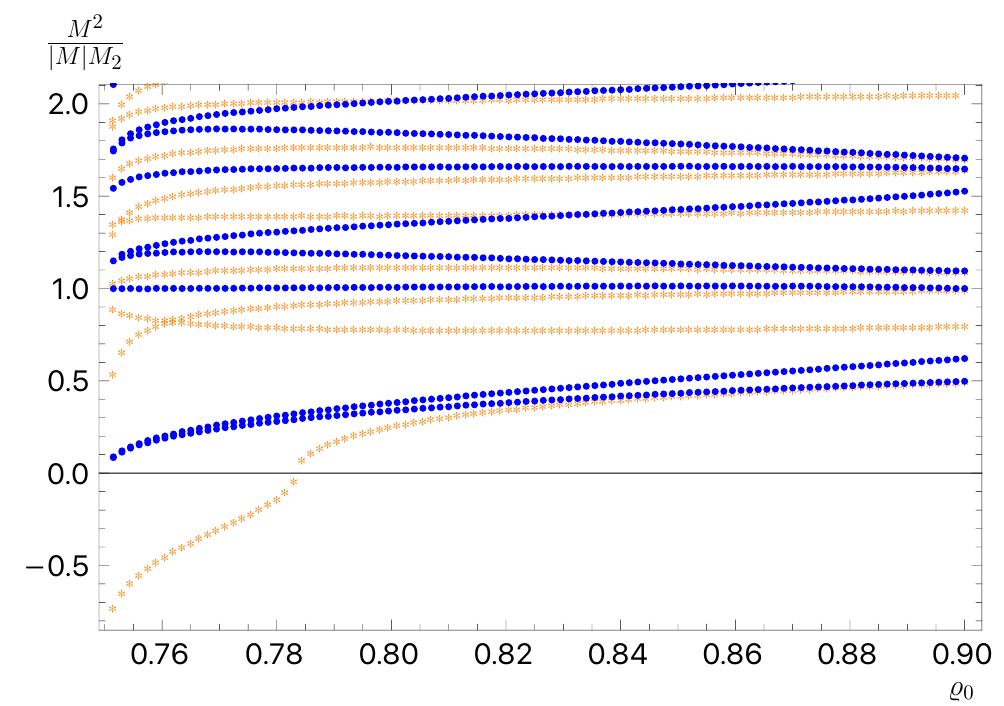}    
   \hfill
	\caption{ In the theory discussed in Sec.~\ref{Sec:Flux-D6}, the classical gravity background is obtained within the half-maximal supergravity theory in $\hat{D}=6$ dimensions, the coset being parametrised by only one scalar field, $\phi$, truncating the rest of the field content to retain only one Abelian, $U(1)$, gauge field, corresponding to the third generator, $T^{(3)}$, of $SU(2)$. One then performs a  dimensional reduction on a circle.
	The left panel shows the dimensionless combination, $\hat{\cal F}={\cal F} \Lambda^{-5}$, of the free energy and the holographic scale, $\Lambda$, as a function of the deformation parameter, ${A}_6^{(3),U}$, for the one one-parameter family of soliton solutions corresponding to confining QFTs. The domain-wall solutions are regular, exist for any value of ${A}_6^{(3),U}$, and have vanishing free energy. 	The right panel shows the spectrum of masses, $M^2/(|M|M_2)$, of gauge-invariant fluctuations computed along the one-parameter line of regular soliton solutions, represented by the position, $\varrho_0$, of the end of space. The blue disks denote the spin-0 states, corresponding to the fluctuations of the three scalars retained in the reduction to $D=5$ dimensions. The orange disks represent the spectrum of scalars computed in the probe approximation, defined by neglecting mixing of the scalars with the trace of the metric.  The masses are normalised so that the lightest spin-2 state has mass $M_2=1$. Figures taken from Ref.~\cite{Fatemiabhari:2026rju}.
  \label{fig:Flux-D6}}
\end{figure}

In the case of the half-maximal supergravity in $\hat{D}=6$ dimensions, the solutions dual to confining theories with magnetic flux, built again along the lines of Ref.~\cite{Anabalon:2021tua} (as in Sect.~\ref{Sec:Flux-D7}, for the maximal supergravity in $\hat{D}=7$ dimensions) have been presented in Ref.~\cite{Fatemiabhari:2024aua}, and their stability analysis presented in Ref.~\cite{Fatemiabhari:2026rju}.  The one scalar, $\phi$, is coupled to gravity, and furthermore one retains in $\hat{D}=6$ dimensions the Abelian $U(1)$ gauge boson connected with the diagonal generator of the gauged $SU(2)$ symmetry. All solutions of interest define gravity backgrounds that are asymptotically AdS$_6$ in the UV, with $\phi\rightarrow 0$. 

It is convenient  to discuss domain-wall and soliton solutions by first performing the circle reduction of the theory. The resulting sigma-model in $D=5$ dimensions consists of three scalars coupled to gravity: besides $\phi$, and the scalar, $\chi$, which controls the volume for the circle, the sixth component of the gauge field is also retained, ${A}_{6}^{(3)}$.\footnote{The superscript, $^{(3)}$, refers to the $T^3$ generator within $SU(2)$, conventionally taken to be proportional to the third Pauli matrix.}  Domain-wall solutions have AdS$_6$ geometry, vanishing $\phi$, and constant  ${A}_{6}^{(3)} \equiv { A}_{6,U}^{(3)}$.  A one-parameter family of soliton solutions exists, in which ${A}_{6}^{(3)}$ asymptotically reaches a constant in the UV. It vanishes at $\varrho\rightarrow \varrho_0$, the end of space in the interior of the geometry, along the holographic direction. The latter is parameterised in terms of the new variable $\varrho$, defined implicitly by a change of variable in respect to $\rho$. The scalar, $\phi$,  has a non-trivial vacuum profile,  but vanishes asymptotically at the conformal boundary. The integration constants are tuned to ensure regularity of the geometry when approaching $\varrho_0$. Once all the constraints are imposed, the value of the dimensionless constant, ${ A}_{6,U}^{(3)}$, labels the one-parameter family of soliton solutions, all other coefficients depending on it in a way that can be solved for numerically.

The first panel of Fig.~\ref{fig:Flux-D6} shows the free energy density, expressed in units of the holographic scale, ${\cal F}\Lambda^{-5}$, of the relevant solutions, as a function of ${ A}_{6,U}^{(3)}$. Regular soliton solutions exist for $\left|{A}_{6,U}^{(3)}\right|<{A}_{6,U}^{(3)}({\rm max})=\frac{1}{2\sqrt{2}}\simeq 0.35$ and have negative definite free energy, ${\cal F}<0$.  Domain-wall solutions exist for any ${A}_{6,U}^{(3)}$, and have vanishing free energy,  ${\cal F}=0$.
The limiting case of the soliton solutions with $\left|{\cal A}_{6,U}^{(3)}\right|=\frac{1}{2\sqrt{2}}$ results in backgrounds that have a singularity at the end of space in the holographic direction, $\varrho\rightarrow\varrho_0=\frac{3}{4}$, and vanishing free energy. The physically realised soliton solutions, dual to confining theories with magnetic flux,  give way to deconfined solutions at ${A}_{6,U}^{(3)}({\rm max})$, at which point a phase transition similar to the one described in Sect.~\ref{Sec:Flux-D7} appears. The free energy is continuous, and its first derivative is discontinuous, as two different solutions have the same free energy at the transition point. At odds with the Van der Waals case presented in Sect.~\ref{Sec:VdW}, there is no evidence of a metastable branch.

The local stability analysis is conducted by computing the spectrum of gauge invariant fluctuations, along the one-parameter families of soliton solutions, and the results are exhibited in the right panel of Fig.~\ref{fig:Flux-D6}. The mass of the scalar bound states are plotted as a function of the value of $\varrho_0$, that sets the end of space, and is in one-to-one correspondence with ${A}_{6,U}^{(3)}$, the asymptotic value of the flux. The two lightest scalar states are never massless, yet their masses are numerically suppressed and close to degenerate across the whole parameter space. The lightest of the two   is well captured by the probe approximation, away from the transition (i.e, for $\varrho_0\gg \frac{3}{4}$,  while the next-to-lightest one is not. The complete calculation shows that both these states become parametrically light close to the transition, and that the probe approximation yields an unphysical tachyon in a finite  range of $\varrho_0$. Both the lightest scalars share some of the defining properties of the dilaton, such as its coupling to the energy-momentum tensor.

The results of the analysis of this theory sum up to a picture that in several respects resemble those obtained in the case of the confining solutions with magnetic fluxes obtained in the maximal supergravity in $\hat{D}=7$ dimensions. In particular, it is possible to identify a confinement/deconfinement transition between regular background solutions, except for the fact that, as one approaches the transition, the soliton solutions approach a singular case. There is no branch of metastable solutions in proximity of the transition. There is a dilaton, with mass suppressed numerically in proximity of the transition. But there are also two striking elements of novelty. There is a second state that is approximately degenerate with the dilaton, and furthermore it couples to the energy-momentum tensor, when approaching the transition. The fact that these effects appear in a region of parameter space in which the soliton background solutions have large curvature near the end of space requires further study.

\subsection{Parametrically light dilaton near critical points in top-down holography} 
\label{Sec:TD-2}

The theory described and analysed in Ref.~\cite{Elander:2025fpk} has a more subtle origin in top-down holography. The solutions of interest can be shown to be regular in maximal supergravity in eleven dimensions. The truncation to $\hat{D}=4$ dimensions involves six scalars coupled to gravity~\cite{Elander:2018gte,Faedo:2017fbv,Elander:2020rgv}, and a further reduction on a circle yields a gravity theory in $D=3$ dimensions, coupled to seven scalar and one $U(1)$ gauge field. The classes of solution of interest are parameterised by two constants. The microscopic origin of these quantities can be traced back in the literature, but is not directly relevant to the discussion we present in these pages, for which purposes the reader can simply think of a non-trivial, strongly coupled field theory in four dimensions, with two independent deformations, called $b_0$ and $\ell \Lambda$, one of which corresponds to the compactification of a dimension on a circle,  a response function called $\langle T_{22}\rangle\ell^3 \alpha^{-1}$, and the free energy ${\mathcal F} \ell \alpha^{-1}$. As we borrow the figures from Ref.~\cite{Elander:2025fpk}, we notice that in this subsection the masses of the gauge-invariant fluctuations are denoted as $m^2$ (rather than $M^2$ as elsewhere in the review), and that the scale $\Lambda$ is not the same as the one of Eq.~(\ref{Eq:Lambda}), but has been defined in a way that is explained in Ref.~\cite{Elander:2025fpk}.

\begin{figure}[t]
\centering  
\includegraphics[width=0.4\textwidth]{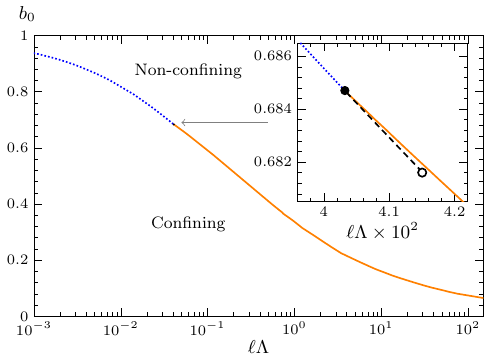}    
\includegraphics[width=0.4\textwidth]{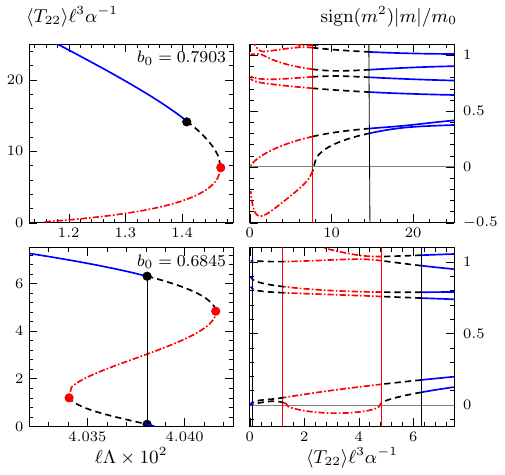}    
   \hfill
	\caption{In the theory discussed in Sec.~\ref{Sec:TD-2}, the classical gravity background is obtained from dimensional reduction on a circle of a special sigma-model consisting of six scalars coupled to gravity in $\hat{D}=4$ dimensions~\cite{Elander:2018gte,Faedo:2017fbv,Elander:2020rgv}. The resulting gravity theory in $D=3$ dimensions consists of seven real scalars and one $U(1)$ gauge field, and the regular solutions of interest describe a two-parameter family of backgrounds.
	The left panel shows the phase diagram of the theory, described by the dimensionless parameters $b_0$ and $\ell\Lambda$. Three lines of first-oder phase transitions are present. The dotted blue line represents a deconfining transition between regular solutions. The orange, continuous line describes a transition between soliton and singular solutions. The black dashed line separates two different classes of soliton solutions corresponding to different confining field theories. It joins the resulting triple point, with $b_0^{\rm triple}\simeq 0.6847$ (black disk), to the critical point, with  $b_0^{\rm CP}\simeq 0.6815$ (white disk).
	The four panels on the  right show the response functions, $\langle T_{22}\rangle \ell^3 \alpha^{-1}$ (left), and scalar
spectrum (right) for two values of $b_0$ large enough to display the phase transitions. In the very the top row is a representative
case chosen well above the triple point, where the first-order phase transition is between confining and non-confining solutions. In the bottom row $b_0 \simeq b_0^{\rm triple}$, slightly below the triple point.
	Figures taken from Ref.~\cite{Elander:2025fpk}.
  \label{fig:TD-2-a}}
\end{figure}
\begin{figure}[t]
\centering  
\begin{picture}(160,120)
\put(0,0){\includegraphics[width=0.4\textwidth]{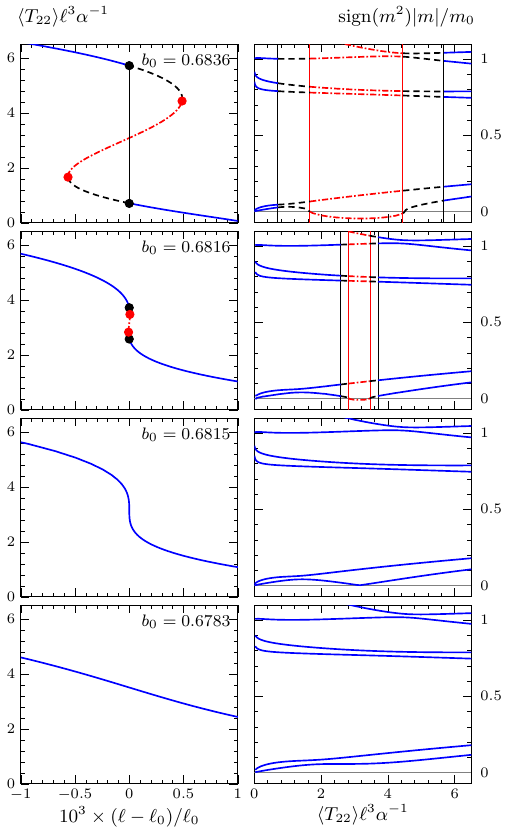}   }
\put(80,80){\includegraphics[width=0.4\textwidth]{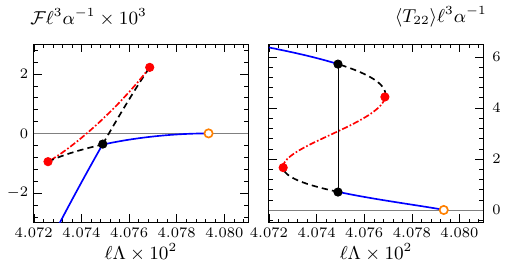}    }
\put(80,0){\includegraphics[width=0.4\textwidth]{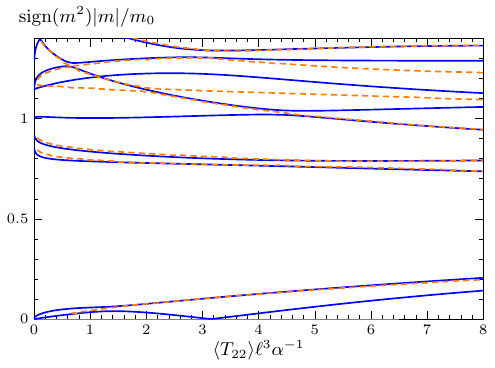}    }
\end{picture}
   \hfill
	\caption{ For the theory discussed in Sec.~\ref{Sec:TD-2}, the eight  panels on the left half of the figure show  the response function, $\langle T_{22}\rangle \ell^3 \alpha^{-1}$, as a function of the dimensionless parameter $\ell \Lambda$, for selected values of $b_0$, chosen to cluster around the line of first-order phase transitions denoted by black dashed lines in the left panel of Fig.~\ref{fig:TD-2-a}, and the mass spectrum of scalars, $m$, expressed in units of the lightest vector mass for the same choices of parameters as  in the  left panels.
The  two panels on the top right half show the free energy and response function expressed in dimensionless units, ${\cal F}\ell\alpha^{-1}$ and $\langle T_{22}\rangle \ell^3 \alpha^{-1}$, as a function of the dimensionless parameter $\ell
	\Lambda$, for a representative value of $b_0^{\rm triple}>b_0=0.6836>b_0^{\rm CP}$, highlighting the swallow-tail behaviour appearing in proximity of the confinement-confinement phase transition, and a further phase transition structure at larger values of $\ell \Lambda$ (orange circle).	The bottom-right panel shows  the mass spectrum of scalar fluctuations computed for
$b_0 = b_0^{\rm CP} \simeq 0.6815$, along confining solutions, comparing the correct calculation (blue solid lines) to the results of the probe approximation	(dashed orange lines).	
  	Figures taken from Ref.~\cite{Elander:2025fpk}.
	  \label{fig:TD-2-b}}
\end{figure}

The phase diagram of this theory is very rich, and is depicted in the left panel of Fig.~\ref{fig:TD-2-a}, in the plane defined by the two dimensionless parameters, $b_0$ and $\ell \Lambda$. There are three lines of first-order phase transitions, meeting at a triple point, with $b_0^{\rm triple}\simeq 0.6847$. Two of the lines emanating from this point separate confining from non-confining theories, as the diagram shows. In the case of the line going towards large values of $b_0>b_0^{\rm triple}$, the first-order transition is strong, all the backgrounds involved are regular, and along the regular branch of confining solutions, when they are stable, there are no particularly interesting features in the spectrum of bound states, as can be seen in the two top-right panels of Fig.~\ref{fig:TD-2-a}. A suppression of the mass and eventually a tachyon manifest themselves only in a region of parameter space far past the transition, as we saw in other top-down constructions, earlier in this section.

Most interesting is the behaviour displayed in the range $b_0^{\rm triple}>b_0\geq b_0^{\rm CP}$, which can be seen in the bottom-right panels of Fig.~\ref{fig:TD-2-a}, and the top-left rows of Fig.~\ref{fig:TD-2-b}. Here one finds that upon varying $\ell \Lambda$ from small to large values, the system undergoes  two phase transitions. The first involves two different types of regular backgrounds corresponding to two separate confined phases of the dual QFT. As the plots show, in this case the analogy with the Van der Waals gas is very close: there is coexistence of phases, with a stable and a metastable phase competing to minimise the free energy, and a third, unstable solution showing a tachyonic state in the spectrum.  Moving down to smaller values of  $b_0$, towards the critical point, the confined/confined transition becomes weaker, and the range of $\ell \Lambda$ over which phases coexists shrinks. At $b=b_0^{\rm CP} \simeq 0.6815$, the free energy and its first derivative are continuous, but the second derivative diverges at the transition. This line of first-order phase transitions ends at a critical point, in a second-order one. Beyond this point, the transition disappears, as can be seen in the bottom-left panels of Fig.~\ref{fig:TD-2-b}. Most interesting, for $b_0=b_0^{\rm CP}$ there is an exactly massless scalar state in the spectrum, to which we will return shortly.

Examining further the plots in Fig.~\ref{fig:TD-2-b}, but moving towards larger values of $\ell \Lambda$,
one finds a second line of first-order transitions. But this transition resembles those described in Sects.~\ref{Sec:Flux-D7} and ~\ref{Sec:Flux-D6}. This is best seen in the two plots on the top right section of Fig.~\ref{fig:TD-2-b}. The second phase transition is between confining and non-confining solutions, and at the transition the curvature, evaluated at the end of space of the background solutions, becomes divergent. Furthermore, in this case there are no metastable branches of solutions. The lightest scalar has mass that approaches zero, but with current information it is not readily possible to assess whether the light mass is a physical effect or a signal of the breakdown of the supergravity approximation. It is worth noting that, as in the case discussed in Sect.~\ref{Sec:Flux-D6}, a second light scalar state appears in proximity of this line of singular confinement/deconfinement phase transitions, with unclear origin.

We return now to the solutions in proximity of the critical point, and analyse the spectrum of bound states in detail, along a line of fixed $b_0=b_0^{\rm CP}$. In the bottom-right panel of Fig.~\ref{fig:TD-2-b}, we compare the spectrum of scalar bound states, as a function of the response function $\langle T_{22}\rangle\ell^3 \alpha^{-1}$, computed along the resulting line of solutions, with the  probe approximation. First, the main result of general value of this whole review is the arising of an isolated, light scalar that becomes exactly massless for $\langle T_{22}\rangle\ell^3 \alpha^{-1}\simeq 3$, which coincides with the position of the critical point at the end of the black dashed line in the phase diagram in Fig.~\ref{fig:TD-2-a}. We interpret this state as a dilaton, on the basis of  the fact that the probe approximation fails completely to capture its very existence, which indicates that it couples to the trace of the energy-momentum tensor, as dictated by scale invariance~\cite{Coleman:1985rnk}.

This striking result is the first and only known example to date of an exactly massless holographic dilaton, emerging in top-down holography, for a controllable calculation within a strongly coupled field theory, for which the mass of the dilaton is stabilised by the presence of a second-order phase transition at the critical end point of a line of first-order transitions.  In proximity of this critical point one can arrange for the dilaton to be arbitrarily light, and retain its coupling to the trace of the energy-momentum tensor. The only limitation of this construction is that the dual field theory is lower dimensional: the search for an example of this type in four dimensions is still ongoing.

\section{Bottom-up examples}
\label{Sec:Bottom}

As we have seen in Sect.~\ref{Sec:Top}, it can be quite challenging to build (or, better, find) top-down holographic models that are relevant to the research programme illustrated in these pages. The space-time dimensionality, the number of sigma-model fields, the coset they describe, the residual symmetries, and the interactions encoded in the potential and sigma-model metric, are all rigidly determined, within the catalogue of known supergravities and their truncations. The ordinary non-linear differential equations that control backgrounds of the domain-wall or soliton type require tuning the integration constants to eliminate curvature and conical singularities. This process makes the non-linear dynamics hard to tame. Only once this is done, and the holographic dual of a confining QFT has been identified, the global and local stability analysis can be performed, to test whether a light dilaton is present in the spectrum of bound states, and whether its mass is influenced by possible zero-temperature phase transitions. 

The bottom-up approach to holography offers a more flexible, pragmatic framework, in which one can fast-forward the process, by cutting across the complications and rigid requirements related to the first part of the top-down programme. In particular, the number of space-time dimensions, the dimensionality of the local operators the dual of which are part of the scalar manifold, and the symmetries of the model, can all be chosen on the basis of phenomenological and simplicity arguments. Rather than exploring the space of microscopic, fundamental realisations of a field-theoretical feature of interest, the bottom-up approach is useful to map out potential correlations between different, derived quantities of phenomenological relevance. In this section we discuss three interesting examples of this approach. We highlight the striking similarities between these models and the classes of top-down theories examined in Sect.~\ref{Sec:Top}. We use the combined information to draw general lessons for future studies.

\subsection{First-order phase transitions in a simplified bottom-up model}
\label{Sec:BU-D6}

\begin{figure}[t]
\centering  
\includegraphics[width=0.8\textwidth]{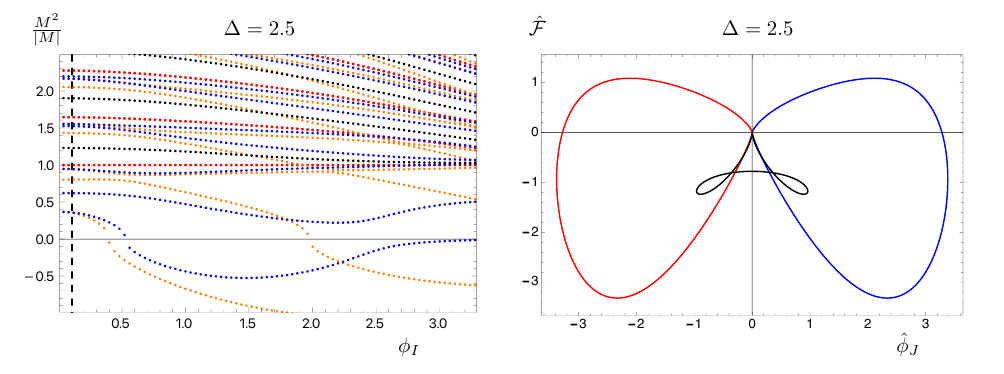}    
\includegraphics[width=0.4\textwidth]{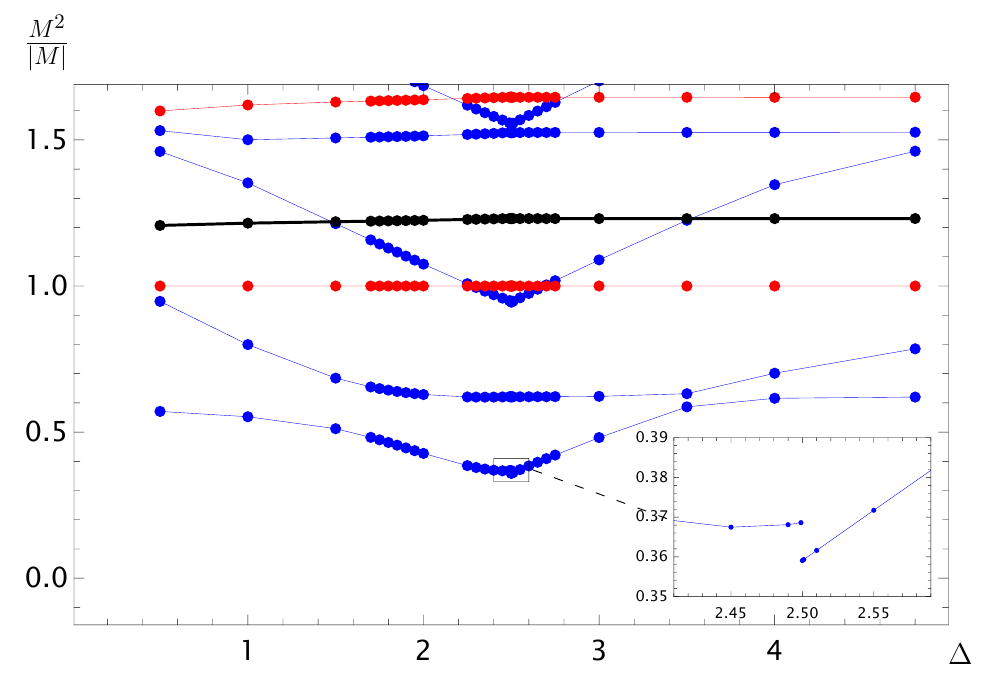}    
   \hfill
	\caption{ In the theory discussed in Sec.~\ref{Sec:BU-D6}, taken from Ref.~\cite{Elander:2022ebt}, the classical gravity background is obtained by compactifying on a circle a simple sigma-model of one scalar, $\phi$, with canonical kinetic term and potential compatible with a quadratic superpotential, coupled to gravity in $\hat{D}=6$ dimensions. 
	 The top-left panel shows the spectrum of masses, $M^2/|M|$, of gauge-invariant fluctuations computed along a one-parameter family of soliton solutions labelled by the asymptotic value of the scalar at the end of space, $\phi_I$. The example displayed assumes that the scaling dimension of the scalar operator dual to $\phi$ is half of the dimension of the space-time of the QFT, $\Delta=5/2$. The blue disks denote the spin-0 bound states, corresponding to the fluctuations of the two scalars retained in the reduction to $D=5$ dimensions, the red disks correspond to the tensor, spin-2 states, the black disks to the spin-1 states, and the orange disks represent the spectrum of scalars computed in the probe approximation, defined by neglecting mixing with the trace of the metric.  The masses are normalised so that the lightest spin-2 state has mass $M_2=1$. The vertical, dashed line denotes the position of the first-order phase transition.
	The top-right panel shows the free energy, $\hat{\cal F}$, normalised to the holographic scale, $\Lambda$, as a function of the deformation parameter $\hat{\phi}_{J}$ (itself expressed in units of $\Lambda$), again for the illustrative choices $\Delta=5/2$. We display the results for three distinct one-parameter families of solutions: the regular soliton solutions, corresponding to confining field theories (black, continuous line), and two classes of singular domain wall solutions (red and blue, respectively). 
	The bottom panel shows the spectrum of masses $M^2/|M|$ of gauge-invariant fluctuations computed along the one-parameter line of soliton solutions, by evaluating the masses at the critical value at which the transition takes place, $\phi_I(c)$, for each value of $\Delta$. The masses are normalised so that the lightest spin-2 state has mass $M_2=1$. The blue disks denote the spin-0 states associated with the two scalars, the black disks with the spin-1 states associated to the gravi-photon, while the red disks correspond to the spin-2 states associated with tensorial fluctuations of the metric.
    Figures taken from Ref.~\cite{Elander:2022ebt}.
  \label{fig:BU-D6}}
\end{figure}

The model proposed and analysed in Ref.~\cite{Elander:2022ebt} is a simple generalisation, constructed  in the bottom-up context, of the three examples of top-down models discussed in Sects.~\ref{Sec:D7}, \ref{Sec:D6}, and~\ref{Sec:D5}~\cite{Elander:2020fmv,Elander:2020ial,Elander:2021wkc}. In all three cases, the (higher-dimensional) sigma-model coupled to gravity consists of one scalar field,  and admits a trivial critical point (local maximum for the scalar potential), leading to background gravity solutions with AdS geometry.  The regular soliton solutions are obtained by assuming that one of the dimensions be a circle, which shrinks to zero size in the interior of the geometry. The sigma-model of scalars are non-trivial, and involve the deformation of the theory by the insertion of  a source for  one composite operator. In the regular soliton solutions the background values of the scalars evolve only over a limited range. In all three top-down cases, large classes of singular domain-wall solutions exist as well, and they are energetically favoured over the soliton ones for large values of the deformation. The three theories show the presence of a first-order phase transition. Along unstable branches of solutions a tachyon is present in the spectrum, which continuously turns into a parametrically light dilaton along the metastable branches of solutions. Along the energetically favoured branch of the soliton solutions, the dual, confining QFT has a spectrum of bound states in which the lightest scalar has approximately half the mass of the lightest tensor. This state is continuously connected with the dilaton, which  appears on the metastable branch. But it is not a particularly light state, and there is no clear indication that it behaves as a dilaton, by coupling to the trace of the energy-momentum tensor.

The purpose of building the bottom-up model Ref.~\cite{Elander:2022ebt}  is to test whether the results obtained in the three top-down theories in Refs.~\cite{Elander:2020fmv,Elander:2020ial,Elander:2021wkc} are of general validity, or rather due to the accidental fact that the dimension of the deforming coupling is, in all three cases, the same, namely $2$. The bottom-up model is hence constructed by fixing the space-time dimensionality of the gravity theory to $\hat{D}=6$, and coupling it to a single scalar, $\phi$, with canonical kinetic term and potential that can be derived from a quadratic superpotential. The potential admits a local maximum for $\phi=0$, which yields a stable solution with AdS$_6$ geometry. The coefficient appearing in the quadratic term of the superpotential is a free parameter, called $\Delta$ in this subsection, which may be identified with either the dimension of the source, if $\Delta<(\hat{D}-1)/2$ or the condensate, if $\Delta>(\hat{D}-1)/2$.

A one-parameter family of regular soliton solutions exists, for each choice of $\Delta>0$.  One of the dimensions is a circle, that shrinks to zero size at a finite value of the holographic direction, while the six-dimensional background is completely smooth and regular. These solutions can be interpreted, holographically,  as confining QFTs in $\hat{D}-2=4$ dimensions. There are also singular domain-wall solutions, in which the scalar field, $\phi$, evolves to large values in the deep IR of the geometry. The global stability analysis can be performed for all values of $\Delta$, in the range $0\leq \Delta \leq \hat{D}-1$, and shows a variety of new features and trends. But for small values of the deforming parameter, which can be identified either with the source term appearing in the UV expansion of the solutions for $\phi$, called $\hat{\phi}_J$ (measured in units of the scale $\Lambda$), or, equivalently, with the constant, $\phi_I$, appearing in the IR expansions, the results are similar to those that emerge in the three aforementioned top-down theories in Refs.~\cite{Elander:2020fmv,Elander:2020ial,Elander:2021wkc}.

The global stability analysis is exemplified by the top-right panel of Fig.~\ref{fig:BU-D6}. For illustrative purposes, we choose to show here $\Delta=5/2$ (half of the dimensions of the QFT defined at the conformal boundary, in the UV), while other choices are catalogued in Ref.~\cite{Elander:2022ebt}. For small values of the deforming parameter, $\hat{\phi}_J$, the regular soliton solutions are those with minimum value free energy density, $\hat{\mathcal F}$ (expressed in units of $\Lambda$). Metastable and unstable branches of soliton solutions exist as well. But singular domain wall solutions are energetically favoured when larger values of  $\hat{\phi}_J$ are considered. A first-order phase transition between the regular soliton and singular domain-wall ones is present, for all choices of $\Delta$.

The local stability analysis, conducted along the branch of soliton solutions by computing the spectrum of fluctuations, is exemplified in the top-left panel of Fig.~\ref{fig:BU-D6}. Along the stable, physically realised branch of soliton solutions, before the phase transition appears, the lightest state is a scalar, but it is not parametrically light, and its mass is well captured by the probe approximation, suggesting that its coupling to the trace of the energy-momentum tensor is suppressed. This state becomes lighter, eventually massless, after entering the metastable branch of soliton solutions, and, furthermore, in this case the probe approximation fails, showing that this state, when light, can be interpreted as a dilaton. Eventually, the scalar becomes tachyonic, along the unphysical, unstable branch of soliton solutions.

These features are general, for all values of $\Delta$, and corroborate the findings of the three top-down models in Refs.~\cite{Elander:2020fmv,Elander:2020ial,Elander:2021wkc}. Repeating the same process by scanning over the allowed values of $\Delta$ allows one to assess the magnitude of the suppression of the mass of the lightest scalar, in units of the mass of the lightest spin-2 bound state, along the stable, physical realised branch of soliton solutions. We illustrate the result in the bottom panel of Fig.~\ref{fig:BU-D6}. The spectra of spin-0, spin-1, and spin-2 states are evaluated at the exact transition point, for each value of $\Delta$. What the figure shows is that there is some modest suppression of the mass of the lightest state when $\Delta=5/2$, which is the case in which the dual composite operator has the same dimensionality as its coupling. Interestingly, also the next-to-lightest scalar is suppressed in this range. But both effects are modest,  neither numerically nor  parametrically enhanced. 

The case in which $\Delta=(\hat{D}-1)/2$, half of the dimension of the dual field theory, coincides with the scalar having (negative) mass squared that saturates the Breitenlohner-Freedman (BF) bound~\cite{Breitenlohner:1982jf}. The main conclusion of the exercise is hence the following: while it is true that, when approaching the value of $\Delta$ associated with the BF bound, the lightest scalar bound state of the dual, confining  theory exhibits some degree of mass suppression, the effect is small, and becomes parametric only along a metastable branch of solutions, beyond a first-order phase transition. The bottom-up model confirms the suggestions in this kind, coming from top-down models, while also highlighting how the dimensionality of the deformation does not play a central role in making a light dilaton appear in the spectrum of bound states. This conclusion does not exclude that interesting physics might appear in the case of other holographic mechanisms to describe confinement. The model does  not have anything to teach us about the cases in which a weaker (second-order) phase transition appears, which is not captured by this bottom-up construction.

\subsection{Bottom-up holographic confinement with magnetic fluxes}
\label{Sec:BU-flux}

\begin{figure}[t]
\centering  
\includegraphics[width=0.4\textwidth]{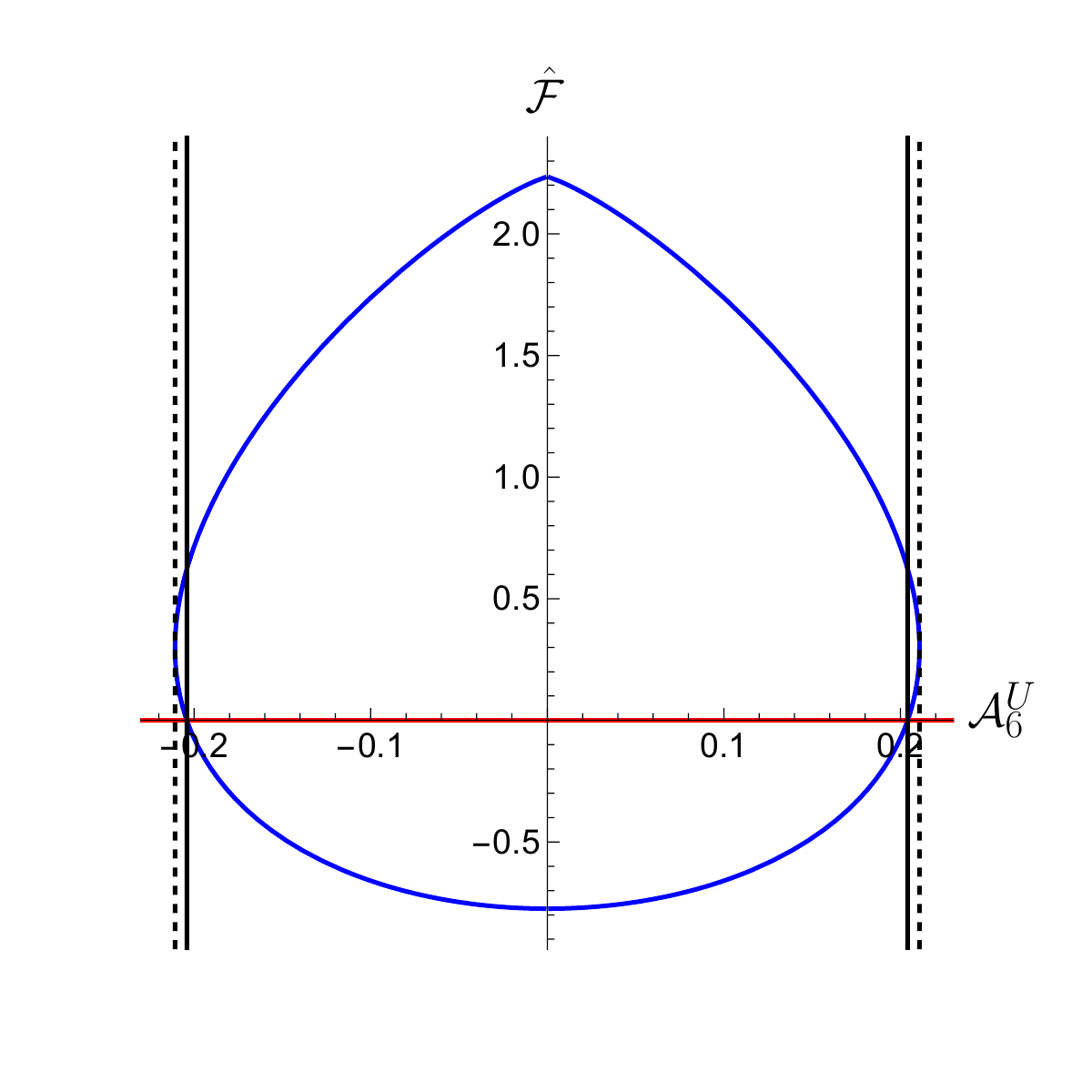}    
\includegraphics[width=0.4\textwidth]{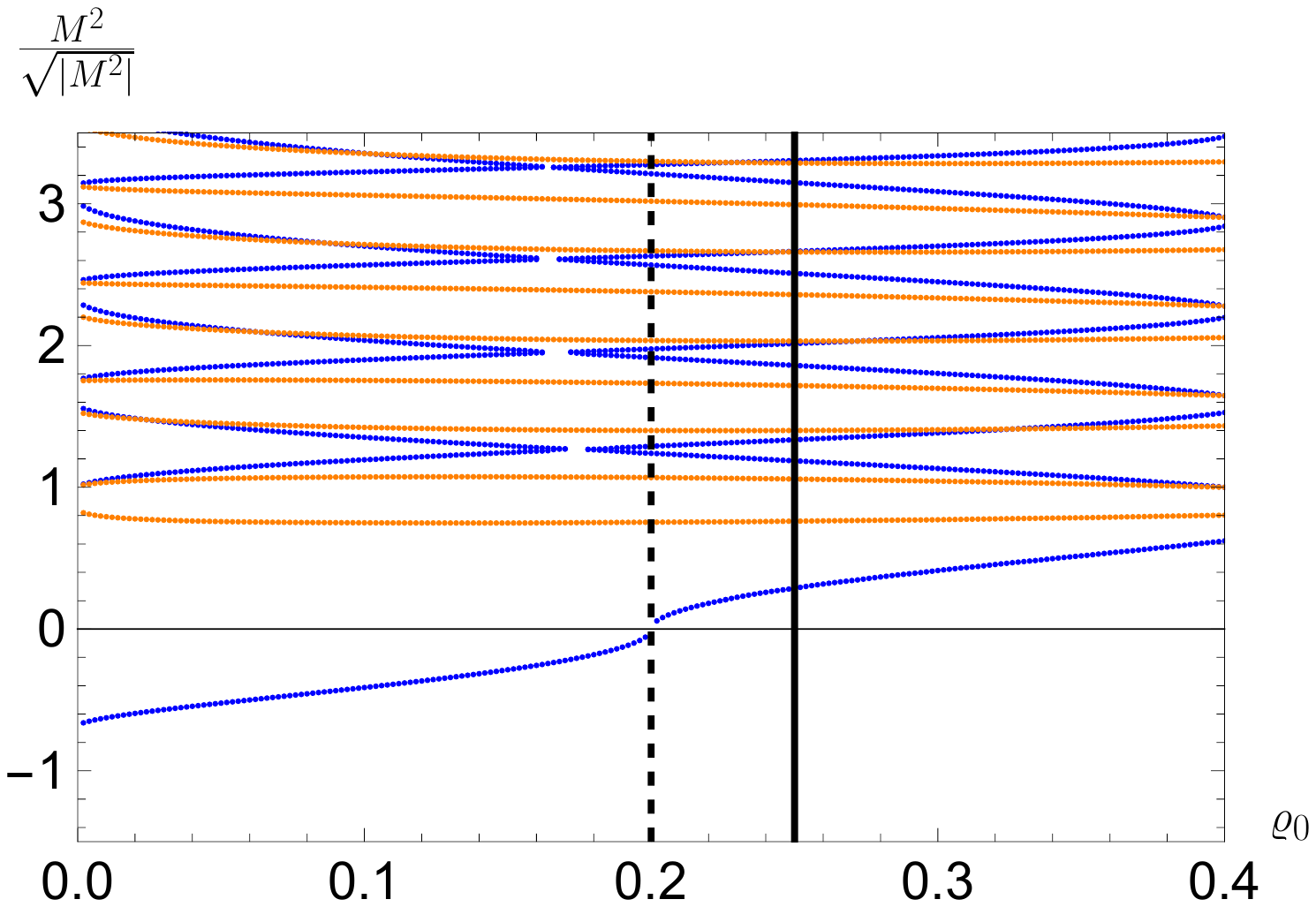}    
   \hfill
	\caption{ In the theory discussed in Sec.~\ref{Sec:BU-flux}, the classical gravity background is obtained by compactifying on a circle a $U(1)$ gauge theory coupled to gravity in $\hat{D}=6$ dimensions, in the presence of a negative cosmological constant. A magnetic flux is controlled by the non-trivial profile of the ${\cal A}_6$ component of the gauge field along the compact dimension. 
	The left panel shows the free energy $\hat{\cal F}$, normalised to the holographic scale, $\Lambda$, as a function of the deformation parameter ${\cal A}_6^U$  for two distinct one-parameter families of solutions.  The regular soliton solutions correspond to confining field theories (blue continuous  line). The domain wall solutions (red line) correspond to the deconfined phase. The vertical lines denote the phase transition, and the vertical dashed line the largest values of $\left|{\cal A}_6^U\right|$  for which soliton solutions exist. The right panel shows the spectrum of masses, $M^2/|M|$, of gauge-invariant fluctuations computed along the one-parameter family of soliton solutions, labelled here by the value of the parameter determining the end of space, $\varrho_0$. The blue disks denote the spin-0 states, corresponding to the fluctuations of the two scalars retained in the reduction to $D=5$ dimensions, the orange disks represent the spectrum of scalars computed in the probe approximation, defined by neglecting mixing with the trace of the metric.  The vertical lines have the same meaning as in the left panel. The masses are normalised so that the lightest spin-2 state has mass $M_2=1$. Figures taken from Ref.~\cite{Fatemiabhari:2024lct}.
  \label{fig:BU-flux}}
\end{figure}

The bottom-up holographic model in Ref.~\cite{Fatemiabhari:2024lct} is a simplified generalisation that captures some of the salient features of the  top-down models discussed in Sects.~\ref{Sec:Flux-D7} and~\ref{Sec:Flux-D6}~\cite{Piai:2026rst,Fatemiabhari:2026rju}. It consists of a gravity theory in $\hat{D}=6$ dimensions, in the presence of a negative cosmological constant and coupled to a $U(1)$ gauge field with canonical action. One of the two classes of solutions of interest are regular domain wall ones in which the gauge field is constant, with ${\cal A}_{\hat{M}}=(0,\,0,\,0,\,0,\,0,\,{\cal A}_{6}^U)$, and the background geometry is AdS$_6$. Conversely, in the soliton solutions, the sixth dimension is a circle that shrinks in the interior of the geometry. It is convenient to parametrise the holographic direction with a new variable, $\varrho$. In this coordinate system,  the space ends at $\varrho\rightarrow \varrho_0$. The last component of the gauge field has a non-trivial profile, that reaches a constant  ${\cal A}_{6}\rightarrow {\cal A}_{6}^U$ asymptotically, but vanishes exactly at the end of space, which is the point at which the circle shrinks.

The global stability analysis is illustrated by the left panel of Fig.~\ref{fig:BU-flux}. The free energy density, $\hat{\cal F}$, expressed in units of the holographic scale, $\Lambda$,  is studied as a function of the dimensionless parameter ${\cal A}_{6}^U$, that controls the magnetic flux. For the domain-wall solutions, which exist for any real value of ${\cal A}_{6}^U$, the free energy vanishes identically. Conversely, the soliton solutions exist only for $\left|{\cal A}_{6}^U\right|<{\cal A}_{6}^U({\rm max})$. Furthermore, for each such value of ${\cal A}_{6}^U$ there are two possible solutions, both regular, that coincide for $\left|{\cal A}_{6}^U\right|={\cal A}_{6}^U({\rm max})$, one of which has positive definite free energy, and is hence always energetically disfavoured. The soliton solutions are energetically favoured for small ${\cal A}_{6}^U({\rm max})$, but a phase transition appears when $\varrho_0=1/4$, past which the AdS$_6$, domain-wall solutions are favoured. There is hence one branch of stable, one of metastable, and one of unstable solutions, among the soliton ones.

The local stability analysis is carried out by studying the spectrum of fluctuations of the soliton solutions, which can be interpreted as bound states of the dual, confining QFT. The main results are shown in the right panel of Fig.~\ref{fig:BU-flux}. The masses, $M^2/|M|$, of the bound states are expressed in units of the mass of the lightest spin-2 bound state, as shown as a function of $\varrho_0$. For large $\varrho_0$, the soliton solutions are stable and favored energetically. In this case, the lightest scalar state has a mass significantly lower than the rest of the spectrum. A parametric suppression of the mass appears past the transition, along the metastable branch. Eventually, this state becomes a tachyon. Interestingly, the probe approximation fails to capture well the mass of this lightest scalar across the whole parameter space, suggesting that even along the stable branch the lightest scalar is, at least approximately, a dilaton, with mass numerically suppressed.

In summary, there are similarities as well as differences with the top-down models in Refs.~\cite{Piai:2026rst,Fatemiabhari:2026rju}. All the branches of solutions, domain-wall and soliton alike, are regular.  A  tachyon appears far away from the phase transition, in parameter space. The lightest scalar state is a dilaton also along the stable branch, not just along the metastable one, which results in some numerical suppression of its mass even in the physically realised part of the phase diagram.  

\subsection{Parametrically light dilaton near critical points in bottom-up holography} 
\label{Sec:BU-2}

\begin{figure}[t]
\centering  
\includegraphics[width=0.48\textwidth]{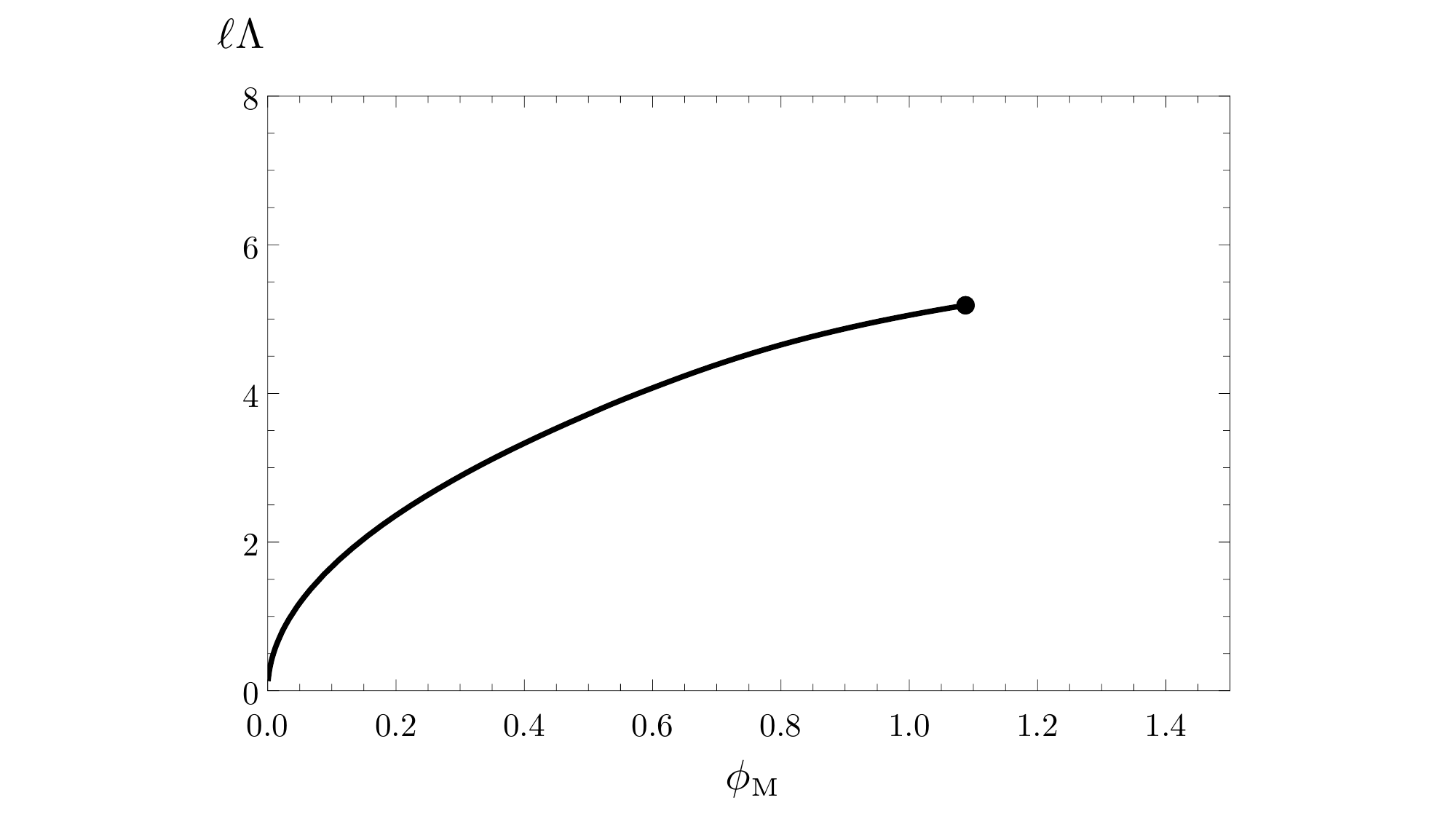}    
\includegraphics[width=0.48\textwidth]{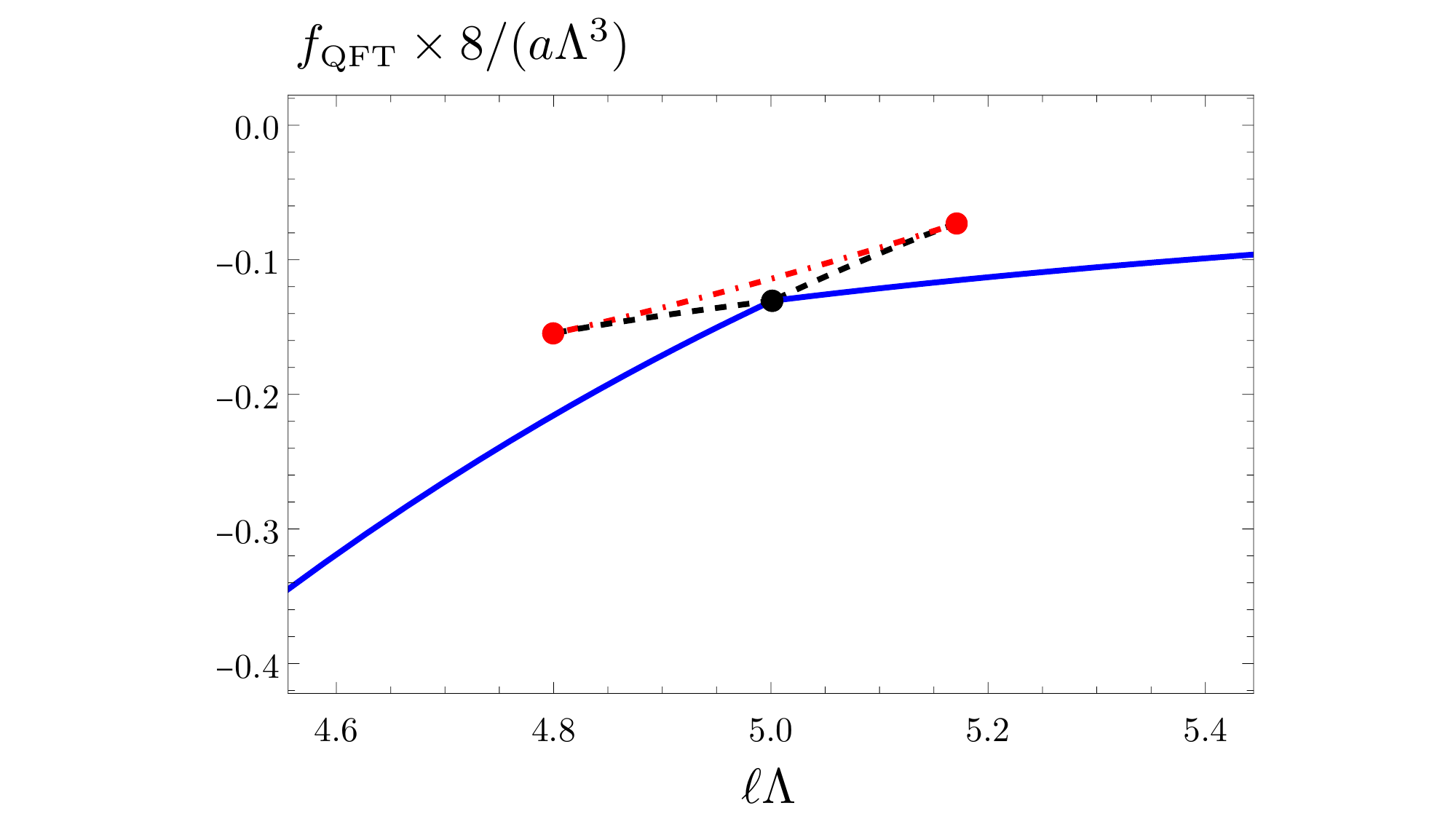}    
   \hfill
	\caption{In the theory discussed in Sec.~\ref{Sec:BU-2}, called Model B in Ref.~\cite{Faedo:2024zib},  the classical gravity background is obtained from dimensional reduction on a circle of a bottom-up sigma-model consisting of one scalar coupled to gravity in $\hat{D}=5$ dimensions. The scalar has potential derived from the  superpotential provided in Ref.~\cite{Bea:2018whf}. In the analysis in Ref.~\cite{Faedo:2024zib}, one of the free parameters is held fixed, $\phi_{Q}=10$,  while a second parameter, $\phi_M$, is allowed to vary.	 The resulting gravity theory in $D=4$ dimensions is described by two real scalars coupled to gravity.	The left panel shows the phase diagram of the theory, described by the dimensionless parameters $ \phi_M$ and $\ell\Lambda$, with a line of phase transitions ending at a critical point, defined by  $(\phi_M^c,\ell^c\Lambda) \simeq (1.088,5.186)$. 
	The right panel shows the free energy density expressed in dimensionless units, $f_{\rm QFT}/(a \Lambda^3)$, as a function of the dimensionless parameter $\ell
	\Lambda$, for a representative value of $\phi_M=0.97$, highlighting the swallow-tail behaviour appearing in proximity of one of the first-order  phase transition points.
		Figures taken from Ref.~\cite{Faedo:2024zib}.  \label{fig:BU-2-a}}
\end{figure}

Finding examples in holography of theories that exhibit a line of zero-temperature, first-order phase transitions that end at a critical point is a non-trivial problem. As briefly discussed in Sect.~\ref{Sec:TD-2}, the supergravity theories that yield such a behaviour are rare, hard to come by, and difficult to work with~\cite{Elander:2025fpk}. The first such example was exposed in a bottom-up background described as Model B in Ref.~\cite{Faedo:2024zib}, built starting from the  bottom-up constructions that had been previously studied, for completely different purposes, in Refs.~\cite{Bea:2020ees,Bea:2021ieq,Ares:2020lbt,Bea:2021zol,Bea:2021zsu,Bea:2022mfb,Escriva:2022yaf}. The gravity theory consists of one scalar coupled to gravity in $\hat{D}=5$ dimensions, with superpotential written as a polynomial of degree six, taken from Ref.~\cite{Bea:2018whf}. One dimension is compactified on a circle, and a two-parameter family of regular soliton solutions, in which the internal space of the geometry closes smoothly, are studied, as a function of the dimensionless parameters $\phi_M$ and $\ell \Lambda$. It should be noted that the scale  $\Lambda$ is not the one defined in Eq.~(\ref{Eq:Lambda}), but provides an alternative definition of the characteristic energy scale of the theory---for details, see the original literature.

The details of the dynamics and of the backgrounds can be found in the literature, and are not essential to this part of the presentation. The salient aspect is the phase diagram,  reproduced in the left panel of Fig.~\ref{fig:BU-2-a}. A line of first-order phase transitions terminates at a critical point. As can be seen in the right panel of the figure, in the proximity of the line the phenomenology is qualitatively the same as for the Van der Waals gas discussed in Sect.~\ref{Sec:VdW}. One stable and one metastable branch of solutions compete energetically to minimise the free energy. The minimum of the free energy density is a continuous function, but at the transition, as one branch takes over as dominant from the other, one finds that the first derivative is discontinuous. This is accompanied by the emergent phenomena associated with phase coexistence. There is also a third branch, that has larger free energy, exists only for a finite range of the control parameter, $\ell \Lambda$, and is never physically realised.

\begin{figure}[t]
\centering  
\includegraphics[width=0.49\textwidth]{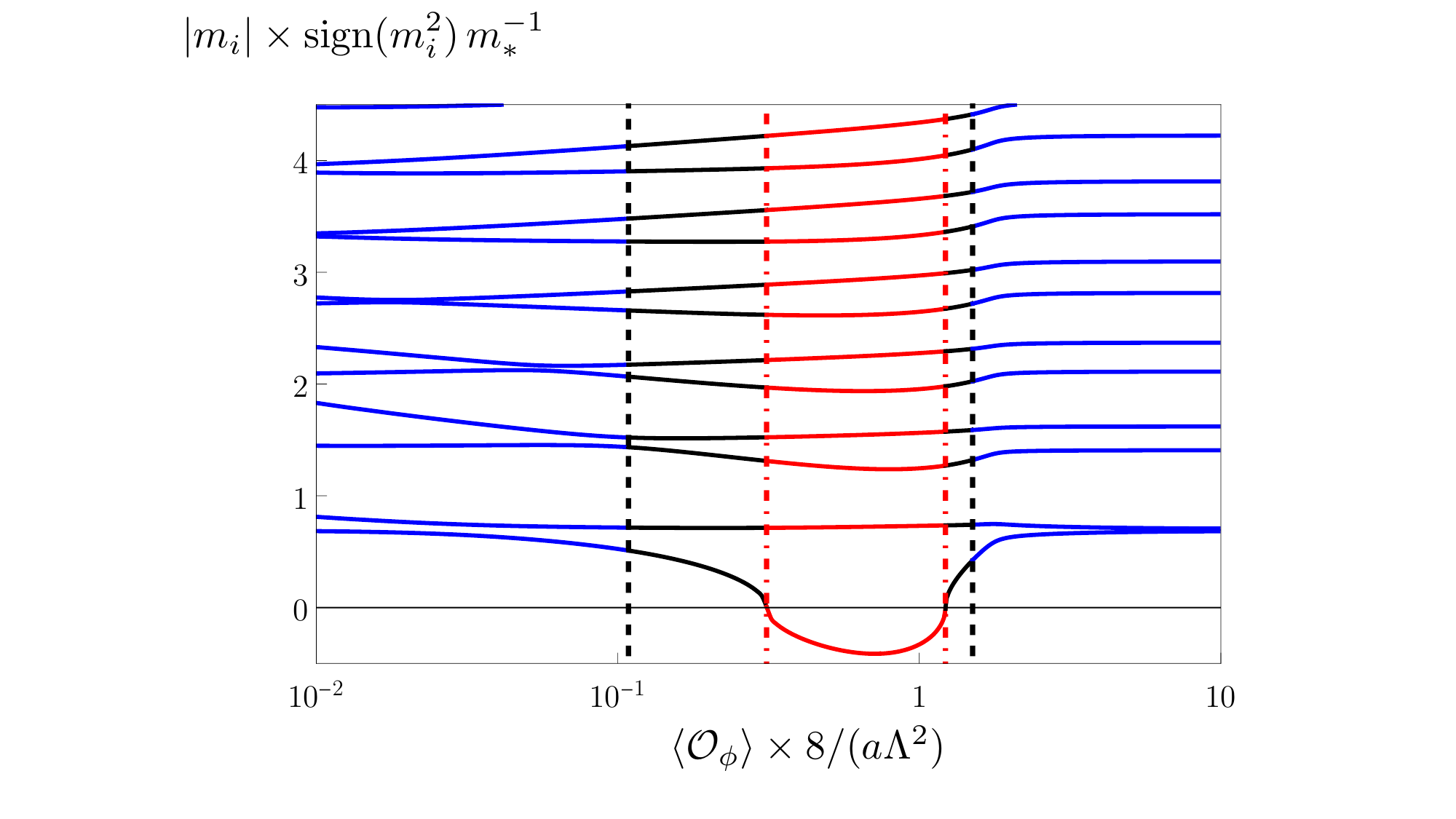}    
\includegraphics[width=0.49\textwidth]{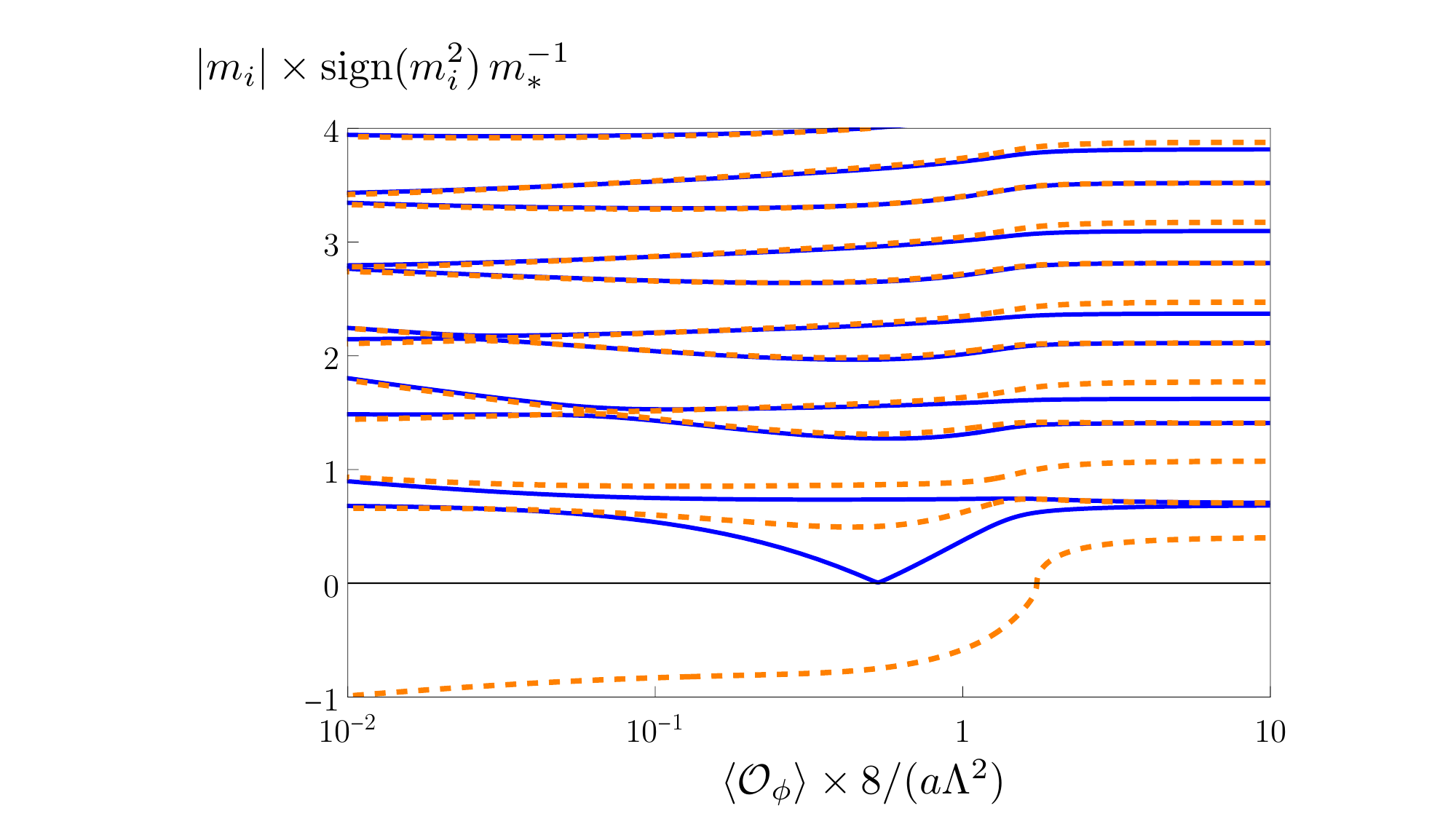}    
   \hfill
	\caption{ 
	As in Fig.~\ref{fig:BU-2-a}, study of the theory discussed in Sec.~\ref{Sec:BU-2}, called Model B in Ref.~\cite{Faedo:2024zib}. 
The left panel shows the mass spectrum of scalars, $m$, expressed in units of the mass of the lightest spin-2 bound state, called $m_{\ast}$ in Ref.~\cite{Faedo:2024zib}, plotted for the same choices of parameter, $\phi_{Q}=10$ and $\phi_M=0.97$, as  in 
the right panel of Fig.~\ref{fig:BU-2-a}, as a function of the response function, $\langle {\cal O}_{\phi} \rangle$. The vertical black dashed lines represent the phase transition, and the vertical red dot-dashed line the appearance of a tachyon in the spectrum, corresponding to the boundaries of the unstable branch of solutions.
	The right panel shows a comparison of the mass of the scalar fluctuations (blue, solid lines) with the results of the probe approximation (orange, dashed lines), for the choice of  $\phi_{Q}=10$ and $\phi_M=\phi_M^c$ at criticality, as a function of the response function, $\langle {\cal O}_{\phi} \rangle$. The minimum of the blue line coincides with the  position of the second-order phase transition.
	Figures taken from Ref.~\cite{Faedo:2024zib}.  \label{fig:BU-2-b}}
\end{figure}

The study of the fluctuations of the theory, exemplified by the left panel of Fig.~\ref{fig:BU-2-b}, shows that along the third branch of solutions described above, a tachyon is present, but this appears to be in a region of parameter space that is removed from the position of the transition. As a result, the spectrum of bound states of the dual, confining field theory, computed for generic values of the parameters, but in the proximity of the line of  first-order transitions, and along the physically realised branches of solutions, is not particularly striking, showing towers of states with masses characterised by the typical scale of the theory, in broad agreement with results from the top-down models discussed in the previous section.

Most interesting is what happens near the critical point, at the end of the line of first-order phase transitions. The right panel of Fig.~\ref{fig:BU-2-b} shows the results for the spectrum of bound states, for a one-parameter subclass of solutions tuned to go through the critical point. A massless scalar state is present at this point, and this scalar acquires a mass moving away from criticality. The probe approximation fails completely to capture the behaviour of this lightest of scalar bound states, but rather yields (incorrectly) a tachyon. This is clear evidence that the light state is a dilaton, as the probe approximation works well for all other states, which happen to not appreciably couple to the trace of the energy-momentum tensor.

Historically, this was the first example demonstrating in holographic models the emergence of a light dilaton in the physical spectrum of a confining theory in the region of parameter space in proximity of a second-order (or weak) phase transition, realising the scenario suggested in the closing remarks of Ref.~\cite{Elander:2021wkc}. This holographic model is not complete, emerging in the context of bottom-up holography. Its non-trivial phase diagram depends on the rather rich structure of the sigma-model action, which has no known microscopic origin. Also, the phenomena connected with confinement are taking place in lower dimensions, again in a context that is reminiscent of QCD$_3$, rather than a four-dimensional QFT. Nevertheless, the findings exposed in this section, and reinforced by the qualitatively identical results found in the top-down model summarised in Sect.~\ref{Sec:TD-2} and Ref.~\cite{Elander:2025fpk}, are clear enough that it is reasonable to expect them to hold in general, beyond the reach of lower-dimensional holography. It would be interesting to construct holographic descriptions of confinement of a four-dimensional QFT, that exhibit a line of first-order phase transitions ending at a critical point, but this has not been found yet. Encouraging  steps have been made towards realising this scenario in other approaches that can tackle the calculability challenges  presented by strongly-coupled, confining gauge theories. For example, see the discussions in the lattice field theory studies in Refs.~\cite{Lucini:2013wsa,Bennett:2022yfa}, and in the lower-dimensional field-theoretical study presented in Ref.~\cite{Cresswell-Hogg:2025kvr}.

\section{Summary and Outlook}
\label{Sec:Outlook}

Table~\ref{tbl1} summarises schematically the main results of the survey of theories and models that we discussed in Sects.~\ref{Sec:Top} and~\ref{Sec:Bottom}. The use of holography as a calculation technique has provided evidence that if there is a line of first-order phase transitions in the parameter space of the confining QFT, and this line ends in a critical point, then in close proximity of the critical point the spectrum of bound states contains a light scalar. The coupling of this scalar to the energy-momentum tensor is induced via mixing of the relevant fluctuation in the gravity dual with the fluctuation of the trace of the metric, and reproduces the expectations for a dilaton, so that it is legitimate to identify this scalar as such. Its mass can be dialled to be arbitrarily light, by choosing the control parameters closer to the critical point. This has been proven explicitly both in top-down and a bottom-up holographic constructions, though in both cases the space-time dimension of the dual confining QFT is lower than four.

This survey can be thought of as a mid-term status update for a long-term research programme that is still ongoing.
Despite its encouraging results, it also leaves several open questions. We list below, in no particular order,
an incomplete selection of such open challenges, that we leave to future research.

\begin{itemize}

\item Is the mechanism for the stabilisation of the mass of the holographic dilaton based on the vicinity of second-order phase transitions viable in any numbers of dimensions, in particular in four?

\item Does the stabilisation of the mass of the holographic dilaton change its nature depending on the features of the holographic, geometric mechanism that describes confinement?

\item Are there first-order transitions that are weak enough to lead to a strong suppression of the holographic dilaton mass, without invoking the presence of second-order transitions?

\item Are there realisations of this mechanism that involve simpler top-down holographic models than those discussed in Sect.~\ref{Sec:TD-2}?

\item Are there realistic implementations of these ideas, that can, for example, address the hierarchy problem of the Higgs and electroweak sector of the standard model of particle physics?

\item Are there gauge theories amenable to numerical study on the lattice that exhibit the same type of behaviour as the holographic theories discussed here?

\end{itemize}

\begin{table*}[t]
\caption{Summary of the examples discussed in the main body of the review. For each one, we identify the theory in which the gravity backgrounds of interest emerge, the type and order of the transition (focusing on the most interesting regime of parameter space, for the physics of the dilaton), and the best case scenario in which a dilaton is present, with suppressed mass. We also provide the relevant references, in this review and to the literature. }\label{tbl1}
\begin{tabular}{|c|c|c|c|c|c|}
\hline
\hline
Theory  & Background &  Transition & Dilaton & Ref. \\ 
\hline
Supergravity, $\hat{D}=7$ & Soliton / Singular DW& First-order& Metastable branch & Sect.~\ref{Sec:D7}~\cite{Elander:2020fmv}\\
Supergravity, $\hat{D}=6$ &Soliton / Singular DW& First-order& Metastable branch  & Sect.~\ref{Sec:D6}~\cite{Elander:2020ial}\\
Supergravity, $\hat{D}=5$ &Soliton / Singular DW& First-order& Metastable branch & Sect.~\ref{Sec:D5}~\cite{Elander:2021wkc}\\
Supergravity, $\hat{D}=7$ &Soliton with flux /  AdS& First-order& Numerical suppression & Sect.~\ref{Sec:Flux-D7}~\cite{Piai:2026rst}\\
Supergravity, $\hat{D}=6$ &Soliton with flux / AdS& First-order& Numerical suppression, doubling & Sect.~\ref{Sec:Flux-D6}~\cite{Fatemiabhari:2026rju}\\
Supergravity, $\hat{D}=4$ & Soliton / Soliton & Critical point & Parametric suppression & Sect.~\ref{Sec:TD-2}~\cite{Elander:2025fpk}\\
Bottom-up, $\hat{D}=6$ & Soliton / Singular DW& First-order& Metastable branch  & Sect.~\ref{Sec:BU-D6}~\cite{Elander:2022ebt}\\
Bottom-up, $\hat{D}=6$ &Soliton with flux /  AdS& First-order& Numerical suppression& Sect.~\ref{Sec:BU-flux}~\cite{Fatemiabhari:2024lct}\\
Bottom-up, $\hat{D}=5$ &Soliton / Soliton & Critical point & Parametric suppression & Sect.~\ref{Sec:BU-2}~\cite{Faedo:2024zib}\\
 \hline
 \hline
\end{tabular}
\end{table*}

{\bf Acknowledgments}

We would like to thank the editors, Giacomo Cacciapaglia, Nick Evans, Ethan Neil, Francesco Sannino, and Kimmo Tuominen, for the invitation to submit this contribution, Roman Zwicky for useful discussions and email exchanges, as well as Anton Faedo, Ali Fatemiabhari, Carlos Hoyos, Carlos Nunez, Ronnie Rodgers, James Rucinski, John Roughley, and Javier Subils, for many valuable discussions that have helped shape the ideas summarised in this review.

The work  of MP has been supported by the STFC Consolidated Grant No. ST/T000813/1, ST/X000648/1,  and UKRI4243. MP received funding from the European Research Council (ERC) under the European Union's Horizon 2020 research and innovation program under Grant Agreement No.~813942. 

{\bf Research Data Access Statement}---No new data has been generated for the purposes of this work.

{\bf Open Access Statement}---For the purpose of open access, the authors have applied a Creative Commons  Attribution (CC BY) licence  to any Author Accepted Manuscript version arising.

\bibliographystyle{JHEP} 

\bibliography{DilatonReview}

\end{document}